\documentclass[%
 reprint,
 superscriptaddress,
 amsmath,amssymb,
 aps,
]{revtex4-2}

\usepackage{tikz}
\usetikzlibrary{quantikz}
\usepackage{graphicx}
\usepackage{dcolumn}
\usepackage{bm}
\usepackage[colorlinks=true,linkcolor=blue,urlcolor=black,citecolor=blue,bookmarksopen=true]{hyperref}
\usepackage{amsmath,physics}
\usepackage{xcolor}
\usepackage[caption=false]{subfig}

\usepackage{float}

\begin{document}

\preprint{APS/123-QED}

\title{Quantum self-consistent equation-of-motion method for computing molecular excitation energies, ionization potentials, and electron affinities on a quantum computer}

\author{Ayush Asthana}
\email{aasthana@vt.edu}
\affiliation{Department of Chemistry, Virginia Tech, Blacksburg, VA 24061, USA}
\author{Ashutosh Kumar}%
\email{akumar1@lanl.gov}
\affiliation{Theoretical Division, Los Alamos National Laboratory, Los Alamos, NM 87545, USA}
\author{Vibin Abraham}
\affiliation{Department of Chemistry, University of Michigan, Ann Arbor, MI 48109, USA}
 \author{Harper Grimsley}
\affiliation{Department of Chemistry, Virginia Tech, Blacksburg, VA 24061, USA}
 \author{Yu Zhang}
\affiliation{Theoretical Division, Los Alamos National Laboratory, Los Alamos, NM 87545, USA}
\author{Lukasz Cincio}
\affiliation{Theoretical Division, Los Alamos National Laboratory, Los Alamos, NM 87545, USA}
\author{Sergei Tretiak}
\affiliation{Theoretical Division, Los Alamos National Laboratory, Los Alamos, NM 87545, USA}
\affiliation{Center for Integrated Nanotechnologies, Los Alamos National Laboratory, Los Alamos, NM 87545, USA}
\author{Pavel A. Dub}
\affiliation{Chemistry Division, Los Alamos National Laboratory, Los Alamos, NM 87545, USA}
 \author{Sophia E. Economou}
 \affiliation{Department of Physics, Virginia Tech, Blacksburg, VA 24061, USA}
  \author{Edwin Barnes}
  \affiliation{Department of Physics, Virginia Tech, Blacksburg, VA 24061, USA}
 \author{Nicholas J. Mayhall}
 \email{nmayhall@vt.edu}
\affiliation{Department of Chemistry, Virginia Tech, Blacksburg, VA 24061, USA}


\begin{abstract}
Near-term quantum computers are expected to facilitate material and chemical research through accurate molecular simulations. 
Several developments have already shown that accurate ground-state energies for small molecules can be evaluated on present-day quantum devices.
Although electronically excited states play a vital role in chemical processes and applications, the search for a reliable and practical approach for routine excited-state calculations on near-term quantum devices is ongoing. 
Inspired by excited-state methods developed for the unitary coupled-cluster theory in quantum chemistry, 
    we present an equation-of-motion-based method to compute excitation energies following the variational quantum eigensolver algorithm for ground-state calculations on a quantum computer. 
We perform numerical simulations on H$_2$, H$_4$, H$_2$O, and LiH molecules to test our quantum self-consistent equation-of-motion (q-sc-EOM) method and compare it to other current state-of-the-art methods.
q-sc-EOM makes use of self-consistent operators to satisfy the vacuum annihilation condition, a critical property for accurate calculations. 
It provides real and size-intensive energy differences corresponding to vertical excitation energies, ionization potentials and electron affinities.
 We also find that q-sc-EOM is more suitable for implementation on NISQ devices as it is expected to be more resilient to noise compared with the currently available methods.
\end{abstract}

\maketitle

\section{Introduction}

Quantum chemistry is expected to be one of the first areas which can have demonstrable quantum advantage in the near term~\cite{tilly2021variational,mcardle2020quantum,cerezo2021variational,peruzzo2014variational,magann2021pulses,kandala2017hardware,cao2019quantum,fedorov2022vqe,magann2021pulses}.
This is owed to the fact that the computational effort required for exact evaluation of electron correlation on a classical computer-whose accurate calculation is essential for a reliable comparison with experimental values-scales factorially with the number of molecular orbitals. 
This unfavourable scaling is expected to reduce drastically when wavefunctions are instead prepared on quantum devices.

For estimation of molecular ground-state properties on noisy intermediate-scale quantum (NISQ) era devices, variational quantum eigensolver (VQE) based algorithms have gained popularity due to their relatively low circuit depth and resilience to noise~\cite{Preskill2018,sharma2020noise}.
This has led to a series of successful demonstrations involving the computation of molecular ground-state energies of small molecules on present-day quantum devices and simulators~\cite{grimsley2019adaptive,tang2021qubit,peruzzo2014variational,kandala2017hardware,huggins2020non,lee2018generalized,ryabinkin2018qubit,o2016scalable,nam2020ground,mccaskey2019quantum,hempel2018quantum,gao2021applications,smart2021quantum,Meitei2020}. 
However, estimation of just the molecular ground-state energy is not sufficient for describing many interesting chemical processes that involve electronic excitations in some form~\cite{nelson2020non}. 
For example, accurate modelling of chemical phenomena such as photochemical reactions, catalytic processes involving transition metal complexes, photosynthesis, solar cell operation, etc. require an accurate simulation of both molecular ground and excited states.
The electronically excited states of such systems are generally strongly correlated and hence, require 
the use of sophisticated quantum chemical theories for their accurate description. A number of methods have been developed in this regard in the last few decades~\cite{Szalay2012, dreuw2015algebraic, Stanton1993,serrano2005quantum,sneskov2012excited,mayhall2014quasidegenerate,nooijen1997new, hemmatiyan2018excited, mazziotti2003extraction}. 
and widely used through several software packages~\cite{cfour,Pyscf}.
The equation-of-motion coupled-cluster (EOM-CC)~\cite{Stanton1993} approach, originally developed by Stanton and Bartlett, is a popular example that is routinely used to calculate molecular excited-state properties such as excitation energies and transition dipole moments~\cite{Vidal2019,goings2014assessment,andersen2022probing, Asthana2019,Halbert2021}. 
EOM-CC has also been extended to calculate energies required to add or remove electrons from the ground-state electronic configuration~\cite{stanton1994analytic,nooijen1995equation,stanton1995perturbative,pieniazek2008charge,liu2020mapping}. 
For example,  IP-EOM-CC~\cite{stanton1994analytic,stanton1995perturbative} and EA-EOM-CC~\cite{nooijen1995equation} approaches have been developed which can compute accurate vertical ionization potentials (IPs) and vertical electron affinities (EAs), respectively.
IPs/EAs are defined as the difference in energy between the ground state and the states obtained by a single electron detachment/attachment process.
Some of the other advantages associated with the EOM-CC formalism are its theoretical rigour, the accuracy and correct scaling behavior of energy differences computed, and the ability to systematically improve the results. 
However, standard quantum chemistry methods like EOM-CC sometimes face challenges in a quantitative determination of excited states and their properties, notably for same-symmetry conical intersections~\cite{thomas2021complex,kohn2007can,yarkony2012nonadiabatic,bernardi1997role} and when the ground state has a prominent multi-reference character~\cite{schmidt1998construction,kohn2013state,mahapatra1998state,roos1996multiconfigurational}.
Since VQE algorithms are expected to provide accurate ground-state wavefunctions, even in the case of strongly correlated systems,
NISQ era devices can help address these challenging problems with practical computational expenses.

We would like to note that methods for the estimation of molecular excited states on a quantum computer based on other popular quantum algorithms have also been proposed. A number of approaches are based on quantum phase estimation algorithm~\cite{kitaev1997quantum,nielsen2002quantum,russo2021evaluating,bauman2020toward,sugisaki2021quantum,santagati2018witnessing,o2019quantum} with new developments for efficient implementation on quantum computers~\cite{sugisaki2021bayesian,sugisaki2022bayesian,poulin2018quantum}. 
Methods based on Krylov subspace diagonalization~\cite{motta2020determining,parrish2019quantum} and quantum annealing~\cite{teplukhin2021computing,sugisaki2022adiabatic} have also been proposed. 
While these methods are theoretically exact (in absence of any noise) and expected to provide a significant computational advantage over exact treatment on a classical computer, they will mostly be useful in fault-tolerant quantum computing and not suitable for NISQ era quantum computers due to their high quantum resource requirements and low tolerance to noise.

Significant effort has been made in developing methods for the calculation of molecular excitation energies within the framework of VQE in the last few years.
These techniques can be broadly classified into circuit optimization and diagonalization-based approaches. 
In the former approach, optimal parametrized circuits are obtained for every excited state, usually by minimizing a cost function involving energies of one or multiple excited states. 
Subspace-search VQE (SS-VQE)~\cite{nakanishi2019subspace}, orthogonal state reduction variational eigensolver (OSRVE)~\cite{xie2022orthogonal}, variational quantum deflation (VQD)~\cite{chan2021molecular, higgott2019variational} and the folded spectrum method~\cite{peruzzo2014variational} are some examples. 
These approaches, however, generally require increased quantum resources, specifically the gate depth. This makes them challenging for near term applications. 
Moreover, there is no guarantee for them to find the entire spectrum when the states are close in energy to one another. 
On the other hand, the diagonalization-based approaches use a classical computer to diagonalize the Hamiltonian in a subspace and can provide several excited states simultaneously. 
In this regard, methods like the Quantum Krylov subspace expansion~\cite{motta2020determining,qdavidson,GarnetNP2020}, the Quantum subspace expansion (QSE)~\cite{colless2018computation,mcclean2020decoding,Takeshita2020,mcclean2017hybrid}, and the quantum equation-of-motion (qEOM)~\cite{qEOM2020} have been developed recently.   
QSE has had significant success in the last few years and has also been extended to capture the missing correlation from large virtual orbital spaces~\cite{Takeshita2020,urbanek2020chemistry}. 
However, it requires an estimate of higher than 2-body reduced density matrices (RDMs), 
prompting the use of cumulant approximations~\cite{Takeshita2020} inspired by developments in quantum chemistry~\cite{mazziotti1998approximate,hemmatiyan2018excited, mazziotti2003extraction}. 
Furthermore, a significant drawback of the QSE approach is the lack of size-intensivity of the computed excitation energies. 
The property of size-intensivity ensures correct scaling of excitation energies computed by a method with increasing size of the system. The violation of this property can lead to errors and even non-physical predictions, for instance, the QSE excitation energies of a ``super molecule'' consisting of two non-interacting systems is not guaranteed to be the same as the excitation energies of the two systems calculated separately (see Fig. \ref{SIQSE}).
This may become a severe limitation when QSE will be applied to larger systems in the future and the underlying ground-state wavefunction is imprecise. 

In search of a size-intensive alternative, the EOM formalism based qEOM method was proposed by Ollitrault et al.~\cite{qEOM2020} for electronic excitation energies (EEs). 
The qEOM method provides good agreement for EEs with the exact results obtained by the full configuration interaction (FCI) method. 
However, the qEOM formalism (in Ref. \cite{qEOM2020}) does not necessarily satisfy the vacuum annihilation condition (VAC), also known as the killer condition~\cite{prasad1985some, szekeres2001killer},
    which ensures that the ground-state wavefunction cannot be de-excited.
This may result in the appearance of large errors when the formalism is extended to calculate properties like IPs and EAs.
Moreover, the qEOM method, just like QSE, requires higher-body RDMs which significantly increases the measurement challenges.

In this work, we propose a generally applicable EOM-based formalism for the calculation of molecular properties like EEs, IPs, and EAs following a VQE ground-state calculation on a quantum computer. 
Our formalism, which we refer to as q-sc-EOM, satisfies the VAC; produces size-intensive and real energy differences between the ground state and the excited states/charged states; does not involve measurements of higher than 2-body RDM-type quantities; is expected to be more resilient to noise than the current diagonalization-based state-of-the-art methods. 

This paper is organized as follows: Section~\ref{theory} discusses the theoretical formalism of q-sc-EOM using self-consistent operators, while the implementation details and circuit design are explained in Sec.~\ref{implementation}. 
Section~\ref{compdet} provides the computational details for the simulation data in this paper. In Sec.~\ref{results}, we discuss the results obtained in this work. Specifically, 
Sec.~\ref{res1} analyses the performance of the q-sc-EOM method in calculating EEs, IPs, and EAs of  H$_2$, LiH, and H$_2$O molecules, while Section~\ref{res2} compares the performance of the q-sc-EOM method with the QSE and qEOM formalisms. 
Finally, the key conclusions from the paper are summarized in Sec.~\ref{conclusion}.

\section{Theory}
The excitation energy of a given excited state can be obtained by the action of the commutator of the Hamiltonian and the corresponding excitation operator acting on the exact ground-state wavefunction. For an arbitrary $k^{\text{th}}$ excited state, this can be expressed as
\begin{align}
\begin{split}
    [\hat{H},\hat{\mathbb{O}}_k]|\Psi_{\text{gr}}\rangle=&\hat{H}\hat{\mathbb{O}}_k|\Psi_{\text{gr}}\rangle-\hat{\mathbb{O}}_k\hat{H}|\Psi_{\text{gr}}\rangle,\\
    =&(E_{k}-E_{\text{gr}})\hat{\mathbb{O}}_k|\Psi_{\text{gr}}\rangle,\label{eommaineq}
\end{split}
\end{align}
where $\ket{\Psi_{\text{gr}}}$ is the ground-state wavefunction, and E$_{\text{gr}}$ and E$_{k}$ refer to the energies of the ground and the $k^{\text{th}}$ excited state, respectively.
$\hat{H}$ is the molecular Hamiltonian, which in the second quantization formalism can be written as 
\begin{align}
    \hat{H}=\sum_{pq}h_{pq}\hat{a}^{\dagger}_p\hat{a}_q+\frac{1}{4}\sum_{pq,rs}\langle pq||rs\rangle \hat{a}^{\dagger}_p\hat{a}^{\dagger}_q\hat{a}_s\hat{a}_r,
\end{align}
where $h_{pq}$ and $\langle pq||rs\rangle$ are the one- and two-electron elements of the Hamiltonian, respectively. $\hat{a}_{p }^{\dagger}$ and $\hat{a}_{p }$ refer to the fermionic creation and annihilation operators (with respect to physical vacuum), respectively.
Following common notations, here, we use indices $\{p,q,r,s\dots\}$ for arbitrary molecular orbitals while $\{a,b,\dots\}$ 
and $\{i,j,\dots\}$ refer to unoccupied and occupied orbitals, respectively in the Hartree-Fock (HF) wavefunction.
The state-transfer operator $\hat{\mathbb{O}}_\text{k}$ is defined as 
\begin{align}
    \hat{\mathbb{O}}_k\ket{\Psi_{\text{gr}}}=\ket{\Psi_k}, 
\end{align}
where $\ket{\Psi_k}$ is the wavefunction of the $k^{\text{th}}$ excited state. 
These operators should ensure an important property, referred to as vacuum annihilation or Killer condition, which expresses that the ground state cannot be de-excited ~\cite{szekeres2001killer,prasad1985some,mertins1996algebraic,weiner1980calculation,hodecker2020unitary,levchenko2004equation}:
\begin{align}
    \hat{\mathbb{O}}^\dagger_k|\Psi_{\text{\text{gr}}}\rangle=0\quad \forall k. \label{VAC1}
\end{align}
In the case of exact operators $\mathbb{O}^\dagger_\text{k}$ and an exact ground-state ($\Psi_{\text{gr}}$), the above equation can be reformulated as 
\begin{align}
    \hat{\mathbb{O}}^\dagger_k|\Psi_{\text{gr}}\rangle=\ket{\Psi_{\text{gr}}}\bra{\Psi_{k}}\ket{\Psi_{\text{gr}}}=0\quad \forall k, \label{VAC2}
\end{align}
where $\hat{\mathbb{O}}^\dagger_k=\ket{\Psi_{\text{gr}}}\bra{\Psi_{k}}$. 
It can be seen that this condition is automatically satisfied for exact state-transfer operators acting on an exact ground-state due to the orthogonality of eigenstates.
However, one needs to ensure that the VAC is satisfied when approximate state-transfer operators are used \cite{prasad1985some,levchenko2004equation}.
The state-transfer operators that satisfy the VAC are referred to as the ``self-consistent'' operators~\cite{prasad1985some}. 

This work utilizes the framework of unitary coupled-cluster (UCC) theory, where the ground-state wavefunction is given by 
\begin{align}
    \ket{\Psi_{\text{UCC}}}=e^{\hat{\sigma}}\ket{\Psi_0},
\end{align}
where $\ket{\Psi_0}$ is the HF wavefunction, and $e^{\hat{\sigma}}$ is a unitary operator.
An approximate form of $\hat{\sigma}$, which is a cluster operator, can be written using single and double excitations as
\begin{align}
\begin{split}
    \hat{\sigma}=& \hat{\sigma}_1+\hat{\sigma}_2,\\
    \hat{\sigma}_1=& \sum_{ia}\sigma_i^a \big(\hat{a}^{\dagger}_a\hat{a}_i-\hat{a}^{\dagger}_i\hat{a}_a\big),\\
    \hat{\sigma}_2=& \sum_{ijab}\sigma_{ij}^{ab} \big(\hat{a}^{\dagger}_a\hat{a}^{\dagger}_b\hat{a}_j\hat{a}_i-\hat{a}^{\dagger}_i\hat{a}^{\dagger}_j\hat{a}_b\hat{a}_a\big).
\end{split}
\end{align}
where $\sigma_i^a$ and $\sigma_{ij}^{ab}$ are amplitudes for the associated excitations. 
In order to capture the dominant electron-correlation effects contributing to electronic excitations,
an excitation manifold can be constructed by including all possible single and double excitations and de-excitations, represented by $\{\hat{G}_I^{\dagger}\}\cup\{\hat{G}_I\}$. Here, $\hat{G}_I$ can refer to any single ($\hat{a}^{\dagger}_a\hat{a}_i$) or double ($\hat{a}^{\dagger}_a\hat{a}^{\dagger}_b\hat{a}_j\hat{a}_i$) excitation operator.
However, to satisfy the VAC, 
the excitation operator manifold is rotated, forming the self-consistent manifold $\{\hat{S}_I^{\dagger}\}\cup\{\hat{S}_I\}$, where 
\begin{align}\label{eq:VACrot}
    \hat{S}_I=e^{\hat{\sigma}}\hat{G}_Ie^{-\hat{\sigma}}.
\end{align}
Similar excitation manifolds can also be constructed using particle number non-conserving excitation operators that are needed for computation of IPs and EAs.
This technique was developed by Mukherjee and co-workers in Refs.~\cite{prasad1985some,datta1993}.
This self-consistent operator manifold can now be used to develop excited-state methods that satisfy the VAC.

Following Eq.~\eqref{eq:VACrot}, the state-transfer operator for electronic excitations, $(\hat{\mathbb{O}}_k)_{\text{EE}}$,  can now be written as a linear combination of all possible operators from the self-consistent excitation manifold, given by
\begin{align}\label{OK_EE}
    (\hat{\mathbb{O}}_{k})_{\text{EE}}=&(\hat{R}^k_1)_{\text{EE}}+(\hat{R}^k_2)_{\text{EE}},
\end{align}
 where $(\hat{R}^k_1)_{\text{EE}}$ and $(\hat{R}^k_2)_{\text{EE}}$ are single and double excitation operators defined as
\begin{align}\label{opsmain}
\begin{split}
    (\hat{R}^k_1)_{\text{EE}}=&\sum_{ia}\big[(A^k)_i^a \hat{S}_i^a-(B^k)_a^i\hat{S}_a^i\big],\\
    (\hat{R}^k_2)_{\text{EE}}=&\sum_{ijab}\big[(A^k)_{ij}^{ab} \hat{S}_{ij}^{ab}-(B^k)^{ij}_{ab}\hat{S}_{ab}^{ij}\big].
\end{split}
\end{align}
Here, $(A^k)_I$ and $(B^k)_{I^{\dagger}}$ are the amplitudes corresponding to the excitation ($I$) and  de-excitation operators ($I^{\dagger}$), respectively,
for the $k^\text{th}$ excited state. 
Here $I$ refers to all possible single 
and double excitations. 
State-transfer operators for singly charged states can also be defined in a similar manner,
\begin{align}\label{OK_ipea}
    (\hat{\mathbb{O}}_{k})_{\text{IP/EA}}=&(\hat{R}^k_1)_{\text{IP/EA}}+(\hat{R}^k_2)_{\text{IP/EA}},
\end{align}
 where $(\hat{R}^k_1)_{\text{IP/EA}}$ and $(\hat{R}^k_2)_{\text{IP/EA}}$ refer to the particle number non-conserving single and double excitation operators, defined as
 \begin{align}\label{opsmainip}
\begin{split}
    (\hat{R}^k_1)_{\text{IP}}=&\sum_{i}\big[(A^k)_i \hat{S}_i-(B^k)^i\hat{S}^i\big],\\
    (\hat{R}^k_2)_{\text{IP}}=&\sum_{ija}\big[(A^k)_{ij}^{a} \hat{S}_{ij}^{a}-(B^k)^{ij}_{a}\hat{S}_{a}^{ij}\big],
\end{split}
\end{align}
and 
  \begin{align}\label{opsmainea}
\begin{split}
    (\hat{R}^k_1)_{\text{EA}}=&\sum_{a}\big[(A^k)^{a} \hat{S}^{a}-(B^k)_{a}\hat{S}_{a}\big],\\
    (\hat{R}^k_2)_{\text{EA}}=&\sum_{iab}\big[(A^k)_{i}^{ab} \hat{S}_{i}^{ab}-(B^k)^{i}_{ab}\hat{S}_{ab}^{i}\big].
\end{split}
\end{align}

\subsection{q-sc-EOM method}\label{theory}

In the VQE algorithm, the unitary evolution operator $U(\theta)$ ($e^{\hat{\sigma}}$ in UCC theory)  is implemented on a quantum computer using a parameterized circuit. The parameters ($\theta$) of the circuit are optimized to variationally minimize the molecular energy and obtain the molecular ground state $\ket{\Psi_{\text{VQE}}}$, such that
\begin{align}
    \ket{\Psi_{\text{VQE}}}=U(\theta)\ket{\Psi_0}.
\end{align}
Projecting Eq.~\eqref{eommaineq} onto the $k^{\text{th}}$ excited state wavefunction and using the state-transfer operator defined in Eq.~\eqref{OK_EE}, the q-sc-EOM excitation energy from the ground state to the $k^\text{th}$ excited state ($E_{0k}$) is given by
\begin{align}
\begin{split}
    E_{0k}=&\frac{\langle \Psi_{\text{VQE}}|[(\hat{\mathbb{O}}^\dagger_k)_{\text{EE}},[\hat{H},(\hat{\mathbb{O}}_k)_{\text{EE}}]]|\Psi_{\text{VQE}}\rangle}{\langle \Psi_{\text{VQE}}|[(\hat{\mathbb{O}}^\dagger_k)_{\text{EE}},(\hat{\mathbb{O}}_k)_{\text{EE}}]|\Psi_{\text{VQE}}\rangle}.
\end{split}\label{qeomeq}
\end{align}
Expressions for IPs and EAs can also be derived using the associated state-transfer operators defined in Eq. \eqref{OK_ipea}.
As discussed in Ref.~\cite{qEOM2020}, Eq.~\eqref{qeomeq} provides size-intensive energy differences.
By inserting the expression for $(\hat{\mathbb{O}}_k)_{\text{EE}}$ from Eq.~\eqref{OK_EE} in Eq.~\eqref{qeomeq}, it can be seen that the final equation for the 
excitation energy for the $k^{\text{th}}$ excited state has a parametric dependence on the amplitudes $(A^k)_I$ and $(B^k)_{I^{\dagger}}$, where $I$ refers to all possible single 
and double excitations. 
A variational minimization of the resulting equation ($\delta E_{0k}$ = 0) with respect to these amplitudes leads to the following secular equation
\begin{align}\label{secular}
\left(\begin{array}{cc}
\mathbf{M} & \mathbf{Q} \\
\mathbf{Q}^{*} & \mathbf{M}^{*}
\end{array}\right)\left(\begin{array}{l}
\mathbf{A}_{k} \\
\mathbf{B}_{k}
\end{array}\right)=E_{0 k}\left(\begin{array}{cc}
\mathbf{V} & \mathbf{W} \\
-\mathbf{W}^{*} & -\mathbf{V}^{*}
\end{array}\right)\left(\begin{array}{l}
\mathbf{A}_{k} \\
\mathbf{B}_{k}
\end{array}\right),
\end{align}
where the matrix elements of matrices $\textbf{M}$, $\textbf{Q}$, $\textbf{V}$, $\textbf{W}$ are defined as
\begin{align}\label{MVQW}
    M_{IJ}= & \langle \Psi_{\text{VQE}}|[\hat{S}_I^{\dagger},[\hat{H},\hat{S}_J]]|\Psi_{\text{VQE}}\rangle,   \\
    V_{IJ}= & \langle \Psi_{\text{VQE}}|[\hat{S}_I^{\dagger},\hat{S}_J]|\Psi_{\text{VQE}}\rangle, \nonumber\\
    Q_{IJ}= & -\langle \Psi_{\text{VQE}}|[\hat{S}_I^{\dagger},[\hat{H},\hat{S}^{\dagger}_J]]|\Psi_{\text{VQE}}\rangle, \nonumber \\
    W_{IJ}= & -\langle \Psi_{\text{VQE}}|[\hat{S}^{\dagger}_I,\hat{S}^{\dagger}_J]|\Psi_{\text{VQE}}\rangle \nonumber .
\end{align}
Upon careful inspection, one can see that the matrices $\textbf{W}$ and $\textbf{Q}$ are zero due to the use of self-consistent operators. 
This is a simplification (compared to the qEOM formalism of Ref.~\cite{qEOM2020}) as it reduces the secular equation to
\begin{align}\label{gen_eig}
    \mathbf{M}\mathbf{A}_k=E_{0k}\mathbf{V}\mathbf{A}_k.
\end{align}
The matrix elements of $\textbf{M}$ and $\textbf{V}$ can be further simplified as
\begin{align}
\begin{split}
    M_{IJ}=&\langle \Psi_{\text{VQE}}|[\hat{S}_I^{\dagger},[\hat{H},\hat{S}_J]]|\Psi_{\text{VQE}}\rangle,\\
    =&\langle \Psi_{\text{VQE}}|[U(\theta)\hat{G}_I^{\dagger}U^{\dagger}(\theta),[\hat{H},U(\theta)\hat{G}_JU^{\dagger}(\theta)]]|\Psi_{\text{VQE}}\rangle,\\
    =&\langle \Psi_{0}|\hat{G}_I^{\dagger}U^\dagger(\theta)\hat{H}U(\theta)\hat{G}_J|\Psi_{0}\rangle-\delta_{IJ}*E_{\text{gr}},
\end{split}\label{eqMeom}
\end{align}
and 
\begin{align}
\begin{split}
    V_{IJ}=&\langle \Psi_{\text{VQE}}|[\hat{S}_I^{\dagger},\hat{S}_J]|\Psi_{\text{VQE}}\rangle,\\
    =&\langle \Psi_{\text{VQE}}|[U(\theta)\hat{G}_I^{\dagger}U^{\dagger}(\theta),U(\theta)\hat{G}_JU^{\dagger}(\theta)]|\Psi_{\text{VQE}}\rangle,\\
    =&\langle \Psi_{0}|[\hat{G}_I^{\dagger},\hat{G}_J]|\Psi_{0}\rangle,\\
    =&\delta_{IJ}.
\end{split}\label{eqVeom}
\end{align}
Thus, in q-sc-EOM the overlap matrix ($\textbf{V}$) is guaranteed to be the identity matrix and does not need to be computed on a quantum computer. 
It has an two key benefits, namely, it leads to a Hermitian formulation for excitation energy and it converts the generalized eigenvalue equation into an eigenvalue equation. 
The formalism developed so far can be written in the form of a concise eigenvalue equation as
\begin{align}\label{finaleq}
\left(\begin{array}{cc}
    \mathbf{M_{SS}}-E_{\text{gr}}*\mathbf{I_{SS}}&\mathbf{M_{SD}}\\
    \mathbf{M_{DS}} & \mathbf{M_{DD}}-E_{\text{gr}}*\mathbf{I_{DD}}
    \end{array}\right)\left(\begin{array}{c}
    \mathbf{A_S}^k\\
    \mathbf{A_D}^k
    \end{array}\right)\\=E_{0k}
    \left(\begin{array}{c}
    \mathbf{A_S}^k\\
    \mathbf{A_D}^k
    \end{array}\right),\nonumber
\end{align}
where $S$ and $D$ refer to single and double excitations, respectively.
Thus, $\textbf{M}_{\textbf{SS}}$ refers to the block of matrix $\textbf{M}$ with two single excitation operators in the double commutator (see Eq.~\eqref{MVQW}), while $\mathbf{I_{XX}}$ is an 
identity matrix of dimension $X$.  
It should be emphasized that, since the above formulation is Hermitian, the eigenvalues obtained using Eq.~\eqref{finaleq} are guaranteed to be real (unlike in EOM-CCSD or qEOM)~\cite{thomas2021complex,liu2018unitary, qEOM2020}. 
This also ensures that this formulation is free from problems related to different left and right eigenfunctions encountered in classical EOM-CC methods~\cite{Stanton1993}.

Finally, each element of matrix $\textbf{M}$ can be computed on a quantum computer using Eq.~\eqref{eqMeom}. 
The resulting eigenvalue equation can then be solved classically to obtain q-sc-EOM EE values. 
Here, Krylov subspace based formalisms, such as the Davidson algorithm~\cite{davidsorq1975theiterative,liu1978simultaneous,tretiak2009representation}, can be used to obtain the excitation energies of a few low-lying excited states while avoiding the explicit evaluation and diagonalization of the $\textbf{M}$ matrix.  
It should be noted that this method is closely related to UCC-based excited-state methods in quantum chemistry, and thus Eq.~\eqref{finaleq} resembles the equation for EEs for UCC-based methods as derived, for example, in Refs.~\cite{liu2018unitary,hodecker2020unitary}.


\subsection{Circuit design and implementation details}\label{implementation}

\begin{figure*}[tbp]
    \hspace*{\fill}%
    \subfloat[ \label{circuitd}]{
        \includegraphics[width = 0.25\textwidth,height=4cm]{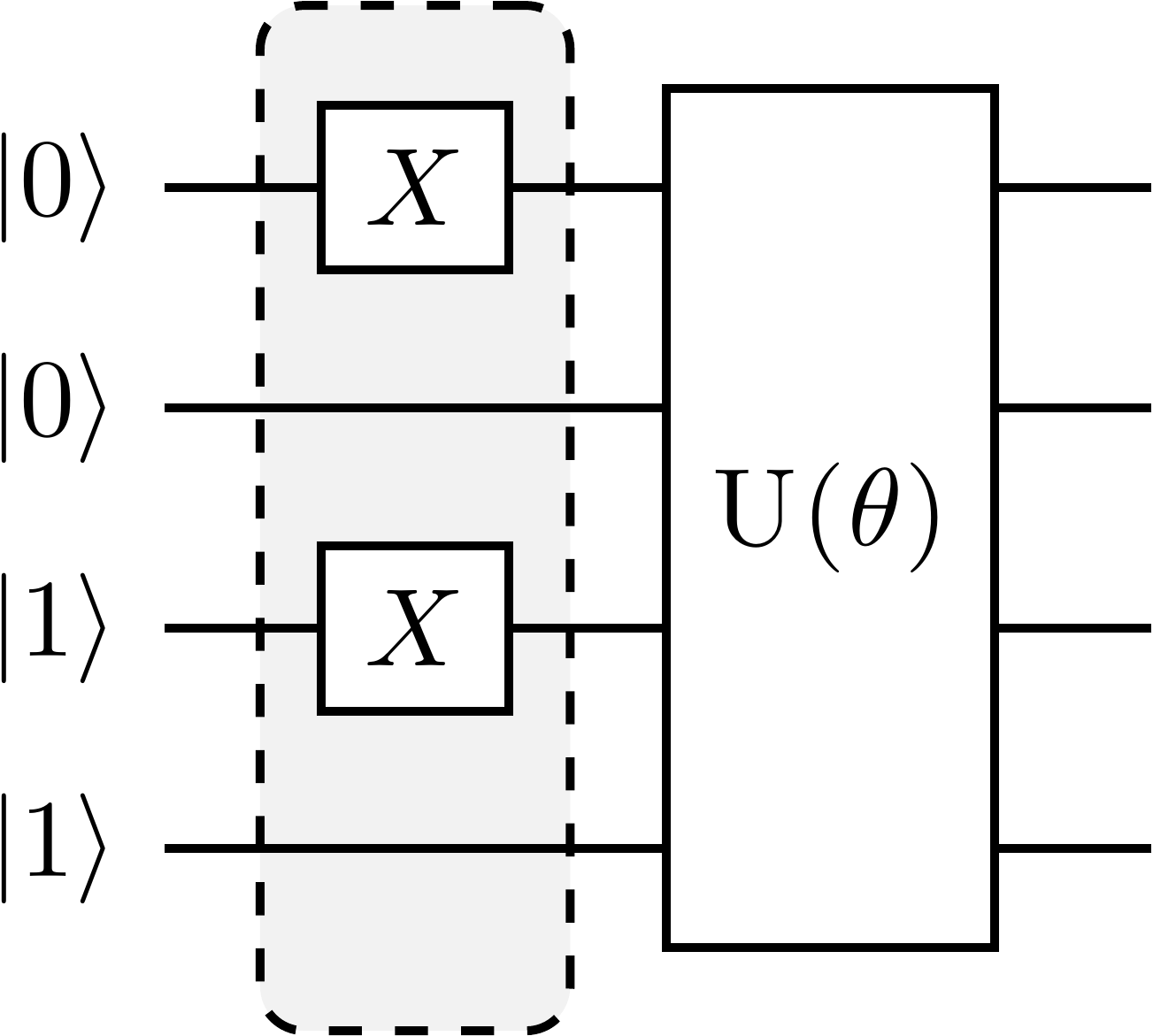}    
    }\hfill
    \subfloat[ \label{circuitoff}]{
    \includegraphics[width = 0.5 \textwidth,height=4cm]{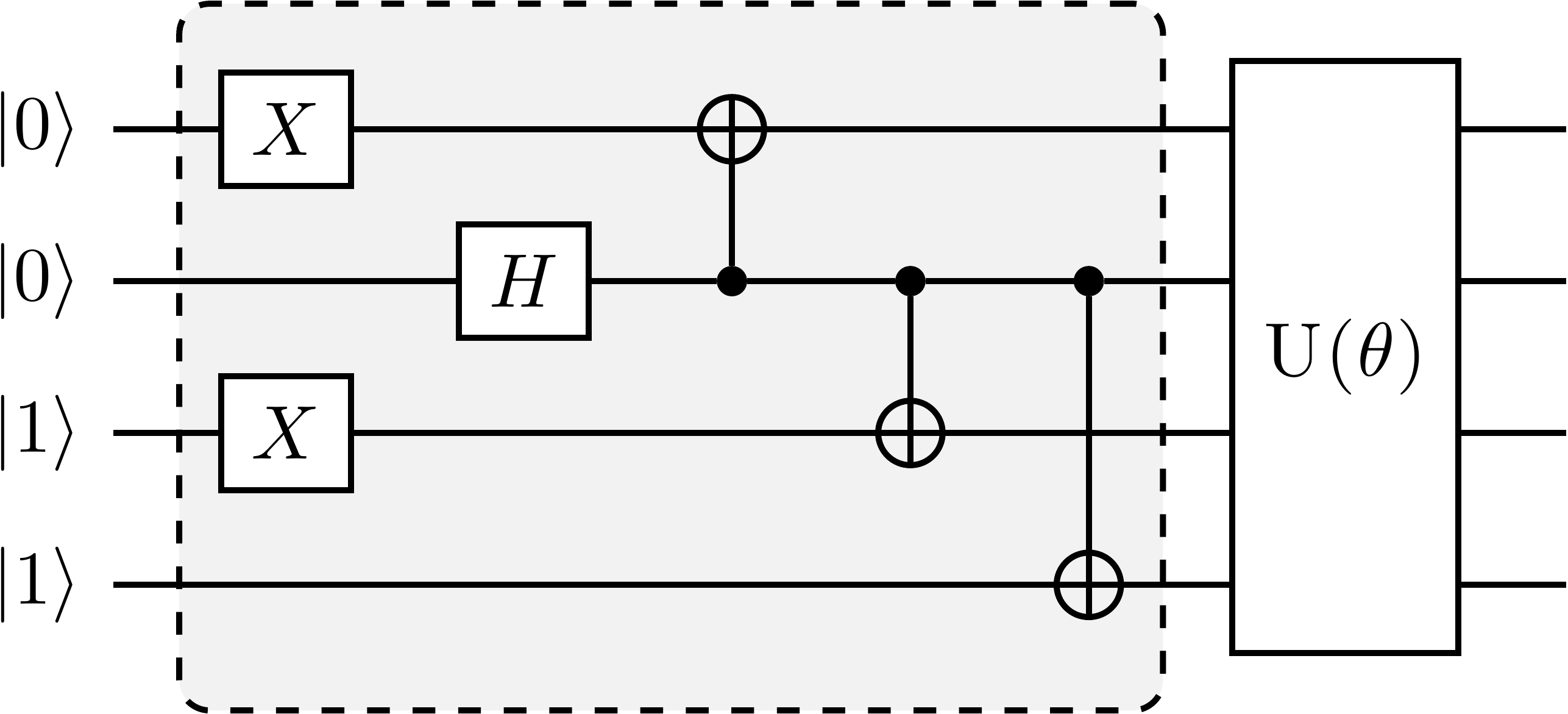}
    }
    \caption{Proposed quantum circuit for the estimation of a representative element of the $\textbf{M}$ matrix for the H$_2$ molecule using (a) an excited Slater determinant as the reference state needed to compute diagonal elements and (b) an entangled state involving two excited Slater determinants as the reference required for the evaluation of off-diagonal elements.}
    \label{circuit}
    \hspace*{\fill}%
\end{figure*}

Here, we discuss our proposed implementation of the q-sc-EOM formalism on a quantum computer. 
The discussion can be divided into two parts: circuit details to evaluate the diagonal and the off-diagonal elements of the matrix $\textbf{M}$ in Eq.~\eqref{eqMeom}. 
The state preparation for the diagonal elements involves the same circuit as the one optimized for the ground state, but it is now applied to a classical state that represents an excited Slater determinant.
We refer to this classical state as the reference state.
Thus, the circuit can be prepared in two steps: the first step is the creation of a reference state, whereas the second step involves the action of the previously optimized ground-state VQE circuit on the newly formed reference state. 
The molecular Hamiltonian is then measured using this prepared state, as done for the ground-state energy evaluation.
To give an example of a state preparation circuit, we consider the H$_2$ molecule in the STO-3G basis and use a singly excited determinant in Eq.~\eqref{eqMeom}. We choose the single-excitation operator for the excitation from 1s$^\beta$ to 2s$^\beta$ (represented by $\ket{\text{0011}}\rightarrow\ket{\text{1001}}$ notation in the qubit representation using the Jordon-Wigner mapping). 
The classical state that corresponds to such an excited Slater determinant can be created by the action of two NOT gates, as shown in the circuit in Fig.~\ref{circuitd}.
U($\theta$) in Fig.~\ref{circuitd} refers to the optimized circuit prepared for the VQE ground-state evaluation, and the shaded region represents the circuit needed to create the reference state.  
Note that the orbitals 1s$^\alpha$, 1s$^\beta$, 2s$^\alpha$, and 2s$^\beta$ are mapped onto the qubits in the bottom-to-top order, such that the lowest energy orbitals are at the bottom. 

The off-diagonal elements of the matrix $\textbf{M}$ can be evaluated using an entangled state of two excited Slater determinants, as shown below. A representative off-diagonal element $M_{I,J}$ can be written in terms of diagonal elements as 
\begin{align}\label{offdiagonalexp}
   \hbox{Re}[M_{I,J}]=M_{I+J,I+J}-\frac{M_{I,I}}{2}-\frac{M_{J,J}}{2},
\end{align}
where the term $M_{I+J,I+J}$ is given by
\begin{align}\label{mijexp}
\begin{split}
&M_{I+J,I+J}=\\&\bra{\Psi_0} \frac{1}{\sqrt 2}(\hat{G}_{I}+\hat{G}_J)^\dagger U(\theta)^\dagger \hat{H}U(\theta)\frac{1}{\sqrt 2}(\hat{G}_{I}+\hat{G}_J)\ket{\Psi_0}.
\end{split}
\end{align}
A similar expression holds for the imaginary part of $M_{I,J}$.
The matrix elements $M_{I,I}$ and $M_{J,J}$ in Eq.~\eqref{offdiagonalexp} are diagonal elements that are evaluated using the method described previously (without the ground-state energy term). 
The element $M_{I+J,I+J}$ can be evaluated using Eq.~\eqref{mijexp}, which involves the creation of the $I+J$ state, application of the unitary $U(\theta)$, and finally, the measurement of the Hamiltonian using this prepared state. 
The state $I+J$ is a superposition state of two classical states $I$ ($\hat{G}_I\ket{\Psi_0}$) and $J$ ($\hat{G}_J\ket{\Psi_0}$).
Notice that both of these states are excited Slater determinants, which can be represented through qubit states in the computational basis.
An entangled state, such as $I+J$, is commonly created by using an ancilla qubit (for example, see Ref.~\cite{huggins2020non}). 
In the case of q-sc-EOM, we can use a simpler method to create this entanglement without adding any ancilla qubit.
Being classical states, $I$ and $J$ are trivial to create on a quantum computer using NOT gates. 
An entangled state can then be created using a Hadamard gate and a series of CNOT gates. For example, the $I$ state can be created using NOT gates and then transformed to the $I+J$ state using Hadamard and CNOT gates.  
We can take the example of two single excitations to demonstrate this, specifically 1s$^\alpha\rightarrow\;$2s$^\alpha$ and 1s$^\beta\rightarrow\;$2s$^\beta$ (denoted as $\ket{\text{0011}}\rightarrow\ket{\text{1001}}$ and $\ket{\text{0011}}\rightarrow\ket{\text{0110}}$ in the qubit representation, respectively). 
Here, we need to create the entangled state $\ket{I+J}=\frac{1}{\sqrt{2}}(\ket{1001}+\ket{0110})$.
The circuit for creating this state and then evaluating $M_{I+J,I+J}$ is shown in Fig.~\ref{circuitoff}.
The shaded region represents the circuit used to create the entangled $\ket{I+J}$ state. 
The gate sequences can be stored for each excitation operator and can be applied to the HF state at the start of the EOM procedure.
A maximum of 7 CNOT gates will be needed to create any entangled state for the off-diagonal terms, when using up to double excitations in Eq.~\eqref{eqMeom}. This is a very small number compared with the total number of CNOT gates required to prepare the ground state using VQE-based algorithms. 
It should be noted that this proposed circuit design loses on the phase factor of excitation operators in $I$ and $J$ classical states, but all our numerical simulations have consistently shown that these phase factors have no effect on the computed eigenvalues of the eigenvalue equation. To preserve the phase factor, an ancilla qubit based creation of entangled $I+J$ states could be used instead. 

 Regarding the resource requirements for running the q-sc-EOM method on a quantum computer, the number of qubits required for the evaluation of each matrix element is the same as that in the computation of the ground state. 
The circuit that is implemented in the ground-state VQE calculation remains unchanged in the generation of the EOM matrix elements as well. The only difference is that the reference state is changed from the HF determinant to an excited Slater determinant or an entangled state involving two excited Slater determinants, as discussed above. 
Thus, unlike qEOM and QSE, where the excitation operators are measured together with the Hamiltonian, the excitation operators act directly on the HF state in q-sc-EOM. 
This provides a notable advantage that q-sc-EOM methods do not need the estimation of higher than 2-body RDMs. 
Such a framework also makes it easier to include higher excitations when required, such as triples, whose inclusion becomes important in higher-accuracy charged-state calculations~\cite{matthews2016new}.
The calculation of elements of the $\textbf{M}$ matrix can also be run in parallel on a quantum computer. 
The evaluation of the $\textbf{M}$ matrix on a quantum computer requires the measurement of $\cal{O}$($o^4v^4$) matrix elements, where $o$ and $v$ are the number of occupied and unoccupied orbitals, respectively.
Despite this scaling, generally, the matrix $\textbf{M}$ is very sparse.
The number of elements in the $\textbf{M}$ matrix can be drastically reduced using this sparsity through the use of spatial symmetry, spin symmetry, etc., which will be a topic of future study.  
The major advantages afforded by the use of quantum computers come from the efficient evaluation of the q-sc-EOM matrix elements and from the accurate ground state wavefunction provided by the VQE-based algorithm.

\section{Computational details}\label{compdet}
 \begin{figure*}
\centering
\includegraphics[width = 1 \textwidth]{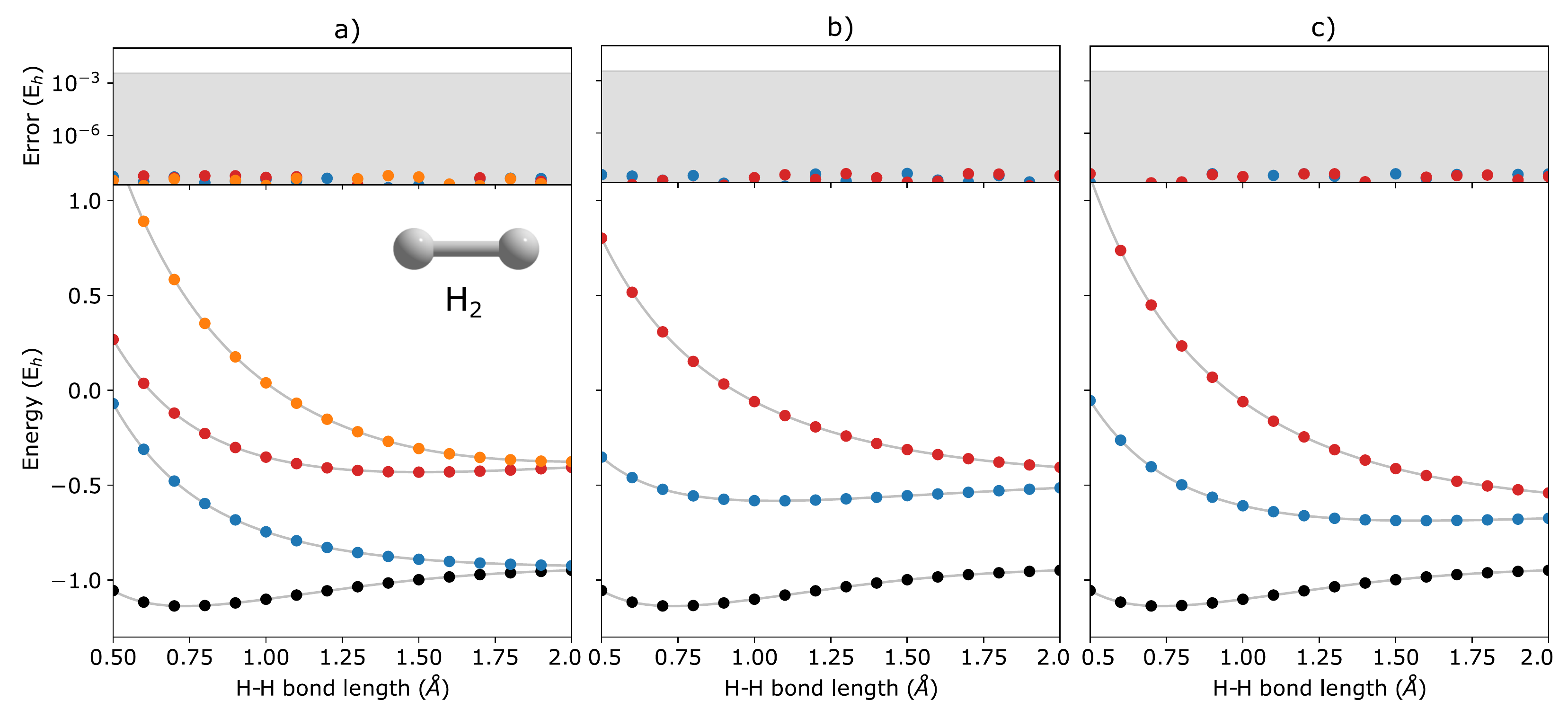}
\caption{q-sc-EOM energies of (a) electronically excited, (b) single electron-detached, and (c) single electron-attached states along with the ground state (black circles) of the H$_2$ molecule as a function of bond length. The gray lines correspond to the FCI results. The deviations from the FCI results are shown in the upper panel, where the shaded region indicates errors below 0.1 eV.}
\label{h2eom}
\end{figure*} 

\begin{figure*}
\centering
\includegraphics[width = 1 \textwidth]{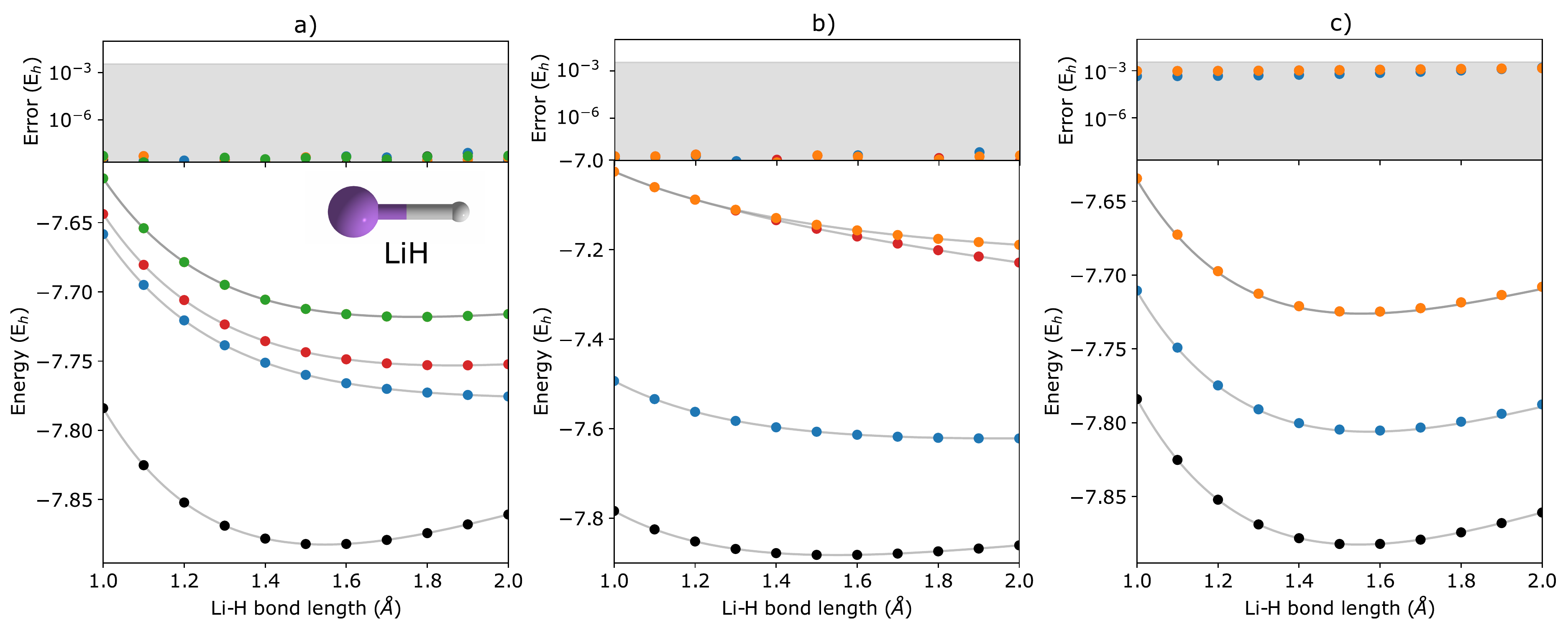}
\caption{q-sc-EOM energies of first few (a) electronically excited, (b) single electron-detached, and (c) single electron-attached states along with the ground state (black circles) of the LiH molecule as a function of bond length. The gray lines correspond to the FCI results. The deviations from the FCI results are shown in the upper panel, where the shaded region indicates errors below 0.1 eV.}
\label{liheom}
\end{figure*} 

\begin{figure*}
\centering
\includegraphics[width =   \textwidth]{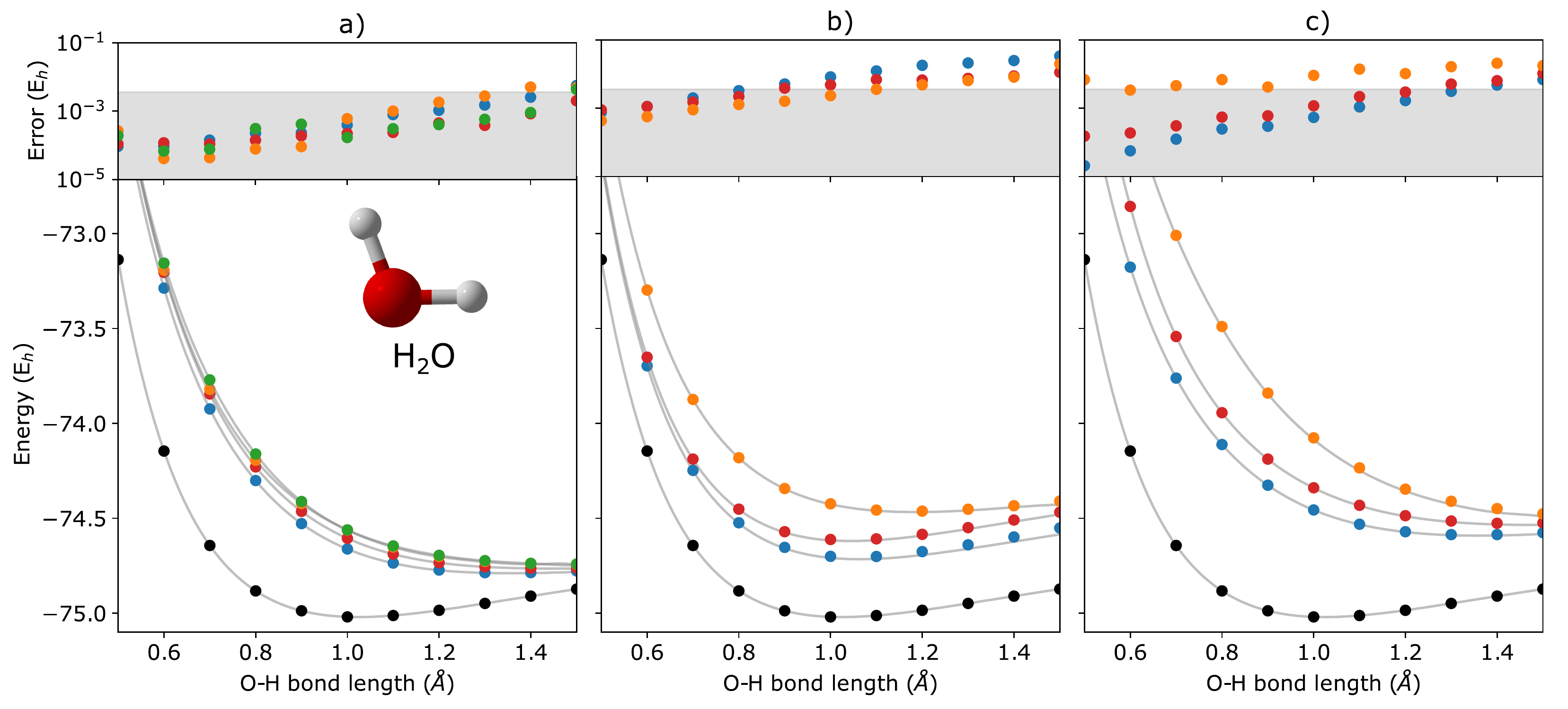}
\caption{q-sc-EOM energies of first few (a) electronically excited, (b) single electron-detached, and (c) single electron-attached states along with the ground state (black circles) of the H$_2$O molecule as a function of the O-H bond lengths where both O-H bonds are stretched symmetrically. The gray lines correspond to the FCI results. The deviations from the FCI results are shown in the upper panel, where the shaded region indicates errors below 0.1 eV.}
\label{h2oeom}
\end{figure*} 

All the computations in this work utilize the STO-3G basis set. One- and two-electron integrals are calculated using the PySCF~\cite{Pyscf} program 
with the HF reference state. The Jordon-Wigner mapping and the transformation of the second-quantized operators to the 
Pauli form are carried out using the OpenFermion~\cite{openfermion} software package. A classical noise-free simulator, where exact unitary operations are applied to the state vector representing the wavefunction, is used to assess the accuracy of the formalism developed in this work.
The ground-state wavefunction is calculated using the fermionic ADAPT-VQE method using the generalized singles and doubles (GSD) operator pool~\cite{grimsley2019adaptive}. We use the gradient convergence criterion with a threshold of $10^{-3}$ for all the ground-state energy calculations.
All the formalisms discussed in this work, namely q-sc-EOM, qEOM, and QSE, utilize the ground-state energy and wavefunction obtained from the ADAPT-VQE simulation.
It should be noted that we have extended the qEOM formulation of Ref.~\cite{qEOM2020}, originally developed for the calculation of EEs, to calculate IPs and EAs for our theoretical investigation.
The EE results obtained using the qEOM approach in this work are verified against Qiskit's qEOM implementation~\cite{aleksandrowicz2019qiskit}.
The code used for generating the data in this work can be found in Ref.~\cite{giteom}.

\section{Results and discussion}\label{results}
We test the accuracy of the q-sc-EOM approach for three small molecules: H$_2$, LiH, and H$_2$O, and compare the results obtained with the exact FCI values. 
Computations are carried out for EEs, IPs, and EAs for these molecules.
The total energies of the electronically excited, single electron-detached and single electron-attached states are computed by adding the EEs, IPs and EAs, calculated using q-sc-EOM, to the ground-state energy obtained using the ADAPT-VQE simulations.
Since the ground-state energy computed using ADAPT-VQE is near-exact for the molecules considered in this study, the errors in the energies of the electronically excited states and the single electron attached/detached states, with respect to the FCI, are almost entirely due to the error in the post-VQE procedure.
For LiH and H$_2$O, we invoke the frozen-core approximation. Thus, the number of qubits required for the q-sc-EOM computation for H$_2$, LiH, and H$_2$O are 4, 10 and 12, respectively. 
We plot energy errors relative to the FCI values and use shading to indicate errors below 0.1~eV, as this value is generally the desired accuracy for these properties, so that they can be quantitatively compared with experiments. We also compare the performance of the q-sc-EOM formalism with QSE and qEOM in subsection~\ref{res2}.
\subsection{EE, IP, and EA calculations with q-sc-EOM}\label{res1}
In Fig.~\ref{h2eom}, we show the energies of the ground state along with first few electronically excited, single electron-detached and single electron-attached states of the H$_2$ molecule calculated using q-sc-EOM, as a function of the inter-hydrogen distance. 
The corresponding FCI results are shown as gray lines. 
The errors in the energies with respect to the exact FCI values are shown in the subgraph on top of each panel. It can be seen that the errors in q-sc-EOM computed energies, or in other words, q-sc-EOM evaluated EEs, IPs, and EAs are essentially identical to FCI, with errors of less than $10^{-8}$ Hartree.
This is because the q-sc-EOM formalism for the H$_2$ molecule using the STO-3G basis set is exact, as the singles and doubles excitations used in Eq.~\eqref{eqMeom} span the full excitation space and the VAC is satisfied. 
Figure~\ref{liheom} shows the energies of ground state along with the first few electronically excited, single electron-detached and single electron-attached states for the LiH molecule as a function of the Li-H bond length in a similar layout as the previous figure. For both EEs and IPs, the q-sc-EOM method gives nearly exact results, and as expected, errors less than  $10^{-8}$ Hartree 
were obtained with respect to the FCI values. However, we start to see the appearance of non-negligible errors for the EA results. 
This is because the computation of EAs for the LiH molecule with q-sc-EOM is no longer exact due to the addition of an electron.
Thus, triple excitation operators should be added to the excitation manifold for an exact treatment for EAs. 
However, q-sc-EOM is still able to produce EAs within the desired accuracy, as seen from the shaded region in the error plot at the top of Fig.~\ref{liheom}c. 
For the H$_2$O molecule,
we investigate the performance of q-sc-EOM as a function of the O-H bond length where both O-H bonds are stretched symmetrically. From Fig.~\ref{h2oeom}, 
one can see that the errors in EEs are within the desired accuracy up to the O-H bond length of 1.4\AA{}. The errors build up as the two O-H bonds are broken further, due to the appearance of strong correlation effects in the wavefunction. 
Classical EOM-based methods, such as EOM-CCSD, show similar trends in errors as well. 
The errors in IPs and EAs are larger compared to EEs and are above our desirable error limit of 0.1 eV.  
This is well-known in classical quantum chemistry, where EOM-based methods require at least an approximate treatment of triple excitations in the EOM framework to reach an accuracy of 0.1 eV relative to FCI values for IPs and EAs~\cite{matthews2016new,hirata2000high,dutta2018lower}. 
For higher accuracy in charged excitations, carefully selected triple excitations can be added to the excitation manifold in q-sc-EOM. This will be a subject of future study.

\subsection{Comparison with QSE and qEOM}\label{res2}

\begin{figure}[t]
\includegraphics[width =  \linewidth]{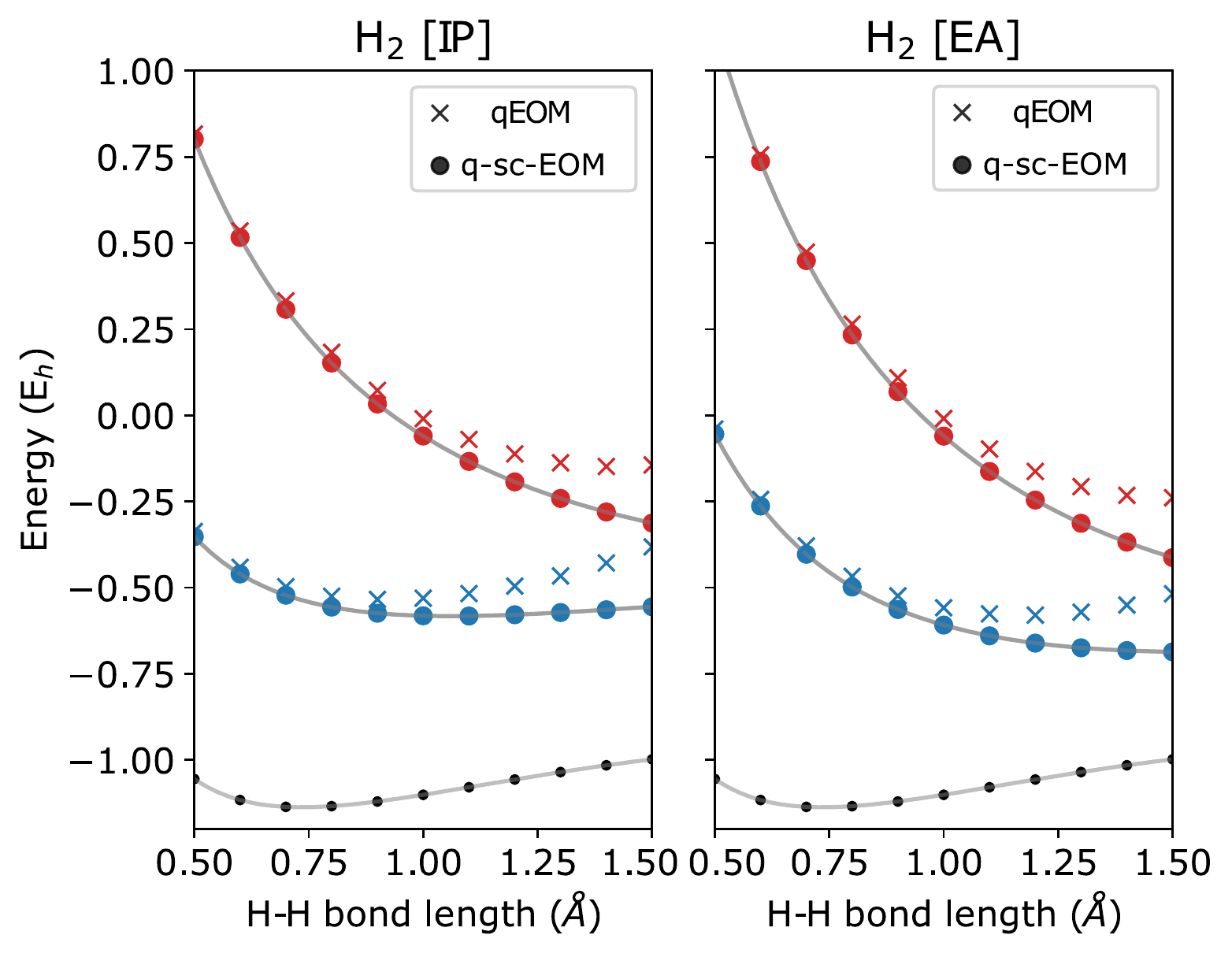}
\caption{Energies of (a) single electron-detached, and (b) single electron-attached states along with the ground state (black circles) of the H$_2$ molecule plotted as a function of the H-H bond length using the STO-3G basis. 
The FCI results are plotted in gray.
}
\label{qeomcompip}
\end{figure} 

\begin{figure}[t]
\includegraphics[width = \linewidth]{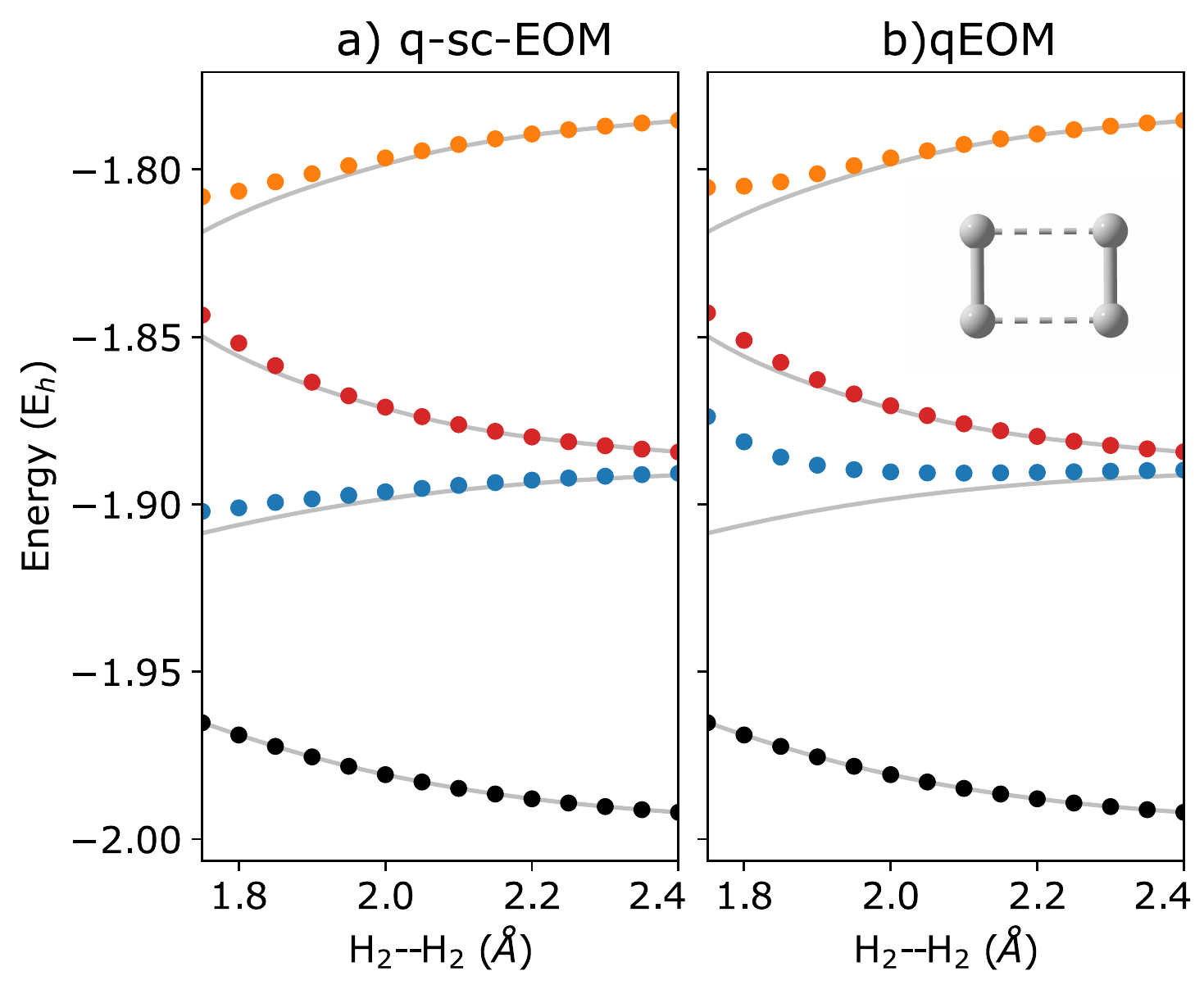}
\caption{Energies of the three lowest excited states along with the ground state (black circles) computed using a) q-sc-EOM and b) qEOM formalisms for the  dissociation of a rectangular geometry of H$_4$ into H$_2\cdots{}$H$_2$ as a function of the H$_2\cdots$H$_2$ separation distance. Both of the H$_2$ molecules have a bond length of 1.5~\AA{}. The FCI results are plotted in gray.}
\label{qeomcompee}
\end{figure}

\begin{figure}[t]
\includegraphics[width =  1.05\linewidth]{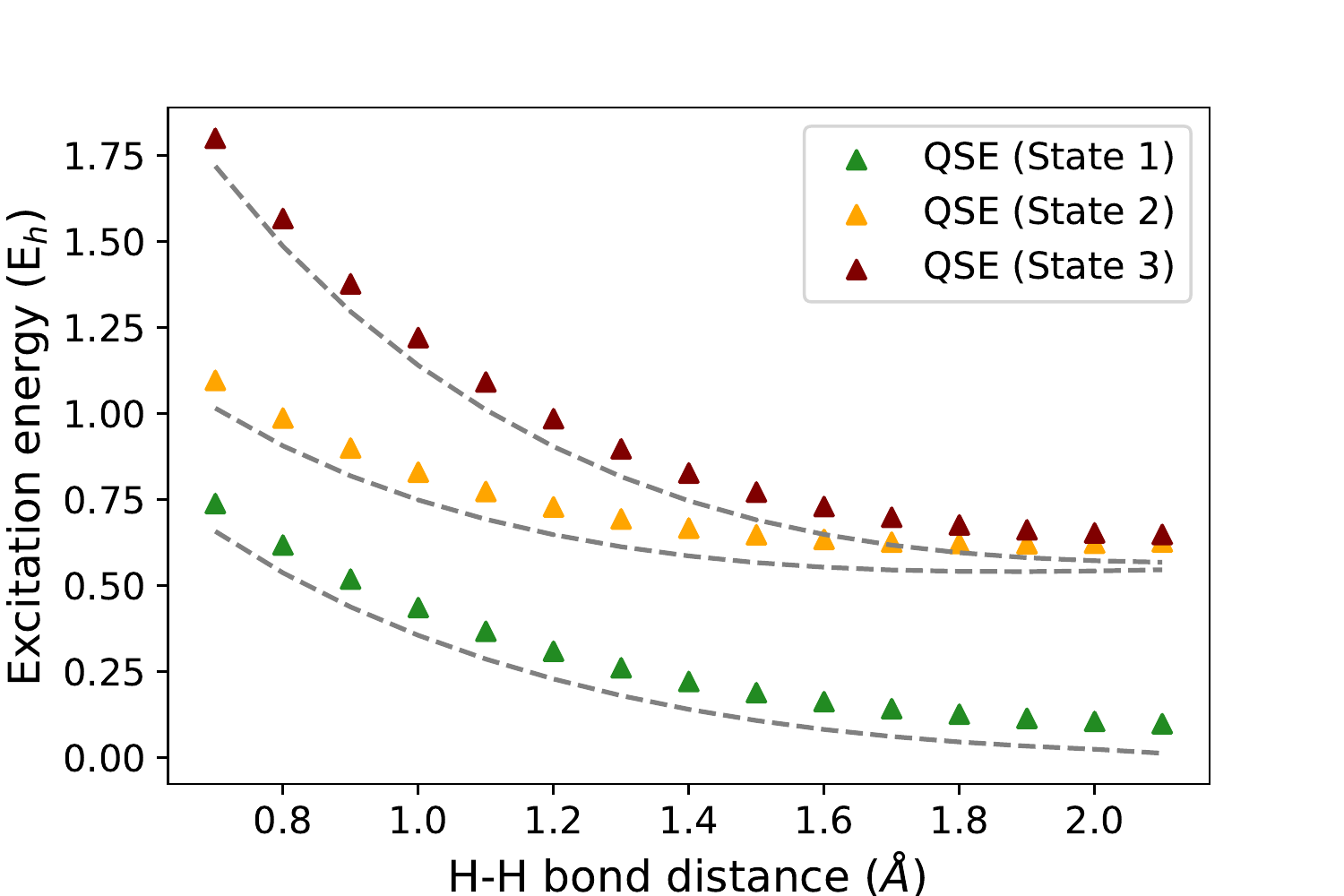}
\caption{Excitation energies of H$_2$ molecule in isolated H$_2$ (grey lines) and H$_2$-H$_4$ system  with H$_2$ not interacting with H$_4$ (H$_2$ and H$_4$ separation taken as 100\AA{}) (triangles) using STO-3G basis set. An inexact ground state is used in the plot computed using ADAPT-VQE stopped after adding 3 operators. The difference in the dashed line with triangles represent the size-intensivity error for each excitation energy using QSE.}
\label{SIQSE}
\end{figure} 

In this subsection, we discuss the connections between the q-sc-EOM, QSE, and qEOM formalisms and compare them in the contexts of a) their accuracy in computing energy differences, and b) quantum resource requirements and sensitivity to noise. 
The EOM formulation in Eq.~\eqref{qeomeq} with excitation operators taken from the excitation manifold represented as $\{\hat{G}_I^{\dagger}\}\cup\{\hat{G}_I\}$ (not the self-consistent excitation manifold) may lead to a violation of the VAC. 
There are two methods discussed in the quantum chemistry literature to impose the VAC in the EOM formalism: the projected operator approach (see Ref. ~\cite{szekeres2001killer}) and the self-consistent operator formalism (see Ref.~\cite{prasad1985some}). If we use projected operators based on Ref. \cite{szekeres2001killer}, we arrive at a QSE-type formalism (as done in Ref.~\cite{fan2021equation}) to calculate excited-state properties.
The q-sc-EOM formalism developed in this work is based on the use of the self-consistent excitation operators to impose
the VAC. It should be noted that this concept has been utilized in the development of different excited-state methods in classical quantum chemistry~\cite{liu2018unitary,liu2022quadratic,hodecker2022theoretical,datta1993consistent}. 
\subsubsection{Accuracy of energy differences}
The qEOM method, which uses the EOM formulation in Eq.~\eqref{secular} with the conventional excitation manifold, $\{\hat{G}_I^{\dagger}\}\cup\{\hat{G}_I\}$ (details can be found in Ref.~\cite{qEOM2020}), may lead to large errors in calculated IPs and EAs due to the violation of the VAC. 
This can be seen through the qEOM-evaluated IPs and EAs added to almost exact ADAPT-VQE ground state of the H$_2$ molecule in the STO-3G basis in Fig.~\ref{qeomcompip}. 
Large deviation from FCI results can be observed in this image.
It should be noted that although single and double excitations span all the possible excitations in the case of H$_2$ molecule in STO-3G basis, the VAC is still not satisfied in qEOM. This is because the excitation manifold is not complete with respect to the exact ground-state when we use the $\{\hat{G}_I^{\dagger}\}\cup\{\hat{G}_I\}$ operator manifold corresponding to electron detached/atached states. 
One way to solve this issue is by increasing the size of the operator manifold. However, this would significantly increase the computational cost.
Figure~\ref{qeomcompee} shows the energies of the three lowest excited states evaluated using qEOM and q-sc-EOM methods for a rectangle geometry H$_4$ molecular system as a function of H$_2\cdots$H$_2$ separation distance.
The H-H bond distance is fixed at 1.5~\AA{} in the two H$_2$ molecules. 
We observe from Fig.~\ref{qeomcompee} that q-sc-EOM is more robust in strongly correlated situations, compared with the qEOM method. 
The first and third excited states of H$_4$ computed using the qEOM method show qualitatively wrong behaviour in the region shown in the figure.

The EEs obtained using QSE are not necessarily size-intensive for an inexact ground state. 
This is due to the inclusion of the identity operator in the operator manifold used in QSE.
We illustrate this point using an H$_2\cdots$H$_4$ molecular system as an example. Fig~\ref{SIQSE} shows the difference in EEs computed for an isolated H$_2$ molecule and an H$_2\cdots$H$_4$ molecular supersystem with no interaction between the H$_2$ and H$_4$ subsystems (the distance between H$_2$ and H$_4$ is taken as 100\AA{}). 
An inexact ground state is taken using an ADAPT-VQE simulation that is stopped after adding 3 operators.
An identity operator is added to the operator manifold of QSE which uses the  operator manifold  represented by $\{\hat{G}_I\}$ in Section~\ref{theory}.
QSE computations give an error of $\sim$81 mH in this test, while the q-sc-EOM method shows the correct behavior. 
In this scenario, the two EEs should be identical for a method that provides size-intensive EEs.
The magnitude of this size-intensivity error in QSE will depend on the accuracy of ground state.
Since it is expected that near-term quantum computers may not provide exact ground-state energies for all molecular cases, these size-intensivity errors may cause problems. 
We note here that, just like q-sc-EOM, qEOM method provides size-intensive EEs, IPs and EAs as well.
\subsubsection{Noise-sensitivity}

\begin{figure*}[tbp]
    \centering
    \subfloat{
        \includegraphics[width = 0.5\textwidth,height=10cm,keepaspectratio]{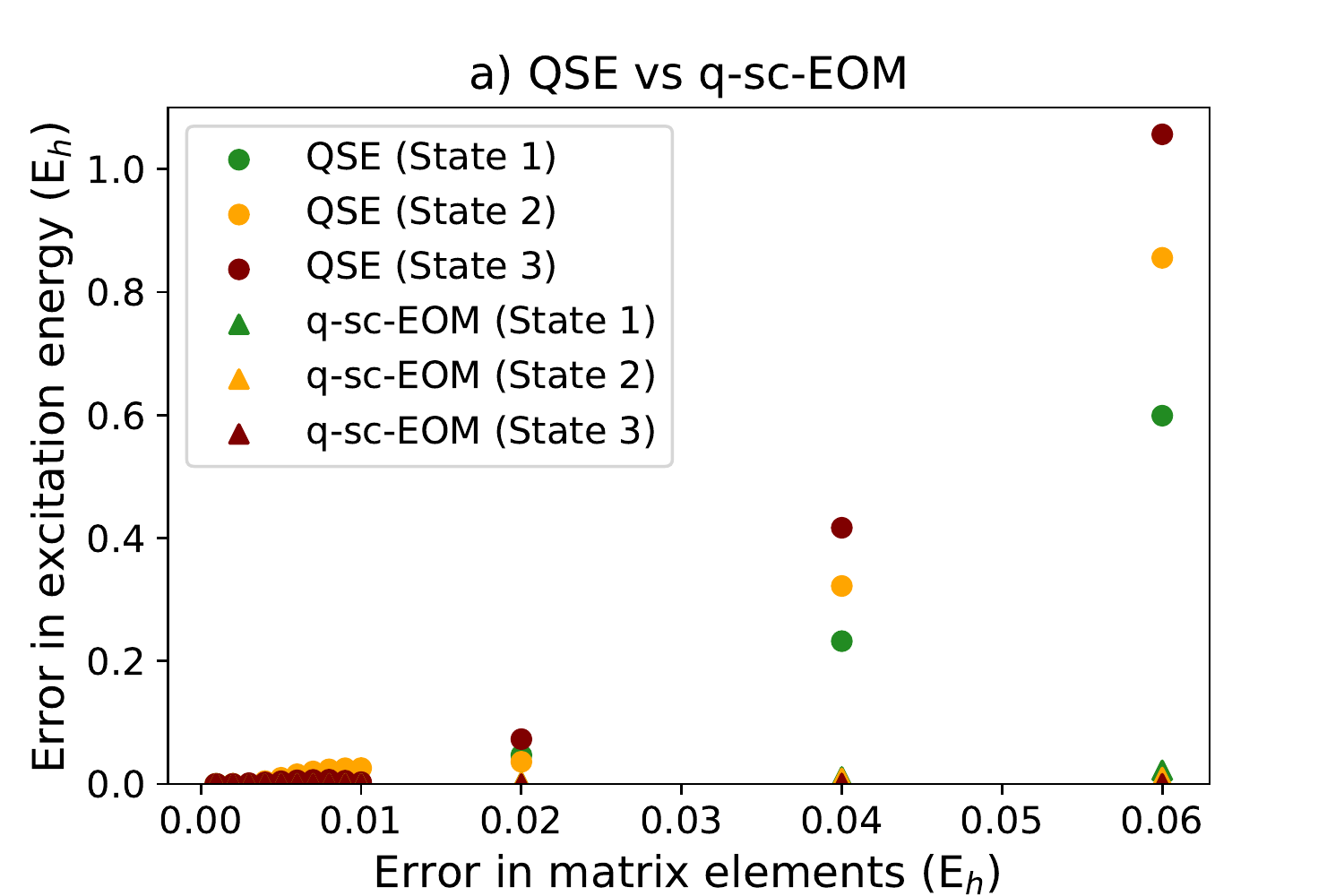}\label{noiseqse} 
    }
    \subfloat{
    \includegraphics[width = 0.5 \textwidth,height=10cm,keepaspectratio]{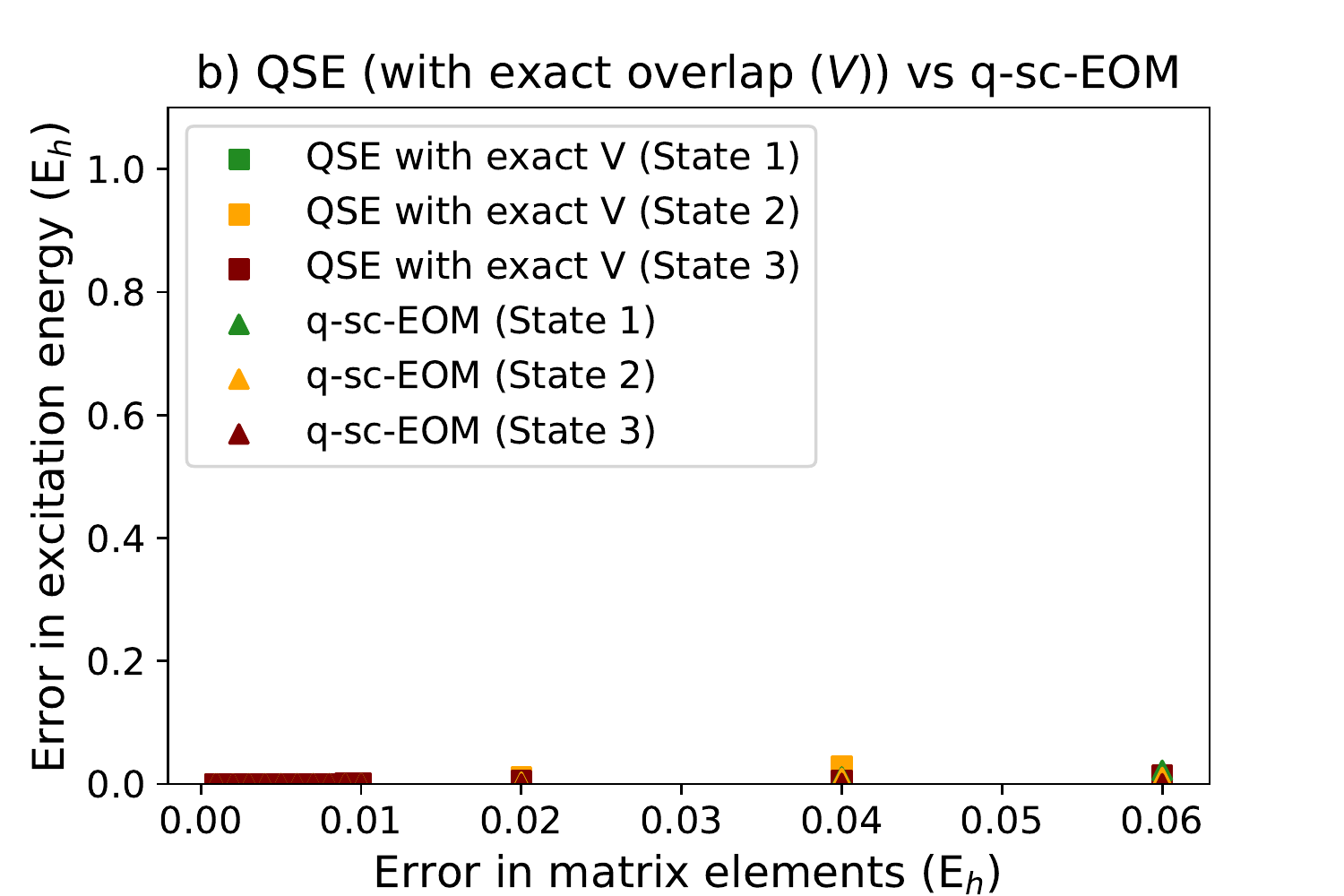}\label{noiseqseexact}
    }
    \caption{(a) Error in the first few excitation energies computed for H$_4$ using the QSE versus q-sc-EOM frameworks plotted against the maximum magnitude of errors in the matrix elements before diagonalization. (b) Error in first few excitation energies computed for H$_4$ using the QSE (with no error in overlap matrix ($V$) versus q-sc-EOM frameworks plotted against the maximum magnitude of errors in the matrix elements before diagonalization. 
    A rectangle geometry H$_4$ molecular system is used for the calculations with 1.5 \AA{} and 2.0 \AA{} as H-H distances.
    This shows that most of the noise sensitivity in QSE arises from noise in the computed overlap matrix, while the overlap matrix in q-sc-EOM is exactly the identity matrix and thus noise-free.}
    \label{noise}
\end{figure*}

Along with the above mentioned thoeretical benefits, q-sc-EOM also has important computational advantages. q-sc-EOM is expected to be more noise-resilient compared with qEOM and QSE. 
A prime reason for this is because q-sc-EOM requires upto 2-RDM type measurements while QSE and qEOM require measurement of higher-body RDMs.
Estimation of higher-body RDMs can significantly increase the noise in measured matrix elements.
This has been noted in Ref. \cite{huggins2021efficient} and shown using simple noise models that error in expectation value will scale by a factor exponential in the number of qubits measured together.
q-sc-EOM strictly requires measurement of upto 2-RDMs or 4 qubit measurements, while QSE and qEOM require measurements of upto 12 qubits at a time.

Additionally, we have also carried out a theoretical study based on matrix perturbation theory to compare the noise-resilience of QSE with q-sc-EOM on a noisy quantum computer.
The analysis is carried out for an interacting  pair of H$_2$ molecules (H$_4$ structure  with a rectangular geometry with 1.5 \AA{} and 2.0 \AA{} bond distances).
We model the effect of noise by adding random errors directly to each matrix element utilized by the two methods.
It should be noted that in real implementations where these matrix elements are measured on quantum computers, there will be multiple sources of errors (gate errors, measurement errors, etc.), which will finally create a net error in each matrix element. 
In our noise study, it is reasonable to add errors directly as we assume that all matrix elements, both in QSE and q-sc-EOM, will have errors of same magnitude when the same quantum resources are used. 
This is an appropriate (and rather conservative) assumption for two reasons: 
First,  all measurements in both q-sc-EOM and QSE utilize the same ground-state quantum circuit (with the difference being that in q-sc-EOM the circuit is applied on different reference states).
Second, q-sc-EOM makes use of at most 2-body RDMs, while QSE requires the estimation of higher-body RDMs as well, which is expected to generate errors of same or higher magnitude in the case of QSE.
A more detailed noise analysis performed on an actual quantum device will be the subject of future studies.

In Fig.~\ref{noiseqse}, we show the performance of QSE compared with q-sc-EOM, where to each matrix element we add random offsets sampled from a uniform distribution inline with matrix perturbation theory. The horizontal axes in the figure correspond to the upper bound of this distribution (i.e., the maximum allowed error).
The EEs are calculated after solving the eigenvalue equation of q-sc-EOM and the generalized eigenvalue equation of QSE (the latter resembles Eq.~\eqref{gen_eig}).
Errors are defined with respect to the EE values obtained from the associated method in the absence of noise.
Figure~\ref{noiseqse} shows the error averaged over 100,000 calculations of the EEs with different random offsets in each case. 
We can observe in the figure that QSE produces much larger errors at the same level of noise compared to q-sc-EOM, thus showing that we can expect q-sc-EOM to be more resilient to noise than QSE.


A key difference to consider between QSE and q-sc-EOM is that the overlap matrix $\textbf{V}$ in q-sc-EOM is exactly known to be the identity matrix and thus does not need to be measured, while in the case of QSE, the overlap matrix must be measured on the quantum computer. Note that the latter is also true for qEOM.
Figure~\ref{noiseqseexact} shows that if the overlap matrix is exactly known in QSE (i.e., we compute it exactly without noise), the noise-resilience of QSE is similar to that of q-sc-EOM. 
This suggests that the knowledge of the exact overlap matrix in q-sc-EOM is critical in providing noise-resilience, whereas methods that measure the overlap matrix and thus solve the generalized eigenvalue problem (such as QSE and qEOM) are expected to be more sensitive to noise. 
This noise-sensivity, we believe, is a direct result of a noise-sensitive matrix inversion step in solving generalized eigenvalue problem. 
A detailed analysis of this problem for general quantum algorithms for ground and excited state estimation will be presented in a future work. 

\section{Conclusion}\label{conclusion}

In this work, we propose a new method, named q-sc-EOM, for calculating molecular excitation energies using a quantum computer. The method can be implemented on top of any quantum variational algorithm used to obtain the ground state of the target molecule. 
Our approach is inspired by excited-state methods developed in quantum chemistry, specifically the ones based on the UCC theory. 
q-sc-EOM has several important benefits compared to current state-of-the-art excited-state methods for NISQ era, with theoretical benefits including (a) q-sc-EOM uses the self-consistent operators that satisfy the vacuum annihilation condition, thus it can be generalized to evaluate accurate vertical excitation energies, ionization potentials, and electron affinities; (b) Energy differences obtained using q-sc-EOM are strictly size-intensive, an important property that ensures their correct scaling with the size of the molecular system; (c) q-sc-EOM is a Hermitian theory, providing guaranteed real energy differences. The major computational benefit of q-sc-EOM is that it can be expected to more resilient to noise because (a) q-sc-EOM does not require higher than 2-body RDMs; and (b) it requires classical solution of eigenvalue equation, bypassing the noise-sensitive step associated inversion of overlap matrix in solving generalized eigenvalue equation in QSE and qEOM.
These benefits provide important theoretical and practical advantages for the computation of excitation energies on near-term quantum devices.

NISQ era devices are expected to be noisy with limited resources.
Thus, to achieve an advantage through these devices in quantum chemistry problems over classical computation, one should use methods that are meaningfully accurate, while at the same time resistant to errors and resource-efficient. 
The q-sc-EOM method proposed in this work is promising in this regard because it exhibits many of the crucial properties of the highly successful EOM-based quantum chemistry methods. At the same time, it remains resource-efficient and is expected to be resilient to noise compared to the currently available diagonalization-based methods. 
Our future studies will combine q-sc-EOM with the recently developed transcorrelated Hamiltonian formalism~\cite{kumar2022accurate,Motta_CTF12_quantum} to obtain quantitatively accurate excited-state properties with minimal utilization of quantum resources, which otherwise generally requires the use of large basis sets.

\section{Author contribution}
A.A, A.K: Conceptualization, Writing - Original Draft; N.J.M, A.A, A.K, V.A: Software, Formal analysis, V. A: Validation; H. G: Software, Methodology; N.J.M, S.E.E, E.B, Y.Z, L.C, S.T, P.A.D: Supervision, Project administration, Funding acquisition; All authors: Writing - Review \& Editing.

\section{Acknowledgement}
This research is supported by the Department of Energy (DOE). E.B. and N.J.M acknowledge Award No.  DE-SC0019199, and S.E.E. acknowledges support from the NSF, award number 1839136. 
A.A. would like to thank Dr. Luke Bertels for helpful discussions.
The authors thank the Advanced Research Computing (ARC) facility at Virginia Tech for the computational infrastructure. A.K., Y.Z., L.C., S.T., and P.A.D thank the support from the Laboratory Directed Research and Development (LDRD) program of Los Alamos National Laboratory (LANL) under project number 20200056DR. LANL is operated by Triad National Security, LLC, for the National Nuclear Security Administration of the U.S. Department of Energy (contract no. 89233218CNA000001).
\bibliography{paper}

\begin{thebibliography}{103}%
\makeatletter
\providecommand \@ifxundefined [1]{%
 \@ifx{#1\undefined}
}%
\providecommand \@ifnum [1]{%
 \ifnum #1\expandafter \@firstoftwo
 \else \expandafter \@secondoftwo
 \fi
}%
\providecommand \@ifx [1]{%
 \ifx #1\expandafter \@firstoftwo
 \else \expandafter \@secondoftwo
 \fi
}%
\providecommand \natexlab [1]{#1}%
\providecommand \enquote  [1]{``#1''}%
\providecommand \bibnamefont  [1]{#1}%
\providecommand \bibfnamefont [1]{#1}%
\providecommand \citenamefont [1]{#1}%
\providecommand \href@noop [0]{\@secondoftwo}%
\providecommand \href [0]{\begingroup \@sanitize@url \@href}%
\providecommand \@href[1]{\@@startlink{#1}\@@href}%
\providecommand \@@href[1]{\endgroup#1\@@endlink}%
\providecommand \@sanitize@url [0]{\catcode `\\12\catcode `\$12\catcode
  `\&12\catcode `\#12\catcode `\^12\catcode `\_12\catcode `\%12\relax}%
\providecommand \@@startlink[1]{}%
\providecommand \@@endlink[0]{}%
\providecommand \url  [0]{\begingroup\@sanitize@url \@url }%
\providecommand \@url [1]{\endgroup\@href {#1}{\urlprefix }}%
\providecommand \urlprefix  [0]{URL }%
\providecommand \Eprint [0]{\href }%
\providecommand \doibase [0]{https://doi.org/}%
\providecommand \selectlanguage [0]{\@gobble}%
\providecommand \bibinfo  [0]{\@secondoftwo}%
\providecommand \bibfield  [0]{\@secondoftwo}%
\providecommand \translation [1]{[#1]}%
\providecommand \BibitemOpen [0]{}%
\providecommand \bibitemStop [0]{}%
\providecommand \bibitemNoStop [0]{.\EOS\space}%
\providecommand \EOS [0]{\spacefactor3000\relax}%
\providecommand \BibitemShut  [1]{\csname bibitem#1\endcsname}%
\let\auto@bib@innerbib\@empty
\bibitem [{\citenamefont {Tilly}\ \emph {et~al.}(2021)\citenamefont {Tilly},
  \citenamefont {Chen}, \citenamefont {Cao}, \citenamefont {Picozzi},
  \citenamefont {Setia}, \citenamefont {Li}, \citenamefont {Grant},
  \citenamefont {Wossnig}, \citenamefont {Rungger}, \citenamefont {Booth} \emph
  {et~al.}}]{tilly2021variational}%
  \BibitemOpen
  \bibfield  {author} {\bibinfo {author} {\bibfnamefont {J.}~\bibnamefont
  {Tilly}}, \bibinfo {author} {\bibfnamefont {H.}~\bibnamefont {Chen}},
  \bibinfo {author} {\bibfnamefont {S.}~\bibnamefont {Cao}}, \bibinfo {author}
  {\bibfnamefont {D.}~\bibnamefont {Picozzi}}, \bibinfo {author} {\bibfnamefont
  {K.}~\bibnamefont {Setia}}, \bibinfo {author} {\bibfnamefont
  {Y.}~\bibnamefont {Li}}, \bibinfo {author} {\bibfnamefont {E.}~\bibnamefont
  {Grant}}, \bibinfo {author} {\bibfnamefont {L.}~\bibnamefont {Wossnig}},
  \bibinfo {author} {\bibfnamefont {I.}~\bibnamefont {Rungger}}, \bibinfo
  {author} {\bibfnamefont {G.~H.}\ \bibnamefont {Booth}}, \emph {et~al.},\
  }\bibfield  {title} {\bibinfo {title} {The variational quantum eigensolver: a
  review of methods and best practices},\ }\href@noop {} {\bibfield  {journal}
  {\bibinfo  {journal} {arXiv preprint arXiv:2111.05176}\ } (\bibinfo {year}
  {2021})}\BibitemShut {NoStop}%
\bibitem [{\citenamefont {McArdle}\ \emph {et~al.}(2020)\citenamefont
  {McArdle}, \citenamefont {Endo}, \citenamefont {Aspuru-Guzik}, \citenamefont
  {Benjamin},\ and\ \citenamefont {Yuan}}]{mcardle2020quantum}%
  \BibitemOpen
  \bibfield  {author} {\bibinfo {author} {\bibfnamefont {S.}~\bibnamefont
  {McArdle}}, \bibinfo {author} {\bibfnamefont {S.}~\bibnamefont {Endo}},
  \bibinfo {author} {\bibfnamefont {A.}~\bibnamefont {Aspuru-Guzik}}, \bibinfo
  {author} {\bibfnamefont {S.~C.}\ \bibnamefont {Benjamin}},\ and\ \bibinfo
  {author} {\bibfnamefont {X.}~\bibnamefont {Yuan}},\ }\bibfield  {title}
  {\bibinfo {title} {Quantum comput. chem.},\ }\href@noop {} {\bibfield
  {journal} {\bibinfo  {journal} {Rev. Modern Phys.}\ }\textbf {\bibinfo
  {volume} {92}},\ \bibinfo {pages} {015003} (\bibinfo {year}
  {2020})}\BibitemShut {NoStop}%
\bibitem [{\citenamefont {Cerezo}\ \emph {et~al.}(2021)\citenamefont {Cerezo},
  \citenamefont {Arrasmith}, \citenamefont {Babbush}, \citenamefont {Benjamin},
  \citenamefont {Endo}, \citenamefont {Fujii}, \citenamefont {McClean},
  \citenamefont {Mitarai}, \citenamefont {Yuan}, \citenamefont {Cincio} \emph
  {et~al.}}]{cerezo2021variational}%
  \BibitemOpen
  \bibfield  {author} {\bibinfo {author} {\bibfnamefont {M.}~\bibnamefont
  {Cerezo}}, \bibinfo {author} {\bibfnamefont {A.}~\bibnamefont {Arrasmith}},
  \bibinfo {author} {\bibfnamefont {R.}~\bibnamefont {Babbush}}, \bibinfo
  {author} {\bibfnamefont {S.~C.}\ \bibnamefont {Benjamin}}, \bibinfo {author}
  {\bibfnamefont {S.}~\bibnamefont {Endo}}, \bibinfo {author} {\bibfnamefont
  {K.}~\bibnamefont {Fujii}}, \bibinfo {author} {\bibfnamefont {J.~R.}\
  \bibnamefont {McClean}}, \bibinfo {author} {\bibfnamefont {K.}~\bibnamefont
  {Mitarai}}, \bibinfo {author} {\bibfnamefont {X.}~\bibnamefont {Yuan}},
  \bibinfo {author} {\bibfnamefont {L.}~\bibnamefont {Cincio}}, \emph
  {et~al.},\ }\bibfield  {title} {\bibinfo {title} {Variational quantum
  algorithms},\ }\href@noop {} {\bibfield  {journal} {\bibinfo  {journal} {Nat.
  Rev. Phys.}\ ,\ \bibinfo {pages} {1}} (\bibinfo {year} {2021})}\BibitemShut
  {NoStop}%
\bibitem [{\citenamefont {Peruzzo}\ \emph {et~al.}(2014)\citenamefont
  {Peruzzo}, \citenamefont {McClean}, \citenamefont {Shadbolt}, \citenamefont
  {Yung}, \citenamefont {Zhou}, \citenamefont {Love}, \citenamefont
  {Aspuru-Guzik},\ and\ \citenamefont {O’brien}}]{peruzzo2014variational}%
  \BibitemOpen
  \bibfield  {author} {\bibinfo {author} {\bibfnamefont {A.}~\bibnamefont
  {Peruzzo}}, \bibinfo {author} {\bibfnamefont {J.}~\bibnamefont {McClean}},
  \bibinfo {author} {\bibfnamefont {P.}~\bibnamefont {Shadbolt}}, \bibinfo
  {author} {\bibfnamefont {M.-H.}\ \bibnamefont {Yung}}, \bibinfo {author}
  {\bibfnamefont {X.-Q.}\ \bibnamefont {Zhou}}, \bibinfo {author}
  {\bibfnamefont {P.~J.}\ \bibnamefont {Love}}, \bibinfo {author}
  {\bibfnamefont {A.}~\bibnamefont {Aspuru-Guzik}},\ and\ \bibinfo {author}
  {\bibfnamefont {J.~L.}\ \bibnamefont {O’brien}},\ }\bibfield  {title}
  {\bibinfo {title} {A variational eigenvalue solver on a photonic quantum
  processor},\ }\href@noop {} {\bibfield  {journal} {\bibinfo  {journal} {Nat.
  Comm.}\ }\textbf {\bibinfo {volume} {5}},\ \bibinfo {pages} {1} (\bibinfo
  {year} {2014})}\BibitemShut {NoStop}%
\bibitem [{\citenamefont {Magann}\ \emph {et~al.}(2021)\citenamefont {Magann},
  \citenamefont {Arenz}, \citenamefont {Grace}, \citenamefont {Ho},
  \citenamefont {Kosut}, \citenamefont {McClean}, \citenamefont {Rabitz},\ and\
  \citenamefont {Sarovar}}]{magann2021pulses}%
  \BibitemOpen
  \bibfield  {author} {\bibinfo {author} {\bibfnamefont {A.~B.}\ \bibnamefont
  {Magann}}, \bibinfo {author} {\bibfnamefont {C.}~\bibnamefont {Arenz}},
  \bibinfo {author} {\bibfnamefont {M.~D.}\ \bibnamefont {Grace}}, \bibinfo
  {author} {\bibfnamefont {T.-S.}\ \bibnamefont {Ho}}, \bibinfo {author}
  {\bibfnamefont {R.~L.}\ \bibnamefont {Kosut}}, \bibinfo {author}
  {\bibfnamefont {J.~R.}\ \bibnamefont {McClean}}, \bibinfo {author}
  {\bibfnamefont {H.~A.}\ \bibnamefont {Rabitz}},\ and\ \bibinfo {author}
  {\bibfnamefont {M.}~\bibnamefont {Sarovar}},\ }\bibfield  {title} {\bibinfo
  {title} {From pulses to circuits and back again: A quantum optimal control
  perspective on variational quantum algorithms},\ }\href@noop {} {\bibfield
  {journal} {\bibinfo  {journal} {PRX Quantum}\ }\textbf {\bibinfo {volume}
  {2}},\ \bibinfo {pages} {010101} (\bibinfo {year} {2021})}\BibitemShut
  {NoStop}%
\bibitem [{\citenamefont {Kandala}\ \emph {et~al.}(2017)\citenamefont
  {Kandala}, \citenamefont {Mezzacapo}, \citenamefont {Temme}, \citenamefont
  {Takita}, \citenamefont {Brink}, \citenamefont {Chow},\ and\ \citenamefont
  {Gambetta}}]{kandala2017hardware}%
  \BibitemOpen
  \bibfield  {author} {\bibinfo {author} {\bibfnamefont {A.}~\bibnamefont
  {Kandala}}, \bibinfo {author} {\bibfnamefont {A.}~\bibnamefont {Mezzacapo}},
  \bibinfo {author} {\bibfnamefont {K.}~\bibnamefont {Temme}}, \bibinfo
  {author} {\bibfnamefont {M.}~\bibnamefont {Takita}}, \bibinfo {author}
  {\bibfnamefont {M.}~\bibnamefont {Brink}}, \bibinfo {author} {\bibfnamefont
  {J.~M.}\ \bibnamefont {Chow}},\ and\ \bibinfo {author} {\bibfnamefont
  {J.~M.}\ \bibnamefont {Gambetta}},\ }\bibfield  {title} {\bibinfo {title}
  {Hardware-efficient variational quantum eigensolver for small molecules and
  quantum magnets},\ }\href@noop {} {\bibfield  {journal} {\bibinfo  {journal}
  {Nature}\ }\textbf {\bibinfo {volume} {549}},\ \bibinfo {pages} {242}
  (\bibinfo {year} {2017})}\BibitemShut {NoStop}%
\bibitem [{\citenamefont {Cao}\ \emph {et~al.}(2019)\citenamefont {Cao},
  \citenamefont {Romero}, \citenamefont {Olson}, \citenamefont {Degroote},
  \citenamefont {Johnson}, \citenamefont {Kieferov{\'a}}, \citenamefont
  {Kivlichan}, \citenamefont {Menke}, \citenamefont {Peropadre}, \citenamefont
  {Sawaya} \emph {et~al.}}]{cao2019quantum}%
  \BibitemOpen
  \bibfield  {author} {\bibinfo {author} {\bibfnamefont {Y.}~\bibnamefont
  {Cao}}, \bibinfo {author} {\bibfnamefont {J.}~\bibnamefont {Romero}},
  \bibinfo {author} {\bibfnamefont {J.~P.}\ \bibnamefont {Olson}}, \bibinfo
  {author} {\bibfnamefont {M.}~\bibnamefont {Degroote}}, \bibinfo {author}
  {\bibfnamefont {P.~D.}\ \bibnamefont {Johnson}}, \bibinfo {author}
  {\bibfnamefont {M.}~\bibnamefont {Kieferov{\'a}}}, \bibinfo {author}
  {\bibfnamefont {I.~D.}\ \bibnamefont {Kivlichan}}, \bibinfo {author}
  {\bibfnamefont {T.}~\bibnamefont {Menke}}, \bibinfo {author} {\bibfnamefont
  {B.}~\bibnamefont {Peropadre}}, \bibinfo {author} {\bibfnamefont {N.~P.}\
  \bibnamefont {Sawaya}}, \emph {et~al.},\ }\bibfield  {title} {\bibinfo
  {title} {Quantum chemistry in the age of quantum computing},\ }\href@noop {}
  {\bibfield  {journal} {\bibinfo  {journal} {Chem. Rev.}\ }\textbf {\bibinfo
  {volume} {119}},\ \bibinfo {pages} {10856} (\bibinfo {year}
  {2019})}\BibitemShut {NoStop}%
\bibitem [{\citenamefont {Fedorov}\ \emph {et~al.}(2022)\citenamefont
  {Fedorov}, \citenamefont {Peng}, \citenamefont {Govind},\ and\ \citenamefont
  {Alexeev}}]{fedorov2022vqe}%
  \BibitemOpen
  \bibfield  {author} {\bibinfo {author} {\bibfnamefont {D.~A.}\ \bibnamefont
  {Fedorov}}, \bibinfo {author} {\bibfnamefont {B.}~\bibnamefont {Peng}},
  \bibinfo {author} {\bibfnamefont {N.}~\bibnamefont {Govind}},\ and\ \bibinfo
  {author} {\bibfnamefont {Y.}~\bibnamefont {Alexeev}},\ }\bibfield  {title}
  {\bibinfo {title} {Vqe method: A short survey and recent developments},\
  }\href@noop {} {\bibfield  {journal} {\bibinfo  {journal} {Mat. Theory}\
  }\textbf {\bibinfo {volume} {6}},\ \bibinfo {pages} {1} (\bibinfo {year}
  {2022})}\BibitemShut {NoStop}%
\bibitem [{\citenamefont {Preskill}(2018)}]{Preskill2018}%
  \BibitemOpen
  \bibfield  {author} {\bibinfo {author} {\bibfnamefont {J.}~\bibnamefont
  {Preskill}},\ }\bibfield  {title} {\bibinfo {title} {Quantum {C}omputing in
  the {NISQ} era and beyond},\ }\href
  {https://doi.org/10.22331/q-2018-08-06-79} {\bibfield  {journal} {\bibinfo
  {journal} {{Quantum}}\ }\textbf {\bibinfo {volume} {2}},\ \bibinfo {pages}
  {79} (\bibinfo {year} {2018})}\BibitemShut {NoStop}%
\bibitem [{\citenamefont {Sharma}\ \emph {et~al.}(2020)\citenamefont {Sharma},
  \citenamefont {Khatri}, \citenamefont {Cerezo},\ and\ \citenamefont
  {Coles}}]{sharma2020noise}%
  \BibitemOpen
  \bibfield  {author} {\bibinfo {author} {\bibfnamefont {K.}~\bibnamefont
  {Sharma}}, \bibinfo {author} {\bibfnamefont {S.}~\bibnamefont {Khatri}},
  \bibinfo {author} {\bibfnamefont {M.}~\bibnamefont {Cerezo}},\ and\ \bibinfo
  {author} {\bibfnamefont {P.~J.}\ \bibnamefont {Coles}},\ }\bibfield  {title}
  {\bibinfo {title} {Noise resilience of variational quantum compiling},\
  }\href@noop {} {\bibfield  {journal} {\bibinfo  {journal} {New J. Phys.}\
  }\textbf {\bibinfo {volume} {22}},\ \bibinfo {pages} {043006} (\bibinfo
  {year} {2020})}\BibitemShut {NoStop}%
\bibitem [{\citenamefont {Grimsley}\ \emph {et~al.}(2019)\citenamefont
  {Grimsley}, \citenamefont {Economou}, \citenamefont {Barnes},\ and\
  \citenamefont {Mayhall}}]{grimsley2019adaptive}%
  \BibitemOpen
  \bibfield  {author} {\bibinfo {author} {\bibfnamefont {H.~R.}\ \bibnamefont
  {Grimsley}}, \bibinfo {author} {\bibfnamefont {S.~E.}\ \bibnamefont
  {Economou}}, \bibinfo {author} {\bibfnamefont {E.}~\bibnamefont {Barnes}},\
  and\ \bibinfo {author} {\bibfnamefont {N.~J.}\ \bibnamefont {Mayhall}},\
  }\bibfield  {title} {\bibinfo {title} {An adaptive variational algorithm for
  exact molecular simulations on a quantum computer},\ }\href@noop {}
  {\bibfield  {journal} {\bibinfo  {journal} {Nat. Comm.}\ }\textbf {\bibinfo
  {volume} {10}},\ \bibinfo {pages} {1} (\bibinfo {year} {2019})}\BibitemShut
  {NoStop}%
\bibitem [{\citenamefont {Tang}\ \emph {et~al.}(2021)\citenamefont {Tang},
  \citenamefont {Shkolnikov}, \citenamefont {Barron}, \citenamefont {Grimsley},
  \citenamefont {Mayhall}, \citenamefont {Barnes},\ and\ \citenamefont
  {Economou}}]{tang2021qubit}%
  \BibitemOpen
  \bibfield  {author} {\bibinfo {author} {\bibfnamefont {H.~L.}\ \bibnamefont
  {Tang}}, \bibinfo {author} {\bibfnamefont {V.~O.}\ \bibnamefont
  {Shkolnikov}}, \bibinfo {author} {\bibfnamefont {G.~S.}\ \bibnamefont
  {Barron}}, \bibinfo {author} {\bibfnamefont {H.~R.}\ \bibnamefont
  {Grimsley}}, \bibinfo {author} {\bibfnamefont {N.~J.}\ \bibnamefont
  {Mayhall}}, \bibinfo {author} {\bibfnamefont {E.}~\bibnamefont {Barnes}},\
  and\ \bibinfo {author} {\bibfnamefont {S.~E.}\ \bibnamefont {Economou}},\
  }\bibfield  {title} {\bibinfo {title} {qubit-adapt-vqe: An adaptive algorithm
  for constructing hardware-efficient ans{\"a}tze on a quantum processor},\
  }\href@noop {} {\bibfield  {journal} {\bibinfo  {journal} {PRX Quantum}\
  }\textbf {\bibinfo {volume} {2}},\ \bibinfo {pages} {020310} (\bibinfo {year}
  {2021})}\BibitemShut {NoStop}%
\bibitem [{\citenamefont {Huggins}\ \emph {et~al.}(2020)\citenamefont
  {Huggins}, \citenamefont {Lee}, \citenamefont {Baek}, \citenamefont
  {O’Gorman},\ and\ \citenamefont {Whaley}}]{huggins2020non}%
  \BibitemOpen
  \bibfield  {author} {\bibinfo {author} {\bibfnamefont {W.~J.}\ \bibnamefont
  {Huggins}}, \bibinfo {author} {\bibfnamefont {J.}~\bibnamefont {Lee}},
  \bibinfo {author} {\bibfnamefont {U.}~\bibnamefont {Baek}}, \bibinfo {author}
  {\bibfnamefont {B.}~\bibnamefont {O’Gorman}},\ and\ \bibinfo {author}
  {\bibfnamefont {K.~B.}\ \bibnamefont {Whaley}},\ }\bibfield  {title}
  {\bibinfo {title} {A non-orthogonal variational quantum eigensolver},\
  }\href@noop {} {\bibfield  {journal} {\bibinfo  {journal} {New J. Phys.}\
  }\textbf {\bibinfo {volume} {22}},\ \bibinfo {pages} {073009} (\bibinfo
  {year} {2020})}\BibitemShut {NoStop}%
\bibitem [{\citenamefont {Lee}\ \emph {et~al.}(2018)\citenamefont {Lee},
  \citenamefont {Huggins}, \citenamefont {Head-Gordon},\ and\ \citenamefont
  {Whaley}}]{lee2018generalized}%
  \BibitemOpen
  \bibfield  {author} {\bibinfo {author} {\bibfnamefont {J.}~\bibnamefont
  {Lee}}, \bibinfo {author} {\bibfnamefont {W.~J.}\ \bibnamefont {Huggins}},
  \bibinfo {author} {\bibfnamefont {M.}~\bibnamefont {Head-Gordon}},\ and\
  \bibinfo {author} {\bibfnamefont {K.~B.}\ \bibnamefont {Whaley}},\ }\bibfield
   {title} {\bibinfo {title} {Generalized unitary coupled cluster wave
  functions for quantum computation},\ }\href@noop {} {\bibfield  {journal}
  {\bibinfo  {journal} {J. Chem. Theory Comput.}\ }\textbf {\bibinfo {volume}
  {15}},\ \bibinfo {pages} {311} (\bibinfo {year} {2018})}\BibitemShut
  {NoStop}%
\bibitem [{\citenamefont {Ryabinkin}\ \emph {et~al.}(2018)\citenamefont
  {Ryabinkin}, \citenamefont {Yen}, \citenamefont {Genin},\ and\ \citenamefont
  {Izmaylov}}]{ryabinkin2018qubit}%
  \BibitemOpen
  \bibfield  {author} {\bibinfo {author} {\bibfnamefont {I.~G.}\ \bibnamefont
  {Ryabinkin}}, \bibinfo {author} {\bibfnamefont {T.-C.}\ \bibnamefont {Yen}},
  \bibinfo {author} {\bibfnamefont {S.~N.}\ \bibnamefont {Genin}},\ and\
  \bibinfo {author} {\bibfnamefont {A.~F.}\ \bibnamefont {Izmaylov}},\
  }\bibfield  {title} {\bibinfo {title} {Qubit coupled cluster method: a
  systematic approach to quantum chemistry on a quantum computer},\ }\href@noop
  {} {\bibfield  {journal} {\bibinfo  {journal} {J. Chem. Theory Comput.}\
  }\textbf {\bibinfo {volume} {14}},\ \bibinfo {pages} {6317} (\bibinfo {year}
  {2018})}\BibitemShut {NoStop}%
\bibitem [{\citenamefont {O’Malley}\ \emph {et~al.}(2016)\citenamefont
  {O’Malley}, \citenamefont {Babbush}, \citenamefont {Kivlichan},
  \citenamefont {Romero}, \citenamefont {McClean}, \citenamefont {Barends},
  \citenamefont {Kelly}, \citenamefont {Roushan}, \citenamefont {Tranter},
  \citenamefont {Ding} \emph {et~al.}}]{o2016scalable}%
  \BibitemOpen
  \bibfield  {author} {\bibinfo {author} {\bibfnamefont {P.~J.~J.}\
  \bibnamefont {O’Malley}}, \bibinfo {author} {\bibfnamefont
  {R.}~\bibnamefont {Babbush}}, \bibinfo {author} {\bibfnamefont {I.~D.}\
  \bibnamefont {Kivlichan}}, \bibinfo {author} {\bibfnamefont {J.}~\bibnamefont
  {Romero}}, \bibinfo {author} {\bibfnamefont {J.~R.}\ \bibnamefont {McClean}},
  \bibinfo {author} {\bibfnamefont {R.}~\bibnamefont {Barends}}, \bibinfo
  {author} {\bibfnamefont {J.}~\bibnamefont {Kelly}}, \bibinfo {author}
  {\bibfnamefont {P.}~\bibnamefont {Roushan}}, \bibinfo {author} {\bibfnamefont
  {A.}~\bibnamefont {Tranter}}, \bibinfo {author} {\bibfnamefont
  {N.}~\bibnamefont {Ding}}, \emph {et~al.},\ }\bibfield  {title} {\bibinfo
  {title} {Scalable quantum simulation of molecular energies},\ }\href@noop {}
  {\bibfield  {journal} {\bibinfo  {journal} {Phys. Rev. X}\ }\textbf {\bibinfo
  {volume} {6}},\ \bibinfo {pages} {031007} (\bibinfo {year}
  {2016})}\BibitemShut {NoStop}%
\bibitem [{\citenamefont {Nam}\ \emph {et~al.}(2020)\citenamefont {Nam},
  \citenamefont {Chen}, \citenamefont {Pisenti}, \citenamefont {Wright},
  \citenamefont {Delaney}, \citenamefont {Maslov}, \citenamefont {Brown},
  \citenamefont {Allen}, \citenamefont {Amini}, \citenamefont {Apisdorf} \emph
  {et~al.}}]{nam2020ground}%
  \BibitemOpen
  \bibfield  {author} {\bibinfo {author} {\bibfnamefont {Y.}~\bibnamefont
  {Nam}}, \bibinfo {author} {\bibfnamefont {J.-S.}\ \bibnamefont {Chen}},
  \bibinfo {author} {\bibfnamefont {N.~C.}\ \bibnamefont {Pisenti}}, \bibinfo
  {author} {\bibfnamefont {K.}~\bibnamefont {Wright}}, \bibinfo {author}
  {\bibfnamefont {C.}~\bibnamefont {Delaney}}, \bibinfo {author} {\bibfnamefont
  {D.}~\bibnamefont {Maslov}}, \bibinfo {author} {\bibfnamefont {K.~R.}\
  \bibnamefont {Brown}}, \bibinfo {author} {\bibfnamefont {S.}~\bibnamefont
  {Allen}}, \bibinfo {author} {\bibfnamefont {J.~M.}\ \bibnamefont {Amini}},
  \bibinfo {author} {\bibfnamefont {J.}~\bibnamefont {Apisdorf}}, \emph
  {et~al.},\ }\bibfield  {title} {\bibinfo {title} {Ground-state energy
  estimation of the water molecule on a trapped-ion quantum computer},\
  }\href@noop {} {\bibfield  {journal} {\bibinfo  {journal} {NPJ Quantum
  Info.}\ }\textbf {\bibinfo {volume} {6}},\ \bibinfo {pages} {1} (\bibinfo
  {year} {2020})}\BibitemShut {NoStop}%
\bibitem [{\citenamefont {McCaskey}\ \emph {et~al.}(2019)\citenamefont
  {McCaskey}, \citenamefont {Parks}, \citenamefont {Jakowski}, \citenamefont
  {Moore}, \citenamefont {Morris}, \citenamefont {Humble},\ and\ \citenamefont
  {Pooser}}]{mccaskey2019quantum}%
  \BibitemOpen
  \bibfield  {author} {\bibinfo {author} {\bibfnamefont {A.~J.}\ \bibnamefont
  {McCaskey}}, \bibinfo {author} {\bibfnamefont {Z.~P.}\ \bibnamefont {Parks}},
  \bibinfo {author} {\bibfnamefont {J.}~\bibnamefont {Jakowski}}, \bibinfo
  {author} {\bibfnamefont {S.~V.}\ \bibnamefont {Moore}}, \bibinfo {author}
  {\bibfnamefont {T.~D.}\ \bibnamefont {Morris}}, \bibinfo {author}
  {\bibfnamefont {T.~S.}\ \bibnamefont {Humble}},\ and\ \bibinfo {author}
  {\bibfnamefont {R.~C.}\ \bibnamefont {Pooser}},\ }\bibfield  {title}
  {\bibinfo {title} {Quantum chemistry as a benchmark for near-term quantum
  computers},\ }\href@noop {} {\bibfield  {journal} {\bibinfo  {journal} {NPJ
  Quantum Info.}\ }\textbf {\bibinfo {volume} {5}},\ \bibinfo {pages} {1}
  (\bibinfo {year} {2019})}\BibitemShut {NoStop}%
\bibitem [{\citenamefont {Hempel}\ \emph {et~al.}(2018)\citenamefont {Hempel},
  \citenamefont {Maier}, \citenamefont {Romero}, \citenamefont {McClean},
  \citenamefont {Monz}, \citenamefont {Shen}, \citenamefont {Jurcevic},
  \citenamefont {Lanyon}, \citenamefont {Love}, \citenamefont {Babbush} \emph
  {et~al.}}]{hempel2018quantum}%
  \BibitemOpen
  \bibfield  {author} {\bibinfo {author} {\bibfnamefont {C.}~\bibnamefont
  {Hempel}}, \bibinfo {author} {\bibfnamefont {C.}~\bibnamefont {Maier}},
  \bibinfo {author} {\bibfnamefont {J.}~\bibnamefont {Romero}}, \bibinfo
  {author} {\bibfnamefont {J.}~\bibnamefont {McClean}}, \bibinfo {author}
  {\bibfnamefont {T.}~\bibnamefont {Monz}}, \bibinfo {author} {\bibfnamefont
  {H.}~\bibnamefont {Shen}}, \bibinfo {author} {\bibfnamefont {P.}~\bibnamefont
  {Jurcevic}}, \bibinfo {author} {\bibfnamefont {B.~P.}\ \bibnamefont
  {Lanyon}}, \bibinfo {author} {\bibfnamefont {P.}~\bibnamefont {Love}},
  \bibinfo {author} {\bibfnamefont {R.}~\bibnamefont {Babbush}}, \emph
  {et~al.},\ }\bibfield  {title} {\bibinfo {title} {Quantum chemistry
  calculations on a trapped-ion quantum simulator},\ }\href@noop {} {\bibfield
  {journal} {\bibinfo  {journal} {Phys. Rev. X}\ }\textbf {\bibinfo {volume}
  {8}},\ \bibinfo {pages} {031022} (\bibinfo {year} {2018})}\BibitemShut
  {NoStop}%
\bibitem [{\citenamefont {Gao}\ \emph {et~al.}(2021)\citenamefont {Gao},
  \citenamefont {Jones}, \citenamefont {Motta}, \citenamefont {Sugawara},
  \citenamefont {Watanabe}, \citenamefont {Kobayashi}, \citenamefont
  {Watanabe}, \citenamefont {Ohnishi}, \citenamefont {Nakamura},\ and\
  \citenamefont {Yamamoto}}]{gao2021applications}%
  \BibitemOpen
  \bibfield  {author} {\bibinfo {author} {\bibfnamefont {Q.}~\bibnamefont
  {Gao}}, \bibinfo {author} {\bibfnamefont {G.~O.}\ \bibnamefont {Jones}},
  \bibinfo {author} {\bibfnamefont {M.}~\bibnamefont {Motta}}, \bibinfo
  {author} {\bibfnamefont {M.}~\bibnamefont {Sugawara}}, \bibinfo {author}
  {\bibfnamefont {H.~C.}\ \bibnamefont {Watanabe}}, \bibinfo {author}
  {\bibfnamefont {T.}~\bibnamefont {Kobayashi}}, \bibinfo {author}
  {\bibfnamefont {E.}~\bibnamefont {Watanabe}}, \bibinfo {author}
  {\bibfnamefont {Y.}~\bibnamefont {Ohnishi}}, \bibinfo {author} {\bibfnamefont
  {H.}~\bibnamefont {Nakamura}},\ and\ \bibinfo {author} {\bibfnamefont
  {N.}~\bibnamefont {Yamamoto}},\ }\bibfield  {title} {\bibinfo {title}
  {Applications of quantum computing for investigations of electronic
  transitions in phenylsulfonyl-carbazole tadf emitters},\ }\href@noop {}
  {\bibfield  {journal} {\bibinfo  {journal} {NPJ Comput. Mat.}\ }\textbf
  {\bibinfo {volume} {7}},\ \bibinfo {pages} {1} (\bibinfo {year}
  {2021})}\BibitemShut {NoStop}%
\bibitem [{\citenamefont {Smart}\ and\ \citenamefont
  {Mazziotti}(2021)}]{smart2021quantum}%
  \BibitemOpen
  \bibfield  {author} {\bibinfo {author} {\bibfnamefont {S.~E.}\ \bibnamefont
  {Smart}}\ and\ \bibinfo {author} {\bibfnamefont {D.~A.}\ \bibnamefont
  {Mazziotti}},\ }\bibfield  {title} {\bibinfo {title} {Quantum solver of
  contracted eigenvalue equations for scalable molecular simulations on quantum
  computing devices},\ }\href@noop {} {\bibfield  {journal} {\bibinfo
  {journal} {Phys. Rev. Lett.}\ }\textbf {\bibinfo {volume} {126}},\ \bibinfo
  {pages} {070504} (\bibinfo {year} {2021})}\BibitemShut {NoStop}%
\bibitem [{\citenamefont {Meitei}\ \emph {et~al.}(2021)\citenamefont {Meitei},
  \citenamefont {Gard}, \citenamefont {Barron}, \citenamefont {Pappas},
  \citenamefont {Economou}, \citenamefont {Barnes},\ and\ \citenamefont
  {Mayhall}}]{Meitei2020}%
  \BibitemOpen
  \bibfield  {author} {\bibinfo {author} {\bibfnamefont {O.~R.}\ \bibnamefont
  {Meitei}}, \bibinfo {author} {\bibfnamefont {B.~T.}\ \bibnamefont {Gard}},
  \bibinfo {author} {\bibfnamefont {G.~S.}\ \bibnamefont {Barron}}, \bibinfo
  {author} {\bibfnamefont {D.~P.}\ \bibnamefont {Pappas}}, \bibinfo {author}
  {\bibfnamefont {S.~E.}\ \bibnamefont {Economou}}, \bibinfo {author}
  {\bibfnamefont {E.}~\bibnamefont {Barnes}},\ and\ \bibinfo {author}
  {\bibfnamefont {N.~J.}\ \bibnamefont {Mayhall}},\ }\bibfield  {title}
  {\bibinfo {title} {Gate-free state preparation for fast variational quantum
  eigensolver simulations},\ }\href
  {https://doi.org/10.1038/s41534-021-00493-0} {\bibfield  {journal} {\bibinfo
  {journal} {NPJ Quantum Info.}\ }\textbf {\bibinfo {volume} {7}},\ \bibinfo
  {pages} {155} (\bibinfo {year} {2021})}\BibitemShut {NoStop}%
\bibitem [{\citenamefont {Nelson}\ \emph {et~al.}(2020)\citenamefont {Nelson},
  \citenamefont {White}, \citenamefont {Bjorgaard}, \citenamefont {Sifain},
  \citenamefont {Zhang}, \citenamefont {Nebgen}, \citenamefont
  {Fernandez-Alberti}, \citenamefont {Mozyrsky}, \citenamefont {Roitberg},\
  and\ \citenamefont {Tretiak}}]{nelson2020non}%
  \BibitemOpen
  \bibfield  {author} {\bibinfo {author} {\bibfnamefont {T.~R.}\ \bibnamefont
  {Nelson}}, \bibinfo {author} {\bibfnamefont {A.~J.}\ \bibnamefont {White}},
  \bibinfo {author} {\bibfnamefont {J.~A.}\ \bibnamefont {Bjorgaard}}, \bibinfo
  {author} {\bibfnamefont {A.~E.}\ \bibnamefont {Sifain}}, \bibinfo {author}
  {\bibfnamefont {Y.}~\bibnamefont {Zhang}}, \bibinfo {author} {\bibfnamefont
  {B.}~\bibnamefont {Nebgen}}, \bibinfo {author} {\bibfnamefont
  {S.}~\bibnamefont {Fernandez-Alberti}}, \bibinfo {author} {\bibfnamefont
  {D.}~\bibnamefont {Mozyrsky}}, \bibinfo {author} {\bibfnamefont {A.~E.}\
  \bibnamefont {Roitberg}},\ and\ \bibinfo {author} {\bibfnamefont
  {S.}~\bibnamefont {Tretiak}},\ }\bibfield  {title} {\bibinfo {title}
  {Non-adiabatic excited-state molecular dynamics: Theory and applications for
  modeling photophysics in extended molecular materials},\ }\href@noop {}
  {\bibfield  {journal} {\bibinfo  {journal} {Chem. Rev.}\ }\textbf {\bibinfo
  {volume} {120}},\ \bibinfo {pages} {2215} (\bibinfo {year}
  {2020})}\BibitemShut {NoStop}%
\bibitem [{\citenamefont {Szalay}\ \emph {et~al.}(2012)\citenamefont {Szalay},
  \citenamefont {Müller}, \citenamefont {Gidofalvi}, \citenamefont {Lischka},\
  and\ \citenamefont {Shepard}}]{Szalay2012}%
  \BibitemOpen
  \bibfield  {author} {\bibinfo {author} {\bibfnamefont {P.~G.}\ \bibnamefont
  {Szalay}}, \bibinfo {author} {\bibfnamefont {T.}~\bibnamefont {Müller}},
  \bibinfo {author} {\bibfnamefont {G.}~\bibnamefont {Gidofalvi}}, \bibinfo
  {author} {\bibfnamefont {H.}~\bibnamefont {Lischka}},\ and\ \bibinfo {author}
  {\bibfnamefont {R.}~\bibnamefont {Shepard}},\ }\bibfield  {title} {\bibinfo
  {title} {Multiconfiguration self-consistent field and multireference
  configuration interaction methods and applications},\ }\href
  {https://doi.org/10.1021/cr200137a} {\bibfield  {journal} {\bibinfo
  {journal} {Chem. Rev.}\ }\textbf {\bibinfo {volume} {112}},\ \bibinfo {pages}
  {108} (\bibinfo {year} {2012})},\ \bibinfo {note} {pMID: 22204633},\ \Eprint
  {https://arxiv.org/abs/https://doi.org/10.1021/cr200137a}
  {https://doi.org/10.1021/cr200137a} \BibitemShut {NoStop}%
\bibitem [{\citenamefont {Dreuw}\ and\ \citenamefont
  {Wormit}(2015)}]{dreuw2015algebraic}%
  \BibitemOpen
  \bibfield  {author} {\bibinfo {author} {\bibfnamefont {A.}~\bibnamefont
  {Dreuw}}\ and\ \bibinfo {author} {\bibfnamefont {M.}~\bibnamefont {Wormit}},\
  }\bibfield  {title} {\bibinfo {title} {The algebraic diagrammatic
  construction scheme for the polarization propagator for the calculation of
  excited states},\ }\href@noop {} {\bibfield  {journal} {\bibinfo  {journal}
  {Wiley Interdisciplinary Reviews: Computational Molecular Science}\ }\textbf
  {\bibinfo {volume} {5}},\ \bibinfo {pages} {82} (\bibinfo {year}
  {2015})}\BibitemShut {NoStop}%
\bibitem [{\citenamefont {Stanton}\ and\ \citenamefont
  {Bartlett}(1993)}]{Stanton1993}%
  \BibitemOpen
  \bibfield  {author} {\bibinfo {author} {\bibfnamefont {J.~F.}\ \bibnamefont
  {Stanton}}\ and\ \bibinfo {author} {\bibfnamefont {R.~J.}\ \bibnamefont
  {Bartlett}},\ }\bibfield  {title} {\bibinfo {title} {The equation of motion
  coupled‐cluster method. a systematic biorthogonal approach to molecular
  excitation energies, transition probabilities, and excited state
  properties},\ }\href {https://doi.org/10.1063/1.464746} {\bibfield  {journal}
  {\bibinfo  {journal} {J. Chem. Phys.}\ }\textbf {\bibinfo {volume} {98}},\
  \bibinfo {pages} {7029} (\bibinfo {year} {1993})},\ \Eprint
  {https://arxiv.org/abs/https://doi.org/10.1063/1.464746}
  {https://doi.org/10.1063/1.464746} \BibitemShut {NoStop}%
\bibitem [{\citenamefont {Serrano-Andr{\'e}s}\ and\ \citenamefont
  {Merch{\'a}n}(2005)}]{serrano2005quantum}%
  \BibitemOpen
  \bibfield  {author} {\bibinfo {author} {\bibfnamefont {L.}~\bibnamefont
  {Serrano-Andr{\'e}s}}\ and\ \bibinfo {author} {\bibfnamefont
  {M.}~\bibnamefont {Merch{\'a}n}},\ }\bibfield  {title} {\bibinfo {title}
  {Quantum chemistry of the excited state: 2005 overview},\ }\href@noop {}
  {\bibfield  {journal} {\bibinfo  {journal} {J. Mol. Struc. Theochem}\
  }\textbf {\bibinfo {volume} {729}},\ \bibinfo {pages} {99} (\bibinfo {year}
  {2005})}\BibitemShut {NoStop}%
\bibitem [{\citenamefont {Sneskov}\ and\ \citenamefont
  {Christiansen}(2012)}]{sneskov2012excited}%
  \BibitemOpen
  \bibfield  {author} {\bibinfo {author} {\bibfnamefont {K.}~\bibnamefont
  {Sneskov}}\ and\ \bibinfo {author} {\bibfnamefont {O.}~\bibnamefont
  {Christiansen}},\ }\bibfield  {title} {\bibinfo {title} {Excited state
  coupled cluster methods},\ }\href@noop {} {\bibfield  {journal} {\bibinfo
  {journal} {Wiley Interdisciplinary Reviews: Computational Molecular Science}\
  }\textbf {\bibinfo {volume} {2}},\ \bibinfo {pages} {566} (\bibinfo {year}
  {2012})}\BibitemShut {NoStop}%
\bibitem [{\citenamefont {Mayhall}\ \emph {et~al.}(2014)\citenamefont
  {Mayhall}, \citenamefont {Goldey},\ and\ \citenamefont
  {Head-Gordon}}]{mayhall2014quasidegenerate}%
  \BibitemOpen
  \bibfield  {author} {\bibinfo {author} {\bibfnamefont {N.~J.}\ \bibnamefont
  {Mayhall}}, \bibinfo {author} {\bibfnamefont {M.}~\bibnamefont {Goldey}},\
  and\ \bibinfo {author} {\bibfnamefont {M.}~\bibnamefont {Head-Gordon}},\
  }\bibfield  {title} {\bibinfo {title} {A quasidegenerate second-order
  perturbation theory approximation to ras-n sf for excited states and strong
  correlations},\ }\href@noop {} {\bibfield  {journal} {\bibinfo  {journal} {J.
  Chem. Theory Comput.}\ }\textbf {\bibinfo {volume} {10}},\ \bibinfo {pages}
  {589} (\bibinfo {year} {2014})}\BibitemShut {NoStop}%
\bibitem [{\citenamefont {Nooijen}\ and\ \citenamefont
  {Bartlett}(1997)}]{nooijen1997new}%
  \BibitemOpen
  \bibfield  {author} {\bibinfo {author} {\bibfnamefont {M.}~\bibnamefont
  {Nooijen}}\ and\ \bibinfo {author} {\bibfnamefont {R.~J.}\ \bibnamefont
  {Bartlett}},\ }\bibfield  {title} {\bibinfo {title} {A new method for excited
  states: Similarity transformed equation-of-motion coupled-cluster theory},\
  }\href@noop {} {\bibfield  {journal} {\bibinfo  {journal} {J. Chem. Phys.}\
  }\textbf {\bibinfo {volume} {106}},\ \bibinfo {pages} {6441} (\bibinfo {year}
  {1997})}\BibitemShut {NoStop}%
\bibitem [{\citenamefont {Hemmatiyan}\ \emph {et~al.}(2018)\citenamefont
  {Hemmatiyan}, \citenamefont {Sajjan}, \citenamefont {Schlimgen},\ and\
  \citenamefont {Mazziotti}}]{hemmatiyan2018excited}%
  \BibitemOpen
  \bibfield  {author} {\bibinfo {author} {\bibfnamefont {S.}~\bibnamefont
  {Hemmatiyan}}, \bibinfo {author} {\bibfnamefont {M.}~\bibnamefont {Sajjan}},
  \bibinfo {author} {\bibfnamefont {A.}~\bibnamefont {Schlimgen}},\ and\
  \bibinfo {author} {\bibfnamefont {D.}~\bibnamefont {Mazziotti}},\ }\bibfield
  {title} {\bibinfo {title} {Excited-state spectra of strongly correlated
  molecules from a reduced-density-matrix approach},\ }\href@noop {} {\bibfield
   {journal} {\bibinfo  {journal} {J. Phys. Chem. Lett.}\ }\textbf {\bibinfo
  {volume} {9}},\ \bibinfo {pages} {5373} (\bibinfo {year} {2018})}\BibitemShut
  {NoStop}%
\bibitem [{\citenamefont {Mazziotti}(2003)}]{mazziotti2003extraction}%
  \BibitemOpen
  \bibfield  {author} {\bibinfo {author} {\bibfnamefont {D.~A.}\ \bibnamefont
  {Mazziotti}},\ }\bibfield  {title} {\bibinfo {title} {Extraction of
  electronic excited states from the ground-state two-particle reduced density
  matrix},\ }\href@noop {} {\bibfield  {journal} {\bibinfo  {journal} {Phys.
  Rev. A}\ }\textbf {\bibinfo {volume} {68}},\ \bibinfo {pages} {052501}
  (\bibinfo {year} {2003})}\BibitemShut {NoStop}%
\bibitem [{\citenamefont {Stanton}\ \emph {et~al.}()\citenamefont {Stanton},
  \citenamefont {Gauss}, \citenamefont {Cheng}, \citenamefont {Harding},
  \citenamefont {Matthews},\ and\ \citenamefont {Szalay}}]{cfour}%
  \BibitemOpen
  \bibfield  {author} {\bibinfo {author} {\bibfnamefont {J.~F.}\ \bibnamefont
  {Stanton}}, \bibinfo {author} {\bibfnamefont {J.}~\bibnamefont {Gauss}},
  \bibinfo {author} {\bibfnamefont {L.}~\bibnamefont {Cheng}}, \bibinfo
  {author} {\bibfnamefont {M.~E.}\ \bibnamefont {Harding}}, \bibinfo {author}
  {\bibfnamefont {D.~A.}\ \bibnamefont {Matthews}},\ and\ \bibinfo {author}
  {\bibfnamefont {P.~G.}\ \bibnamefont {Szalay}},\ }\href@noop {} {\bibinfo
  {title} {{CFOUR, Coupled-Cluster techniques for Computational Chemistry, a
  quantum-chemical program package}}},\ \bibinfo {note} {{W}ith contributions
  from {A}. {A}sthana, {A}.{A}. {A}uer, {R}.{J}. {B}artlett, {U}. {B}enedikt,
  {C}. {B}erger, {D}.{E}. {B}ernholdt, {S}. {B}laschke, {Y}. {J}. {B}omble,
  {S}. {B}urger, {O}. {C}hristiansen, {D}. {D}atta, {F}. {E}ngel, {R}. {F}aber,
  {J}. {G}reiner, {M}. {H}eckert, {O}. {H}eun, {M}. Hilgenberg, {C}. {H}uber,
  {T}.-{C}. {J}agau, {D}. {J}onsson, {J}. {J}us{\'e}lius, {T}. Kirsch,
  {M}.-{P}. {K}itsaras, {K}. {K}lein, {G}.{M}. {K}opper, {W}.{J}. {L}auderdale,
  {F}. {L}ipparini, {J}. {L}iu, {T}. {M}etzroth, {L}.{A}. {M}{\"u}ck, {D}.{P}.
  {O}'{N}eill, {T}. {N}ottoli, {J}. {O}swald, {D}.{R}. {P}rice, {E}.
  {P}rochnow, {C}. {P}uzzarini, {K}. {R}uud, {F}. {S}chiffmann, {W}.
  {S}chwalbach, {C}. {S}immons, {S}. {S}topkowicz, {A}. {T}ajti, {J}.
  {V}{\'a}zquez, {F}. {W}ang, {J}.{D}. {W}atts, {C}. {Z}hang, {X}. {Z}heng, and
  the integral packages {MOLECULE} ({J}. {A}lml{\"o}f and {P}.{R}. {T}aylor),
  {PROPS} ({P}.{R}. {T}aylor), {ABACUS} ({T}. {H}elgaker, {H}.{J}. {A}a.
  {J}ensen, {P}. {J}{\o}rgensen, and {J}. {O}lsen), and {ECP} routines by {A}.
  {V}. {M}itin and {C}. van {W}{\"u}llen. {F}or the current version, see
  http://www.cfour.de.}\BibitemShut {Stop}%
\bibitem [{\citenamefont {Sun}\ \emph {et~al.}(2018)\citenamefont {Sun},
  \citenamefont {Berkelbach}, \citenamefont {Blunt}, \citenamefont {Booth},
  \citenamefont {Guo}, \citenamefont {Li}, \citenamefont {Liu}, \citenamefont
  {McClain}, \citenamefont {Sayfutyarova}, \citenamefont {Sharma} \emph
  {et~al.}}]{Pyscf}%
  \BibitemOpen
  \bibfield  {author} {\bibinfo {author} {\bibfnamefont {Q.}~\bibnamefont
  {Sun}}, \bibinfo {author} {\bibfnamefont {T.~C.}\ \bibnamefont {Berkelbach}},
  \bibinfo {author} {\bibfnamefont {N.~S.}\ \bibnamefont {Blunt}}, \bibinfo
  {author} {\bibfnamefont {G.~H.}\ \bibnamefont {Booth}}, \bibinfo {author}
  {\bibfnamefont {S.}~\bibnamefont {Guo}}, \bibinfo {author} {\bibfnamefont
  {Z.}~\bibnamefont {Li}}, \bibinfo {author} {\bibfnamefont {J.}~\bibnamefont
  {Liu}}, \bibinfo {author} {\bibfnamefont {J.~D.}\ \bibnamefont {McClain}},
  \bibinfo {author} {\bibfnamefont {E.~R.}\ \bibnamefont {Sayfutyarova}},
  \bibinfo {author} {\bibfnamefont {S.}~\bibnamefont {Sharma}}, \emph
  {et~al.},\ }\bibfield  {title} {\bibinfo {title} {Pyscf: the python-based
  simulations of chemistry framework},\ }\href@noop {} {\bibfield  {journal}
  {\bibinfo  {journal} {Wiley Interdisciplinary Reviews: Computational
  Molecular Science}\ }\textbf {\bibinfo {volume} {8}},\ \bibinfo {pages}
  {e1340} (\bibinfo {year} {2018})}\BibitemShut {NoStop}%
\bibitem [{\citenamefont {Vidal}\ \emph {et~al.}(2019)\citenamefont {Vidal},
  \citenamefont {Feng}, \citenamefont {Epifanovsky}, \citenamefont {Krylov},\
  and\ \citenamefont {Coriani}}]{Vidal2019}%
  \BibitemOpen
  \bibfield  {author} {\bibinfo {author} {\bibfnamefont {M.~L.}\ \bibnamefont
  {Vidal}}, \bibinfo {author} {\bibfnamefont {X.}~\bibnamefont {Feng}},
  \bibinfo {author} {\bibfnamefont {E.}~\bibnamefont {Epifanovsky}}, \bibinfo
  {author} {\bibfnamefont {A.~I.}\ \bibnamefont {Krylov}},\ and\ \bibinfo
  {author} {\bibfnamefont {S.}~\bibnamefont {Coriani}},\ }\bibfield  {title}
  {\bibinfo {title} {New and efficient equation-of-motion coupled-cluster
  framework for core-excited and core-ionized states},\ }\href
  {https://doi.org/10.1021/acs.jctc.9b00039} {\bibfield  {journal} {\bibinfo
  {journal} {J. Chem. Theory Comput.}\ }\textbf {\bibinfo {volume} {15}},\
  \bibinfo {pages} {3117} (\bibinfo {year} {2019})},\ \bibinfo {note} {pMID:
  30964297},\ \Eprint
  {https://arxiv.org/abs/https://doi.org/10.1021/acs.jctc.9b00039}
  {https://doi.org/10.1021/acs.jctc.9b00039} \BibitemShut {NoStop}%
\bibitem [{\citenamefont {Goings}\ \emph {et~al.}(2014)\citenamefont {Goings},
  \citenamefont {Caricato}, \citenamefont {Frisch},\ and\ \citenamefont
  {Li}}]{goings2014assessment}%
  \BibitemOpen
  \bibfield  {author} {\bibinfo {author} {\bibfnamefont {J.~J.}\ \bibnamefont
  {Goings}}, \bibinfo {author} {\bibfnamefont {M.}~\bibnamefont {Caricato}},
  \bibinfo {author} {\bibfnamefont {M.~J.}\ \bibnamefont {Frisch}},\ and\
  \bibinfo {author} {\bibfnamefont {X.}~\bibnamefont {Li}},\ }\bibfield
  {title} {\bibinfo {title} {Assessment of low-scaling approximations to the
  equation of motion coupled-cluster singles and doubles equations},\
  }\href@noop {} {\bibfield  {journal} {\bibinfo  {journal} {J. Chem. Phys.}\
  }\textbf {\bibinfo {volume} {141}},\ \bibinfo {pages} {164116} (\bibinfo
  {year} {2014})}\BibitemShut {NoStop}%
\bibitem [{\citenamefont {Andersen}\ \emph {et~al.}(2022)\citenamefont
  {Andersen}, \citenamefont {Nanda}, \citenamefont {Krylov},\ and\
  \citenamefont {Coriani}}]{andersen2022probing}%
  \BibitemOpen
  \bibfield  {author} {\bibinfo {author} {\bibfnamefont {J.~H.}\ \bibnamefont
  {Andersen}}, \bibinfo {author} {\bibfnamefont {K.~D.}\ \bibnamefont {Nanda}},
  \bibinfo {author} {\bibfnamefont {A.~I.}\ \bibnamefont {Krylov}},\ and\
  \bibinfo {author} {\bibfnamefont {S.}~\bibnamefont {Coriani}},\ }\bibfield
  {title} {\bibinfo {title} {Probing molecular chirality of ground and
  electronically excited states in the uv--vis and x-ray regimes: An eom-ccsd
  study},\ }\href@noop {} {\bibfield  {journal} {\bibinfo  {journal} {J. Chem.
  Theory Comput.}\ }\textbf {\bibinfo {volume} {18}},\ \bibinfo {pages} {1748}
  (\bibinfo {year} {2022})}\BibitemShut {NoStop}%
\bibitem [{\citenamefont {Asthana}\ \emph {et~al.}(2019)\citenamefont
  {Asthana}, \citenamefont {Liu},\ and\ \citenamefont {Cheng}}]{Asthana2019}%
  \BibitemOpen
  \bibfield  {author} {\bibinfo {author} {\bibfnamefont {A.}~\bibnamefont
  {Asthana}}, \bibinfo {author} {\bibfnamefont {J.}~\bibnamefont {Liu}},\ and\
  \bibinfo {author} {\bibfnamefont {L.}~\bibnamefont {Cheng}},\ }\bibfield
  {title} {\bibinfo {title} {Exact two-component equation-of-motion
  coupled-cluster singles and doubles method using atomic mean-field spin-orbit
  integrals},\ }\href@noop {} {\bibfield  {journal} {\bibinfo  {journal} {J.
  Chem. Phys.}\ }\textbf {\bibinfo {volume} {150}},\ \bibinfo {pages} {074102}
  (\bibinfo {year} {2019})}\BibitemShut {NoStop}%
\bibitem [{\citenamefont {Halbert}\ \emph {et~al.}(2021)\citenamefont
  {Halbert}, \citenamefont {Vidal}, \citenamefont {Shee}, \citenamefont
  {Coriani},\ and\ \citenamefont {Severo Pereira~Gomes}}]{Halbert2021}%
  \BibitemOpen
  \bibfield  {author} {\bibinfo {author} {\bibfnamefont {L.}~\bibnamefont
  {Halbert}}, \bibinfo {author} {\bibfnamefont {M.~L.}\ \bibnamefont {Vidal}},
  \bibinfo {author} {\bibfnamefont {A.}~\bibnamefont {Shee}}, \bibinfo {author}
  {\bibfnamefont {S.}~\bibnamefont {Coriani}},\ and\ \bibinfo {author}
  {\bibfnamefont {A.}~\bibnamefont {Severo Pereira~Gomes}},\ }\bibfield
  {title} {\bibinfo {title} {Relativistic eom-ccsd for core-excited and
  core-ionized state energies based on the four-component
  dirac–coulomb(-gaunt) hamiltonian},\ }\href
  {https://doi.org/10.1021/acs.jctc.0c01203} {\bibfield  {journal} {\bibinfo
  {journal} {J. Chem. Theory Comput.}\ }\textbf {\bibinfo {volume} {17}},\
  \bibinfo {pages} {3583} (\bibinfo {year} {2021})},\ \bibinfo {note} {pMID:
  33944570},\ \Eprint
  {https://arxiv.org/abs/https://doi.org/10.1021/acs.jctc.0c01203}
  {https://doi.org/10.1021/acs.jctc.0c01203} \BibitemShut {NoStop}%
\bibitem [{\citenamefont {Stanton}\ and\ \citenamefont
  {Gauss}(1994)}]{stanton1994analytic}%
  \BibitemOpen
  \bibfield  {author} {\bibinfo {author} {\bibfnamefont {J.~F.}\ \bibnamefont
  {Stanton}}\ and\ \bibinfo {author} {\bibfnamefont {J.}~\bibnamefont
  {Gauss}},\ }\bibfield  {title} {\bibinfo {title} {Analytic energy derivatives
  for ionized states described by the equation-of-motion coupled cluster
  method},\ }\href@noop {} {\bibfield  {journal} {\bibinfo  {journal} {J. Chem.
  Phys.}\ }\textbf {\bibinfo {volume} {101}},\ \bibinfo {pages} {8938}
  (\bibinfo {year} {1994})}\BibitemShut {NoStop}%
\bibitem [{\citenamefont {Nooijen}\ and\ \citenamefont
  {Bartlett}(1995)}]{nooijen1995equation}%
  \BibitemOpen
  \bibfield  {author} {\bibinfo {author} {\bibfnamefont {M.}~\bibnamefont
  {Nooijen}}\ and\ \bibinfo {author} {\bibfnamefont {R.~J.}\ \bibnamefont
  {Bartlett}},\ }\bibfield  {title} {\bibinfo {title} {Equation of motion
  coupled cluster method for electron attachment},\ }\href@noop {} {\bibfield
  {journal} {\bibinfo  {journal} {J. Chem. Phys.}\ }\textbf {\bibinfo {volume}
  {102}},\ \bibinfo {pages} {3629} (\bibinfo {year} {1995})}\BibitemShut
  {NoStop}%
\bibitem [{\citenamefont {Stanton}\ and\ \citenamefont
  {Gauss}(1995)}]{stanton1995perturbative}%
  \BibitemOpen
  \bibfield  {author} {\bibinfo {author} {\bibfnamefont {J.~F.}\ \bibnamefont
  {Stanton}}\ and\ \bibinfo {author} {\bibfnamefont {J.}~\bibnamefont
  {Gauss}},\ }\bibfield  {title} {\bibinfo {title} {Perturbative treatment of
  the similarity transformed hamiltonian in equation-of-motion coupled-cluster
  approximations},\ }\href@noop {} {\bibfield  {journal} {\bibinfo  {journal}
  {J. Chem. Phys.}\ }\textbf {\bibinfo {volume} {103}},\ \bibinfo {pages}
  {1064} (\bibinfo {year} {1995})}\BibitemShut {NoStop}%
\bibitem [{\citenamefont {Pieniazek}\ \emph {et~al.}(2008)\citenamefont
  {Pieniazek}, \citenamefont {Bradforth},\ and\ \citenamefont
  {Krylov}}]{pieniazek2008charge}%
  \BibitemOpen
  \bibfield  {author} {\bibinfo {author} {\bibfnamefont {P.~A.}\ \bibnamefont
  {Pieniazek}}, \bibinfo {author} {\bibfnamefont {S.~E.}\ \bibnamefont
  {Bradforth}},\ and\ \bibinfo {author} {\bibfnamefont {A.~I.}\ \bibnamefont
  {Krylov}},\ }\bibfield  {title} {\bibinfo {title} {Charge localization and
  jahn--teller distortions in the benzene dimer cation},\ }\href@noop {}
  {\bibfield  {journal} {\bibinfo  {journal} {J. Chem. Phys.}\ }\textbf
  {\bibinfo {volume} {129}},\ \bibinfo {pages} {074104} (\bibinfo {year}
  {2008})}\BibitemShut {NoStop}%
\bibitem [{\citenamefont {Liu}\ \emph {et~al.}(2020)\citenamefont {Liu},
  \citenamefont {Zhang}, \citenamefont {Ciborowski}, \citenamefont {Asthana},
  \citenamefont {Cheng},\ and\ \citenamefont {Bowen}}]{liu2020mapping}%
  \BibitemOpen
  \bibfield  {author} {\bibinfo {author} {\bibfnamefont {G.}~\bibnamefont
  {Liu}}, \bibinfo {author} {\bibfnamefont {C.}~\bibnamefont {Zhang}}, \bibinfo
  {author} {\bibfnamefont {S.~M.}\ \bibnamefont {Ciborowski}}, \bibinfo
  {author} {\bibfnamefont {A.}~\bibnamefont {Asthana}}, \bibinfo {author}
  {\bibfnamefont {L.}~\bibnamefont {Cheng}},\ and\ \bibinfo {author}
  {\bibfnamefont {K.~H.}\ \bibnamefont {Bowen}},\ }\bibfield  {title} {\bibinfo
  {title} {Mapping the electronic structure of the uranium (vi) dinitride
  molecule, un2},\ }\href@noop {} {\bibfield  {journal} {\bibinfo  {journal}
  {J. Phys. Chem. A}\ }\textbf {\bibinfo {volume} {124}},\ \bibinfo {pages}
  {6486} (\bibinfo {year} {2020})}\BibitemShut {NoStop}%
\bibitem [{\citenamefont {Thomas}\ \emph {et~al.}(2021)\citenamefont {Thomas},
  \citenamefont {Hampe}, \citenamefont {Stopkowicz},\ and\ \citenamefont
  {Gauss}}]{thomas2021complex}%
  \BibitemOpen
  \bibfield  {author} {\bibinfo {author} {\bibfnamefont {S.}~\bibnamefont
  {Thomas}}, \bibinfo {author} {\bibfnamefont {F.}~\bibnamefont {Hampe}},
  \bibinfo {author} {\bibfnamefont {S.}~\bibnamefont {Stopkowicz}},\ and\
  \bibinfo {author} {\bibfnamefont {J.}~\bibnamefont {Gauss}},\ }\bibfield
  {title} {\bibinfo {title} {Complex ground-state and excitation energies in
  coupled-cluster theory},\ }\href@noop {} {\bibfield  {journal} {\bibinfo
  {journal} {Mol. Phys.}\ }\textbf {\bibinfo {volume} {119}},\ \bibinfo {pages}
  {e1968056} (\bibinfo {year} {2021})}\BibitemShut {NoStop}%
\bibitem [{\citenamefont {K{\"o}hn}\ and\ \citenamefont
  {Tajti}(2007)}]{kohn2007can}%
  \BibitemOpen
  \bibfield  {author} {\bibinfo {author} {\bibfnamefont {A.}~\bibnamefont
  {K{\"o}hn}}\ and\ \bibinfo {author} {\bibfnamefont {A.}~\bibnamefont
  {Tajti}},\ }\bibfield  {title} {\bibinfo {title} {Can coupled-cluster theory
  treat conical intersections?},\ }\href@noop {} {\bibfield  {journal}
  {\bibinfo  {journal} {J. Chem. Phys.}\ }\textbf {\bibinfo {volume} {127}},\
  \bibinfo {pages} {044105} (\bibinfo {year} {2007})}\BibitemShut {NoStop}%
\bibitem [{\citenamefont {Yarkony}(2012)}]{yarkony2012nonadiabatic}%
  \BibitemOpen
  \bibfield  {author} {\bibinfo {author} {\bibfnamefont {D.~R.}\ \bibnamefont
  {Yarkony}},\ }\bibfield  {title} {\bibinfo {title} {Nonadiabatic quantum
  chemistry past, present, and future},\ }\href@noop {} {\bibfield  {journal}
  {\bibinfo  {journal} {Chem. Rev.}\ }\textbf {\bibinfo {volume} {112}},\
  \bibinfo {pages} {481} (\bibinfo {year} {2012})}\BibitemShut {NoStop}%
\bibitem [{\citenamefont {Bernardi}\ \emph {et~al.}(1997)\citenamefont
  {Bernardi}, \citenamefont {Olivucci},\ and\ \citenamefont
  {Robb}}]{bernardi1997role}%
  \BibitemOpen
  \bibfield  {author} {\bibinfo {author} {\bibfnamefont {F.}~\bibnamefont
  {Bernardi}}, \bibinfo {author} {\bibfnamefont {M.}~\bibnamefont {Olivucci}},\
  and\ \bibinfo {author} {\bibfnamefont {M.~A.}\ \bibnamefont {Robb}},\
  }\bibfield  {title} {\bibinfo {title} {The role of conical intersections and
  excited state reaction paths in photochemical pericyclic reactions},\
  }\href@noop {} {\bibfield  {journal} {\bibinfo  {journal} {J. Photochem.
  Photobiol. A: Chem.}\ }\textbf {\bibinfo {volume} {105}},\ \bibinfo {pages}
  {365} (\bibinfo {year} {1997})}\BibitemShut {NoStop}%
\bibitem [{\citenamefont {Schmidt}\ and\ \citenamefont
  {Gordon}(1998)}]{schmidt1998construction}%
  \BibitemOpen
  \bibfield  {author} {\bibinfo {author} {\bibfnamefont {M.~W.}\ \bibnamefont
  {Schmidt}}\ and\ \bibinfo {author} {\bibfnamefont {M.~S.}\ \bibnamefont
  {Gordon}},\ }\bibfield  {title} {\bibinfo {title} {The construction and
  interpretation of mcscf wavefunctions},\ }\href@noop {} {\bibfield  {journal}
  {\bibinfo  {journal} {Ann. Rev. Phys. Chem.}\ }\textbf {\bibinfo {volume}
  {49}},\ \bibinfo {pages} {233} (\bibinfo {year} {1998})}\BibitemShut
  {NoStop}%
\bibitem [{\citenamefont {K{\"o}hn}\ \emph {et~al.}(2013)\citenamefont
  {K{\"o}hn}, \citenamefont {Hanauer}, \citenamefont {Mueck}, \citenamefont
  {Jagau},\ and\ \citenamefont {Gauss}}]{kohn2013state}%
  \BibitemOpen
  \bibfield  {author} {\bibinfo {author} {\bibfnamefont {A.}~\bibnamefont
  {K{\"o}hn}}, \bibinfo {author} {\bibfnamefont {M.}~\bibnamefont {Hanauer}},
  \bibinfo {author} {\bibfnamefont {L.~A.}\ \bibnamefont {Mueck}}, \bibinfo
  {author} {\bibfnamefont {T.-C.}\ \bibnamefont {Jagau}},\ and\ \bibinfo
  {author} {\bibfnamefont {J.}~\bibnamefont {Gauss}},\ }\bibfield  {title}
  {\bibinfo {title} {State-specific multireference coupled-cluster theory},\
  }\href@noop {} {\bibfield  {journal} {\bibinfo  {journal} {Wiley
  Interdisciplinary Reviews: Computational Molecular Science}\ }\textbf
  {\bibinfo {volume} {3}},\ \bibinfo {pages} {176} (\bibinfo {year}
  {2013})}\BibitemShut {NoStop}%
\bibitem [{\citenamefont {Mahapatra}\ \emph {et~al.}(1998)\citenamefont
  {Mahapatra}, \citenamefont {Datta}, \citenamefont {Bandyopadhyay},\ and\
  \citenamefont {Mukherjee}}]{mahapatra1998state}%
  \BibitemOpen
  \bibfield  {author} {\bibinfo {author} {\bibfnamefont {U.~S.}\ \bibnamefont
  {Mahapatra}}, \bibinfo {author} {\bibfnamefont {B.}~\bibnamefont {Datta}},
  \bibinfo {author} {\bibfnamefont {B.}~\bibnamefont {Bandyopadhyay}},\ and\
  \bibinfo {author} {\bibfnamefont {D.}~\bibnamefont {Mukherjee}},\ }\bibfield
  {title} {\bibinfo {title} {State-specific multi-reference coupled cluster
  formulations: Two paradigms},\ }in\ \href@noop {} {\emph {\bibinfo
  {booktitle} {Adv. Quantum Chem.}}},\ Vol.~\bibinfo {volume} {30}\ (\bibinfo
  {publisher} {Elsevier},\ \bibinfo {year} {1998})\ pp.\ \bibinfo {pages}
  {163--193}\BibitemShut {NoStop}%
\bibitem [{\citenamefont {Roos}\ \emph {et~al.}(1996)\citenamefont {Roos},
  \citenamefont {Andersson}, \citenamefont {Fulscher}, \citenamefont
  {Malmqvist}, \citenamefont {SerranoAndres}, \citenamefont {Pierloot},\ and\
  \citenamefont {Merch{\'a}n}}]{roos1996multiconfigurational}%
  \BibitemOpen
  \bibfield  {author} {\bibinfo {author} {\bibfnamefont {B.~O.}\ \bibnamefont
  {Roos}}, \bibinfo {author} {\bibfnamefont {K.}~\bibnamefont {Andersson}},
  \bibinfo {author} {\bibfnamefont {M.~P.}\ \bibnamefont {Fulscher}}, \bibinfo
  {author} {\bibfnamefont {P.-A.}\ \bibnamefont {Malmqvist}}, \bibinfo {author}
  {\bibfnamefont {L.}~\bibnamefont {SerranoAndres}}, \bibinfo {author}
  {\bibfnamefont {K.}~\bibnamefont {Pierloot}},\ and\ \bibinfo {author}
  {\bibfnamefont {M.}~\bibnamefont {Merch{\'a}n}},\ }\bibfield  {title}
  {\bibinfo {title} {Multiconfigurational perturbation theory: Applications in
  electronic spectroscopy},\ }\href@noop {} {\bibfield  {journal} {\bibinfo
  {journal} {Adv. Chem. Phys., vol xciii}\ }\textbf {\bibinfo {volume} {93}},\
  \bibinfo {pages} {219} (\bibinfo {year} {1996})}\BibitemShut {NoStop}%
\bibitem [{\citenamefont {Kitaev}(1997)}]{kitaev1997quantum}%
  \BibitemOpen
  \bibfield  {author} {\bibinfo {author} {\bibfnamefont {A.~Y.}\ \bibnamefont
  {Kitaev}},\ }\bibfield  {title} {\bibinfo {title} {Quantum computations:
  algorithms and error correction},\ }\href@noop {} {\bibfield  {journal}
  {\bibinfo  {journal} {Russ. Math. Sur.}\ }\textbf {\bibinfo {volume} {52}},\
  \bibinfo {pages} {1191} (\bibinfo {year} {1997})}\BibitemShut {NoStop}%
\bibitem [{\citenamefont {Nielsen}\ and\ \citenamefont
  {Chuang}(2002)}]{nielsen2002quantum}%
  \BibitemOpen
  \bibfield  {author} {\bibinfo {author} {\bibfnamefont {M.~A.}\ \bibnamefont
  {Nielsen}}\ and\ \bibinfo {author} {\bibfnamefont {I.}~\bibnamefont
  {Chuang}},\ }\href@noop {} {\bibinfo {title} {Quantum computation and quantum
  information}} (\bibinfo {year} {2002})\BibitemShut {NoStop}%
\bibitem [{\citenamefont {Russo}\ \emph {et~al.}(2021)\citenamefont {Russo},
  \citenamefont {Rudinger}, \citenamefont {Morrison},\ and\ \citenamefont
  {Baczewski}}]{russo2021evaluating}%
  \BibitemOpen
  \bibfield  {author} {\bibinfo {author} {\bibfnamefont {A.~E.}\ \bibnamefont
  {Russo}}, \bibinfo {author} {\bibfnamefont {K.~M.}\ \bibnamefont {Rudinger}},
  \bibinfo {author} {\bibfnamefont {B.~C.}\ \bibnamefont {Morrison}},\ and\
  \bibinfo {author} {\bibfnamefont {A.~D.}\ \bibnamefont {Baczewski}},\
  }\bibfield  {title} {\bibinfo {title} {Evaluating energy differences on a
  quantum computer with robust phase estimation},\ }\href@noop {} {\bibfield
  {journal} {\bibinfo  {journal} {Phys. Rev. Lett.}\ }\textbf {\bibinfo
  {volume} {126}},\ \bibinfo {pages} {210501} (\bibinfo {year}
  {2021})}\BibitemShut {NoStop}%
\bibitem [{\citenamefont {Bauman}\ \emph {et~al.}(2020)\citenamefont {Bauman},
  \citenamefont {Liu}, \citenamefont {Bylaska}, \citenamefont {Krishnamoorthy},
  \citenamefont {Low}, \citenamefont {Granade}, \citenamefont {Wiebe},
  \citenamefont {Baker}, \citenamefont {Peng}, \citenamefont {Roetteler} \emph
  {et~al.}}]{bauman2020toward}%
  \BibitemOpen
  \bibfield  {author} {\bibinfo {author} {\bibfnamefont {N.~P.}\ \bibnamefont
  {Bauman}}, \bibinfo {author} {\bibfnamefont {H.}~\bibnamefont {Liu}},
  \bibinfo {author} {\bibfnamefont {E.~J.}\ \bibnamefont {Bylaska}}, \bibinfo
  {author} {\bibfnamefont {S.}~\bibnamefont {Krishnamoorthy}}, \bibinfo
  {author} {\bibfnamefont {G.~H.}\ \bibnamefont {Low}}, \bibinfo {author}
  {\bibfnamefont {C.~E.}\ \bibnamefont {Granade}}, \bibinfo {author}
  {\bibfnamefont {N.}~\bibnamefont {Wiebe}}, \bibinfo {author} {\bibfnamefont
  {N.~A.}\ \bibnamefont {Baker}}, \bibinfo {author} {\bibfnamefont
  {B.}~\bibnamefont {Peng}}, \bibinfo {author} {\bibfnamefont {M.}~\bibnamefont
  {Roetteler}}, \emph {et~al.},\ }\bibfield  {title} {\bibinfo {title} {Toward
  quantum computing for high-energy excited states in molecular systems:
  quantum phase estimations of core-level states},\ }\href@noop {} {\bibfield
  {journal} {\bibinfo  {journal} {J. Chem. Theory Comput.}\ }\textbf {\bibinfo
  {volume} {17}},\ \bibinfo {pages} {201} (\bibinfo {year} {2020})}\BibitemShut
  {NoStop}%
\bibitem [{\citenamefont {Sugisaki}\ \emph
  {et~al.}(2021{\natexlab{a}})\citenamefont {Sugisaki}, \citenamefont {Toyota},
  \citenamefont {Sato}, \citenamefont {Shiomi},\ and\ \citenamefont
  {Takui}}]{sugisaki2021quantum}%
  \BibitemOpen
  \bibfield  {author} {\bibinfo {author} {\bibfnamefont {K.}~\bibnamefont
  {Sugisaki}}, \bibinfo {author} {\bibfnamefont {K.}~\bibnamefont {Toyota}},
  \bibinfo {author} {\bibfnamefont {K.}~\bibnamefont {Sato}}, \bibinfo {author}
  {\bibfnamefont {D.}~\bibnamefont {Shiomi}},\ and\ \bibinfo {author}
  {\bibfnamefont {T.}~\bibnamefont {Takui}},\ }\bibfield  {title} {\bibinfo
  {title} {Quantum algorithm for the direct calculations of vertical ionization
  energies},\ }\href@noop {} {\bibfield  {journal} {\bibinfo  {journal} {J.
  Phys. Chem. Lett.}\ }\textbf {\bibinfo {volume} {12}},\ \bibinfo {pages}
  {2880} (\bibinfo {year} {2021}{\natexlab{a}})}\BibitemShut {NoStop}%
\bibitem [{\citenamefont {Santagati}\ \emph {et~al.}(2018)\citenamefont
  {Santagati}, \citenamefont {Wang}, \citenamefont {Gentile}, \citenamefont
  {Paesani}, \citenamefont {Wiebe}, \citenamefont {McClean}, \citenamefont
  {Morley-Short}, \citenamefont {Shadbolt}, \citenamefont {Bonneau},
  \citenamefont {Silverstone} \emph {et~al.}}]{santagati2018witnessing}%
  \BibitemOpen
  \bibfield  {author} {\bibinfo {author} {\bibfnamefont {R.}~\bibnamefont
  {Santagati}}, \bibinfo {author} {\bibfnamefont {J.}~\bibnamefont {Wang}},
  \bibinfo {author} {\bibfnamefont {A.~A.}\ \bibnamefont {Gentile}}, \bibinfo
  {author} {\bibfnamefont {S.}~\bibnamefont {Paesani}}, \bibinfo {author}
  {\bibfnamefont {N.}~\bibnamefont {Wiebe}}, \bibinfo {author} {\bibfnamefont
  {J.~R.}\ \bibnamefont {McClean}}, \bibinfo {author} {\bibfnamefont
  {S.}~\bibnamefont {Morley-Short}}, \bibinfo {author} {\bibfnamefont {P.~J.}\
  \bibnamefont {Shadbolt}}, \bibinfo {author} {\bibfnamefont {D.}~\bibnamefont
  {Bonneau}}, \bibinfo {author} {\bibfnamefont {J.~W.}\ \bibnamefont
  {Silverstone}}, \emph {et~al.},\ }\bibfield  {title} {\bibinfo {title}
  {Witnessing eigenstates for quantum simulation of hamiltonian spectra},\
  }\href@noop {} {\bibfield  {journal} {\bibinfo  {journal} {Sc. Adv.}\
  }\textbf {\bibinfo {volume} {4}},\ \bibinfo {pages} {eaap9646} (\bibinfo
  {year} {2018})}\BibitemShut {NoStop}%
\bibitem [{\citenamefont {O’Brien}\ \emph {et~al.}(2019)\citenamefont
  {O’Brien}, \citenamefont {Tarasinski},\ and\ \citenamefont
  {Terhal}}]{o2019quantum}%
  \BibitemOpen
  \bibfield  {author} {\bibinfo {author} {\bibfnamefont {T.~E.}\ \bibnamefont
  {O’Brien}}, \bibinfo {author} {\bibfnamefont {B.}~\bibnamefont
  {Tarasinski}},\ and\ \bibinfo {author} {\bibfnamefont {B.~M.}\ \bibnamefont
  {Terhal}},\ }\bibfield  {title} {\bibinfo {title} {Quantum phase estimation
  of multiple eigenvalues for small-scale (noisy) experiments},\ }\href@noop {}
  {\bibfield  {journal} {\bibinfo  {journal} {New J. Phys.}\ }\textbf {\bibinfo
  {volume} {21}},\ \bibinfo {pages} {023022} (\bibinfo {year}
  {2019})}\BibitemShut {NoStop}%
\bibitem [{\citenamefont {Sugisaki}\ \emph
  {et~al.}(2021{\natexlab{b}})\citenamefont {Sugisaki}, \citenamefont {Sakai},
  \citenamefont {Toyota}, \citenamefont {Sato}, \citenamefont {Shiomi},\ and\
  \citenamefont {Takui}}]{sugisaki2021bayesian}%
  \BibitemOpen
  \bibfield  {author} {\bibinfo {author} {\bibfnamefont {K.}~\bibnamefont
  {Sugisaki}}, \bibinfo {author} {\bibfnamefont {C.}~\bibnamefont {Sakai}},
  \bibinfo {author} {\bibfnamefont {K.}~\bibnamefont {Toyota}}, \bibinfo
  {author} {\bibfnamefont {K.}~\bibnamefont {Sato}}, \bibinfo {author}
  {\bibfnamefont {D.}~\bibnamefont {Shiomi}},\ and\ \bibinfo {author}
  {\bibfnamefont {T.}~\bibnamefont {Takui}},\ }\bibfield  {title} {\bibinfo
  {title} {Bayesian phase difference estimation: a general quantum algorithm
  for the direct calculation of energy gaps},\ }\href@noop {} {\bibfield
  {journal} {\bibinfo  {journal} {Phys. Chem. Chem. Phys.}\ }\textbf {\bibinfo
  {volume} {23}},\ \bibinfo {pages} {20152} (\bibinfo {year}
  {2021}{\natexlab{b}})}\BibitemShut {NoStop}%
\bibitem [{\citenamefont {Sugisaki}\ \emph
  {et~al.}(2022{\natexlab{a}})\citenamefont {Sugisaki}, \citenamefont
  {Prasannaa}, \citenamefont {Ohshima}, \citenamefont {Katagiri}, \citenamefont
  {Mochizuki}, \citenamefont {Sahoo},\ and\ \citenamefont
  {Das}}]{sugisaki2022bayesian}%
  \BibitemOpen
  \bibfield  {author} {\bibinfo {author} {\bibfnamefont {K.}~\bibnamefont
  {Sugisaki}}, \bibinfo {author} {\bibfnamefont {V.}~\bibnamefont {Prasannaa}},
  \bibinfo {author} {\bibfnamefont {S.}~\bibnamefont {Ohshima}}, \bibinfo
  {author} {\bibfnamefont {T.}~\bibnamefont {Katagiri}}, \bibinfo {author}
  {\bibfnamefont {Y.}~\bibnamefont {Mochizuki}}, \bibinfo {author}
  {\bibfnamefont {B.}~\bibnamefont {Sahoo}},\ and\ \bibinfo {author}
  {\bibfnamefont {B.}~\bibnamefont {Das}},\ }\bibfield  {title} {\bibinfo
  {title} {Bayesian phase difference estimation algorithm for direct
  calculation of fine structure splitting: accelerated simulation of
  relativistic and quantum many-body effects},\ }\href@noop {} {\bibfield
  {journal} {\bibinfo  {journal} {arXiv preprint arXiv:2212.02058}\ } (\bibinfo
  {year} {2022}{\natexlab{a}})}\BibitemShut {NoStop}%
\bibitem [{\citenamefont {Poulin}\ \emph {et~al.}(2018)\citenamefont {Poulin},
  \citenamefont {Kitaev}, \citenamefont {Steiger}, \citenamefont {Hastings},\
  and\ \citenamefont {Troyer}}]{poulin2018quantum}%
  \BibitemOpen
  \bibfield  {author} {\bibinfo {author} {\bibfnamefont {D.}~\bibnamefont
  {Poulin}}, \bibinfo {author} {\bibfnamefont {A.}~\bibnamefont {Kitaev}},
  \bibinfo {author} {\bibfnamefont {D.~S.}\ \bibnamefont {Steiger}}, \bibinfo
  {author} {\bibfnamefont {M.~B.}\ \bibnamefont {Hastings}},\ and\ \bibinfo
  {author} {\bibfnamefont {M.}~\bibnamefont {Troyer}},\ }\bibfield  {title}
  {\bibinfo {title} {Quantum algorithm for spectral measurement with a lower
  gate count},\ }\href@noop {} {\bibfield  {journal} {\bibinfo  {journal}
  {Phys. Rev. Lett.}\ }\textbf {\bibinfo {volume} {121}},\ \bibinfo {pages}
  {010501} (\bibinfo {year} {2018})}\BibitemShut {NoStop}%
\bibitem [{\citenamefont {Motta}\ \emph
  {et~al.}(2020{\natexlab{a}})\citenamefont {Motta}, \citenamefont {Sun},
  \citenamefont {Tan}, \citenamefont {O’Rourke}, \citenamefont {Ye},
  \citenamefont {Minnich}, \citenamefont {Brand{\~a}o},\ and\ \citenamefont
  {Chan}}]{motta2020determining}%
  \BibitemOpen
  \bibfield  {author} {\bibinfo {author} {\bibfnamefont {M.}~\bibnamefont
  {Motta}}, \bibinfo {author} {\bibfnamefont {C.}~\bibnamefont {Sun}}, \bibinfo
  {author} {\bibfnamefont {A.~T.}\ \bibnamefont {Tan}}, \bibinfo {author}
  {\bibfnamefont {M.~J.}\ \bibnamefont {O’Rourke}}, \bibinfo {author}
  {\bibfnamefont {E.}~\bibnamefont {Ye}}, \bibinfo {author} {\bibfnamefont
  {A.~J.}\ \bibnamefont {Minnich}}, \bibinfo {author} {\bibfnamefont {F.~G.}\
  \bibnamefont {Brand{\~a}o}},\ and\ \bibinfo {author} {\bibfnamefont
  {G.~K.-L.}\ \bibnamefont {Chan}},\ }\bibfield  {title} {\bibinfo {title}
  {Determining eigenstates and thermal states on a quantum computer using
  quantum imaginary time evolution},\ }\href@noop {} {\bibfield  {journal}
  {\bibinfo  {journal} {Nat. Phys.}\ }\textbf {\bibinfo {volume} {16}},\
  \bibinfo {pages} {205} (\bibinfo {year} {2020}{\natexlab{a}})}\BibitemShut
  {NoStop}%
\bibitem [{\citenamefont {Parrish}\ and\ \citenamefont
  {McMahon}(2019)}]{parrish2019quantum}%
  \BibitemOpen
  \bibfield  {author} {\bibinfo {author} {\bibfnamefont {R.~M.}\ \bibnamefont
  {Parrish}}\ and\ \bibinfo {author} {\bibfnamefont {P.~L.}\ \bibnamefont
  {McMahon}},\ }\bibfield  {title} {\bibinfo {title} {Quantum filter
  diagonalization: Quantum eigendecomposition without full quantum phase
  estimation},\ }\href@noop {} {\bibfield  {journal} {\bibinfo  {journal}
  {arXiv preprint arXiv:1909.08925}\ } (\bibinfo {year} {2019})}\BibitemShut
  {NoStop}%
\bibitem [{\citenamefont {Teplukhin}\ \emph {et~al.}(2021)\citenamefont
  {Teplukhin}, \citenamefont {Kendrick}, \citenamefont {Mniszewski},
  \citenamefont {Zhang}, \citenamefont {Kumar}, \citenamefont {Negre},
  \citenamefont {Anisimov}, \citenamefont {Tretiak},\ and\ \citenamefont
  {Dub}}]{teplukhin2021computing}%
  \BibitemOpen
  \bibfield  {author} {\bibinfo {author} {\bibfnamefont {A.}~\bibnamefont
  {Teplukhin}}, \bibinfo {author} {\bibfnamefont {B.~K.}\ \bibnamefont
  {Kendrick}}, \bibinfo {author} {\bibfnamefont {S.~M.}\ \bibnamefont
  {Mniszewski}}, \bibinfo {author} {\bibfnamefont {Y.}~\bibnamefont {Zhang}},
  \bibinfo {author} {\bibfnamefont {A.}~\bibnamefont {Kumar}}, \bibinfo
  {author} {\bibfnamefont {C.~F.}\ \bibnamefont {Negre}}, \bibinfo {author}
  {\bibfnamefont {P.~M.}\ \bibnamefont {Anisimov}}, \bibinfo {author}
  {\bibfnamefont {S.}~\bibnamefont {Tretiak}},\ and\ \bibinfo {author}
  {\bibfnamefont {P.~A.}\ \bibnamefont {Dub}},\ }\bibfield  {title} {\bibinfo
  {title} {Computing molecular excited states on a d-wave quantum annealer},\
  }\href@noop {} {\bibfield  {journal} {\bibinfo  {journal} {Scientific
  reports}\ }\textbf {\bibinfo {volume} {11}},\ \bibinfo {pages} {1} (\bibinfo
  {year} {2021})}\BibitemShut {NoStop}%
\bibitem [{\citenamefont {Sugisaki}\ \emph
  {et~al.}(2022{\natexlab{b}})\citenamefont {Sugisaki}, \citenamefont {Toyota},
  \citenamefont {Sato}, \citenamefont {Shiomi},\ and\ \citenamefont
  {Takui}}]{sugisaki2022adiabatic}%
  \BibitemOpen
  \bibfield  {author} {\bibinfo {author} {\bibfnamefont {K.}~\bibnamefont
  {Sugisaki}}, \bibinfo {author} {\bibfnamefont {K.}~\bibnamefont {Toyota}},
  \bibinfo {author} {\bibfnamefont {K.}~\bibnamefont {Sato}}, \bibinfo {author}
  {\bibfnamefont {D.}~\bibnamefont {Shiomi}},\ and\ \bibinfo {author}
  {\bibfnamefont {T.}~\bibnamefont {Takui}},\ }\bibfield  {title} {\bibinfo
  {title} {Adiabatic state preparation of correlated wave functions with
  nonlinear scheduling functions and broken-symmetry wave functions},\
  }\href@noop {} {\bibfield  {journal} {\bibinfo  {journal} {Comm. Chem.}\
  }\textbf {\bibinfo {volume} {5}},\ \bibinfo {pages} {1} (\bibinfo {year}
  {2022}{\natexlab{b}})}\BibitemShut {NoStop}%
\bibitem [{\citenamefont {Nakanishi}\ \emph {et~al.}(2019)\citenamefont
  {Nakanishi}, \citenamefont {Mitarai},\ and\ \citenamefont
  {Fujii}}]{nakanishi2019subspace}%
  \BibitemOpen
  \bibfield  {author} {\bibinfo {author} {\bibfnamefont {K.~M.}\ \bibnamefont
  {Nakanishi}}, \bibinfo {author} {\bibfnamefont {K.}~\bibnamefont {Mitarai}},\
  and\ \bibinfo {author} {\bibfnamefont {K.}~\bibnamefont {Fujii}},\ }\bibfield
   {title} {\bibinfo {title} {Subspace-search variational quantum eigensolver
  for excited states},\ }\href@noop {} {\bibfield  {journal} {\bibinfo
  {journal} {Phys. Rev. Res.}\ }\textbf {\bibinfo {volume} {1}},\ \bibinfo
  {pages} {033062} (\bibinfo {year} {2019})}\BibitemShut {NoStop}%
\bibitem [{\citenamefont {Xie}\ \emph {et~al.}(2022)\citenamefont {Xie},
  \citenamefont {Liu},\ and\ \citenamefont {Zhao}}]{xie2022orthogonal}%
  \BibitemOpen
  \bibfield  {author} {\bibinfo {author} {\bibfnamefont {Q.-X.}\ \bibnamefont
  {Xie}}, \bibinfo {author} {\bibfnamefont {S.}~\bibnamefont {Liu}},\ and\
  \bibinfo {author} {\bibfnamefont {Y.}~\bibnamefont {Zhao}},\ }\bibfield
  {title} {\bibinfo {title} {Orthogonal state reduction variational eigensolver
  for the excited-state calculations on quantum computers},\ }\href@noop {}
  {\bibfield  {journal} {\bibinfo  {journal} {J. Chem. Theory Comput.}\ }
  (\bibinfo {year} {2022})}\BibitemShut {NoStop}%
\bibitem [{\citenamefont {Chan}\ \emph {et~al.}(2021)\citenamefont {Chan},
  \citenamefont {Fitzpatrick}, \citenamefont {Segarra-Mart{\'\i}},
  \citenamefont {Bearpark},\ and\ \citenamefont {Tew}}]{chan2021molecular}%
  \BibitemOpen
  \bibfield  {author} {\bibinfo {author} {\bibfnamefont {H.~H.~S.}\
  \bibnamefont {Chan}}, \bibinfo {author} {\bibfnamefont {N.}~\bibnamefont
  {Fitzpatrick}}, \bibinfo {author} {\bibfnamefont {J.}~\bibnamefont
  {Segarra-Mart{\'\i}}}, \bibinfo {author} {\bibfnamefont {M.~J.}\ \bibnamefont
  {Bearpark}},\ and\ \bibinfo {author} {\bibfnamefont {D.~P.}\ \bibnamefont
  {Tew}},\ }\bibfield  {title} {\bibinfo {title} {Molecular excited state
  calculations with adaptive wavefunctions on a quantum eigensolver emulation:
  reducing circuit depth and separating spin states},\ }\href@noop {}
  {\bibfield  {journal} {\bibinfo  {journal} {Phys. Chem. Chem. Phys.}\
  }\textbf {\bibinfo {volume} {23}},\ \bibinfo {pages} {26438} (\bibinfo {year}
  {2021})}\BibitemShut {NoStop}%
\bibitem [{\citenamefont {Higgott}\ \emph {et~al.}(2019)\citenamefont
  {Higgott}, \citenamefont {Wang},\ and\ \citenamefont
  {Brierley}}]{higgott2019variational}%
  \BibitemOpen
  \bibfield  {author} {\bibinfo {author} {\bibfnamefont {O.}~\bibnamefont
  {Higgott}}, \bibinfo {author} {\bibfnamefont {D.}~\bibnamefont {Wang}},\ and\
  \bibinfo {author} {\bibfnamefont {S.}~\bibnamefont {Brierley}},\ }\bibfield
  {title} {\bibinfo {title} {Variational quantum computation of excited
  states},\ }\href@noop {} {\bibfield  {journal} {\bibinfo  {journal}
  {Quantum}\ }\textbf {\bibinfo {volume} {3}},\ \bibinfo {pages} {156}
  (\bibinfo {year} {2019})}\BibitemShut {NoStop}%
\bibitem [{\citenamefont {Tkachenko}\ \emph {et~al.}(2022)\citenamefont
  {Tkachenko}, \citenamefont {Zhang}, \citenamefont {Cincio}, \citenamefont
  {Boldyrev}, \citenamefont {Tretiak},\ and\ \citenamefont {Dub}}]{qdavidson}%
  \BibitemOpen
  \bibfield  {author} {\bibinfo {author} {\bibfnamefont {N.~V.}\ \bibnamefont
  {Tkachenko}}, \bibinfo {author} {\bibfnamefont {Y.}~\bibnamefont {Zhang}},
  \bibinfo {author} {\bibfnamefont {L.}~\bibnamefont {Cincio}}, \bibinfo
  {author} {\bibfnamefont {A.~I.}\ \bibnamefont {Boldyrev}}, \bibinfo {author}
  {\bibfnamefont {S.}~\bibnamefont {Tretiak}},\ and\ \bibinfo {author}
  {\bibfnamefont {P.~A.}\ \bibnamefont {Dub}},\ }\bibfield  {title} {\bibinfo
  {title} {Quantum davidson algorithm for excited states},\ }\href@noop {}
  {\bibfield  {journal} {\bibinfo  {journal} {arxiv preprint arXiv:2204.10741}\
  } (\bibinfo {year} {2022})}\BibitemShut {NoStop}%
\bibitem [{\citenamefont {Motta}\ \emph
  {et~al.}(2020{\natexlab{b}})\citenamefont {Motta}, \citenamefont {Sun},
  \citenamefont {Tan}, \citenamefont {O’Rourke}, \citenamefont {Ye},
  \citenamefont {Minnich}, \citenamefont {Brandão},\ and\ \citenamefont
  {Chan}}]{GarnetNP2020}%
  \BibitemOpen
  \bibfield  {author} {\bibinfo {author} {\bibfnamefont {M.}~\bibnamefont
  {Motta}}, \bibinfo {author} {\bibfnamefont {C.}~\bibnamefont {Sun}}, \bibinfo
  {author} {\bibfnamefont {A.~T.~K.}\ \bibnamefont {Tan}}, \bibinfo {author}
  {\bibfnamefont {M.~J.}\ \bibnamefont {O’Rourke}}, \bibinfo {author}
  {\bibfnamefont {E.}~\bibnamefont {Ye}}, \bibinfo {author} {\bibfnamefont
  {A.~J.}\ \bibnamefont {Minnich}}, \bibinfo {author} {\bibfnamefont {F.~G.
  S.~L.}\ \bibnamefont {Brandão}},\ and\ \bibinfo {author} {\bibfnamefont
  {G.~K.-L.}\ \bibnamefont {Chan}},\ }\bibfield  {title} {\bibinfo {title}
  {Determining eigenstates and thermal states on a quantum computer using
  quantum imaginary time evolution},\ }\href
  {https://doi.org/10.1038/s41567-019-0704-4} {\bibfield  {journal} {\bibinfo
  {journal} {Nat. Phys.}\ }\textbf {\bibinfo {volume} {16}},\ \bibinfo {pages}
  {205} (\bibinfo {year} {2020}{\natexlab{b}})}\BibitemShut {NoStop}%
\bibitem [{\citenamefont {Colless}\ \emph {et~al.}(2018)\citenamefont
  {Colless}, \citenamefont {Ramasesh}, \citenamefont {Dahlen}, \citenamefont
  {Blok}, \citenamefont {Kimchi-Schwartz}, \citenamefont {McClean},
  \citenamefont {Carter}, \citenamefont {De~Jong},\ and\ \citenamefont
  {Siddiqi}}]{colless2018computation}%
  \BibitemOpen
  \bibfield  {author} {\bibinfo {author} {\bibfnamefont {J.~I.}\ \bibnamefont
  {Colless}}, \bibinfo {author} {\bibfnamefont {V.~V.}\ \bibnamefont
  {Ramasesh}}, \bibinfo {author} {\bibfnamefont {D.}~\bibnamefont {Dahlen}},
  \bibinfo {author} {\bibfnamefont {M.~S.}\ \bibnamefont {Blok}}, \bibinfo
  {author} {\bibfnamefont {M.~E.}\ \bibnamefont {Kimchi-Schwartz}}, \bibinfo
  {author} {\bibfnamefont {J.~R.}\ \bibnamefont {McClean}}, \bibinfo {author}
  {\bibfnamefont {J.}~\bibnamefont {Carter}}, \bibinfo {author} {\bibfnamefont
  {W.~A.}\ \bibnamefont {De~Jong}},\ and\ \bibinfo {author} {\bibfnamefont
  {I.}~\bibnamefont {Siddiqi}},\ }\bibfield  {title} {\bibinfo {title}
  {Computation of molecular spectra on a quantum processor with an
  error-resilient algorithm},\ }\href@noop {} {\bibfield  {journal} {\bibinfo
  {journal} {Phys. Rev. X}\ }\textbf {\bibinfo {volume} {8}},\ \bibinfo {pages}
  {011021} (\bibinfo {year} {2018})}\BibitemShut {NoStop}%
\bibitem [{\citenamefont {McClean}\ \emph
  {et~al.}(2020{\natexlab{a}})\citenamefont {McClean}, \citenamefont {Jiang},
  \citenamefont {Rubin}, \citenamefont {Babbush},\ and\ \citenamefont
  {Neven}}]{mcclean2020decoding}%
  \BibitemOpen
  \bibfield  {author} {\bibinfo {author} {\bibfnamefont {J.~R.}\ \bibnamefont
  {McClean}}, \bibinfo {author} {\bibfnamefont {Z.}~\bibnamefont {Jiang}},
  \bibinfo {author} {\bibfnamefont {N.~C.}\ \bibnamefont {Rubin}}, \bibinfo
  {author} {\bibfnamefont {R.}~\bibnamefont {Babbush}},\ and\ \bibinfo {author}
  {\bibfnamefont {H.}~\bibnamefont {Neven}},\ }\bibfield  {title} {\bibinfo
  {title} {Decoding quantum errors with subspace expansions},\ }\href@noop {}
  {\bibfield  {journal} {\bibinfo  {journal} {Nat. Comm.}\ }\textbf {\bibinfo
  {volume} {11}},\ \bibinfo {pages} {1} (\bibinfo {year}
  {2020}{\natexlab{a}})}\BibitemShut {NoStop}%
\bibitem [{\citenamefont {Takeshita}\ \emph {et~al.}(2020)\citenamefont
  {Takeshita}, \citenamefont {Rubin}, \citenamefont {Jiang}, \citenamefont
  {Lee}, \citenamefont {Babbush},\ and\ \citenamefont
  {McClean}}]{Takeshita2020}%
  \BibitemOpen
  \bibfield  {author} {\bibinfo {author} {\bibfnamefont {T.}~\bibnamefont
  {Takeshita}}, \bibinfo {author} {\bibfnamefont {N.~C.}\ \bibnamefont
  {Rubin}}, \bibinfo {author} {\bibfnamefont {Z.}~\bibnamefont {Jiang}},
  \bibinfo {author} {\bibfnamefont {E.}~\bibnamefont {Lee}}, \bibinfo {author}
  {\bibfnamefont {R.}~\bibnamefont {Babbush}},\ and\ \bibinfo {author}
  {\bibfnamefont {J.~R.}\ \bibnamefont {McClean}},\ }\bibfield  {title}
  {\bibinfo {title} {Increasing the representation accuracy of quantum
  simulations of chemistry without extra quantum resources},\ }\href
  {https://doi.org/10.1103/PhysRevX.10.011004} {\bibfield  {journal} {\bibinfo
  {journal} {Phys. Rev. X}\ }\textbf {\bibinfo {volume} {10}},\ \bibinfo
  {pages} {011004} (\bibinfo {year} {2020})}\BibitemShut {NoStop}%
\bibitem [{\citenamefont {McClean}\ \emph {et~al.}(2017)\citenamefont
  {McClean}, \citenamefont {Kimchi-Schwartz}, \citenamefont {Carter},\ and\
  \citenamefont {De~Jong}}]{mcclean2017hybrid}%
  \BibitemOpen
  \bibfield  {author} {\bibinfo {author} {\bibfnamefont {J.~R.}\ \bibnamefont
  {McClean}}, \bibinfo {author} {\bibfnamefont {M.~E.}\ \bibnamefont
  {Kimchi-Schwartz}}, \bibinfo {author} {\bibfnamefont {J.}~\bibnamefont
  {Carter}},\ and\ \bibinfo {author} {\bibfnamefont {W.~A.}\ \bibnamefont
  {De~Jong}},\ }\bibfield  {title} {\bibinfo {title} {Hybrid quantum-classical
  hierarchy for mitigation of decoherence and determination of excited
  states},\ }\href@noop {} {\bibfield  {journal} {\bibinfo  {journal} {Phys.
  Rev. A}\ }\textbf {\bibinfo {volume} {95}},\ \bibinfo {pages} {042308}
  (\bibinfo {year} {2017})}\BibitemShut {NoStop}%
\bibitem [{\citenamefont {Ollitrault}\ \emph {et~al.}(2020)\citenamefont
  {Ollitrault}, \citenamefont {Kandala}, \citenamefont {Chen}, \citenamefont
  {Barkoutsos}, \citenamefont {Mezzacapo}, \citenamefont {Pistoia},
  \citenamefont {Sheldon}, \citenamefont {Woerner}, \citenamefont {Gambetta},\
  and\ \citenamefont {Tavernelli}}]{qEOM2020}%
  \BibitemOpen
  \bibfield  {author} {\bibinfo {author} {\bibfnamefont {P.~J.}\ \bibnamefont
  {Ollitrault}}, \bibinfo {author} {\bibfnamefont {A.}~\bibnamefont {Kandala}},
  \bibinfo {author} {\bibfnamefont {C.-F.}\ \bibnamefont {Chen}}, \bibinfo
  {author} {\bibfnamefont {P.~K.}\ \bibnamefont {Barkoutsos}}, \bibinfo
  {author} {\bibfnamefont {A.}~\bibnamefont {Mezzacapo}}, \bibinfo {author}
  {\bibfnamefont {M.}~\bibnamefont {Pistoia}}, \bibinfo {author} {\bibfnamefont
  {S.}~\bibnamefont {Sheldon}}, \bibinfo {author} {\bibfnamefont
  {S.}~\bibnamefont {Woerner}}, \bibinfo {author} {\bibfnamefont {J.~M.}\
  \bibnamefont {Gambetta}},\ and\ \bibinfo {author} {\bibfnamefont
  {I.}~\bibnamefont {Tavernelli}},\ }\bibfield  {title} {\bibinfo {title}
  {Quantum equation of motion for computing molecular excitation energies on a
  noisy quantum processor},\ }\href@noop {} {\bibfield  {journal} {\bibinfo
  {journal} {Phys. Rev. Res.}\ }\textbf {\bibinfo {volume} {2}},\ \bibinfo
  {pages} {043140} (\bibinfo {year} {2020})}\BibitemShut {NoStop}%
\bibitem [{\citenamefont {Urbanek}\ \emph {et~al.}(2020)\citenamefont
  {Urbanek}, \citenamefont {Camps}, \citenamefont {Van~Beeumen},\ and\
  \citenamefont {De~Jong}}]{urbanek2020chemistry}%
  \BibitemOpen
  \bibfield  {author} {\bibinfo {author} {\bibfnamefont {M.}~\bibnamefont
  {Urbanek}}, \bibinfo {author} {\bibfnamefont {D.}~\bibnamefont {Camps}},
  \bibinfo {author} {\bibfnamefont {R.}~\bibnamefont {Van~Beeumen}},\ and\
  \bibinfo {author} {\bibfnamefont {W.~A.}\ \bibnamefont {De~Jong}},\
  }\bibfield  {title} {\bibinfo {title} {Chemistry on quantum computers with
  virtual quantum subspace expansion},\ }\href@noop {} {\bibfield  {journal}
  {\bibinfo  {journal} {J. Chem. Theory Comput.}\ }\textbf {\bibinfo {volume}
  {16}},\ \bibinfo {pages} {5425} (\bibinfo {year} {2020})}\BibitemShut
  {NoStop}%
\bibitem [{\citenamefont {Mazziotti}(1998)}]{mazziotti1998approximate}%
  \BibitemOpen
  \bibfield  {author} {\bibinfo {author} {\bibfnamefont {D.~A.}\ \bibnamefont
  {Mazziotti}},\ }\bibfield  {title} {\bibinfo {title} {Approximate solution
  for electron correlation through the use of schwinger probes},\ }\href@noop
  {} {\bibfield  {journal} {\bibinfo  {journal} {Chem. Phys. Lett.}\ }\textbf
  {\bibinfo {volume} {289}},\ \bibinfo {pages} {419} (\bibinfo {year}
  {1998})}\BibitemShut {NoStop}%
\bibitem [{\citenamefont {Prasad}\ \emph {et~al.}(1985)\citenamefont {Prasad},
  \citenamefont {Pal},\ and\ \citenamefont {Mukherjee}}]{prasad1985some}%
  \BibitemOpen
  \bibfield  {author} {\bibinfo {author} {\bibfnamefont {M.~D.}\ \bibnamefont
  {Prasad}}, \bibinfo {author} {\bibfnamefont {S.}~\bibnamefont {Pal}},\ and\
  \bibinfo {author} {\bibfnamefont {D.}~\bibnamefont {Mukherjee}},\ }\bibfield
  {title} {\bibinfo {title} {Some aspects of self-consistent propagator
  theories},\ }\href@noop {} {\bibfield  {journal} {\bibinfo  {journal} {Phys.
  Rev. A}\ }\textbf {\bibinfo {volume} {31}},\ \bibinfo {pages} {1287}
  (\bibinfo {year} {1985})}\BibitemShut {NoStop}%
\bibitem [{\citenamefont {Szekeres}\ \emph {et~al.}(2001)\citenamefont
  {Szekeres}, \citenamefont {Szabados}, \citenamefont {K{\'a}llay},\ and\
  \citenamefont {Surj{\'a}n}}]{szekeres2001killer}%
  \BibitemOpen
  \bibfield  {author} {\bibinfo {author} {\bibfnamefont {Z.}~\bibnamefont
  {Szekeres}}, \bibinfo {author} {\bibfnamefont {{\'A}.}~\bibnamefont
  {Szabados}}, \bibinfo {author} {\bibfnamefont {M.}~\bibnamefont
  {K{\'a}llay}},\ and\ \bibinfo {author} {\bibfnamefont {P.~R.}\ \bibnamefont
  {Surj{\'a}n}},\ }\bibfield  {title} {\bibinfo {title} {On the “killer
  condition” in the equation-of-motion method: ionization potentials from
  multi-reference wave functions},\ }\href@noop {} {\bibfield  {journal}
  {\bibinfo  {journal} {Phys. Chem. Chem. Phys.}\ }\textbf {\bibinfo {volume}
  {3}},\ \bibinfo {pages} {696} (\bibinfo {year} {2001})}\BibitemShut {NoStop}%
\bibitem [{\citenamefont {Mertins}\ \emph {et~al.}(1996)\citenamefont
  {Mertins}, \citenamefont {Schirmer},\ and\ \citenamefont
  {Tarantelli}}]{mertins1996algebraic}%
  \BibitemOpen
  \bibfield  {author} {\bibinfo {author} {\bibfnamefont {F.}~\bibnamefont
  {Mertins}}, \bibinfo {author} {\bibfnamefont {J.}~\bibnamefont {Schirmer}},\
  and\ \bibinfo {author} {\bibfnamefont {A.}~\bibnamefont {Tarantelli}},\
  }\bibfield  {title} {\bibinfo {title} {Algebraic propagator approaches and
  intermediate-state representations. ii. the equation-of-motion methods for n,
  n$\pm$1, and n$\pm$2 electrons},\ }\href@noop {} {\bibfield  {journal}
  {\bibinfo  {journal} {Phys. Rev. A}\ }\textbf {\bibinfo {volume} {53}},\
  \bibinfo {pages} {2153} (\bibinfo {year} {1996})}\BibitemShut {NoStop}%
\bibitem [{\citenamefont {Weiner}\ and\ \citenamefont
  {Goscinski}(1980)}]{weiner1980calculation}%
  \BibitemOpen
  \bibfield  {author} {\bibinfo {author} {\bibfnamefont {B.}~\bibnamefont
  {Weiner}}\ and\ \bibinfo {author} {\bibfnamefont {O.}~\bibnamefont
  {Goscinski}},\ }\bibfield  {title} {\bibinfo {title} {Calculation of optimal
  generalized antisymmetrized geminal-power
  (projected—bardeen-cooper-schrieffer) functions and their associated
  excitation spectrum},\ }\href@noop {} {\bibfield  {journal} {\bibinfo
  {journal} {Phys. Rev. A}\ }\textbf {\bibinfo {volume} {22}},\ \bibinfo
  {pages} {2374} (\bibinfo {year} {1980})}\BibitemShut {NoStop}%
\bibitem [{\citenamefont {Hodecker}\ and\ \citenamefont
  {Dreuw}(2020)}]{hodecker2020unitary}%
  \BibitemOpen
  \bibfield  {author} {\bibinfo {author} {\bibfnamefont {M.}~\bibnamefont
  {Hodecker}}\ and\ \bibinfo {author} {\bibfnamefont {A.}~\bibnamefont
  {Dreuw}},\ }\bibfield  {title} {\bibinfo {title} {Unitary coupled cluster
  ground-and excited-state molecular properties},\ }\href@noop {} {\bibfield
  {journal} {\bibinfo  {journal} {J. Chem. Phys.}\ }\textbf {\bibinfo {volume}
  {153}},\ \bibinfo {pages} {084112} (\bibinfo {year} {2020})}\BibitemShut
  {NoStop}%
\bibitem [{\citenamefont {Levchenko}\ and\ \citenamefont
  {Krylov}(2004)}]{levchenko2004equation}%
  \BibitemOpen
  \bibfield  {author} {\bibinfo {author} {\bibfnamefont {S.~V.}\ \bibnamefont
  {Levchenko}}\ and\ \bibinfo {author} {\bibfnamefont {A.~I.}\ \bibnamefont
  {Krylov}},\ }\bibfield  {title} {\bibinfo {title} {Equation-of-motion
  spin-flip coupled-cluster model with single and double substitutions: Theory
  and application to cyclobutadiene},\ }\href@noop {} {\bibfield  {journal}
  {\bibinfo  {journal} {J. Chem. Phys.}\ }\textbf {\bibinfo {volume} {120}},\
  \bibinfo {pages} {175} (\bibinfo {year} {2004})}\BibitemShut {NoStop}%
\bibitem [{\citenamefont {Datta}\ \emph
  {et~al.}(1993{\natexlab{a}})\citenamefont {Datta}, \citenamefont
  {Mukhopadhyay},\ and\ \citenamefont {Mukherjee}}]{datta1993}%
  \BibitemOpen
  \bibfield  {author} {\bibinfo {author} {\bibfnamefont {B.}~\bibnamefont
  {Datta}}, \bibinfo {author} {\bibfnamefont {D.}~\bibnamefont
  {Mukhopadhyay}},\ and\ \bibinfo {author} {\bibfnamefont {D.}~\bibnamefont
  {Mukherjee}},\ }\bibfield  {title} {\bibinfo {title} {Consistent propagator
  theory based on the extended coupled-cluster parametrization of the ground
  state},\ }\href@noop {} {\bibfield  {journal} {\bibinfo  {journal} {Phys.
  Rev. A}\ }\textbf {\bibinfo {volume} {47}},\ \bibinfo {pages} {3632}
  (\bibinfo {year} {1993}{\natexlab{a}})}\BibitemShut {NoStop}%
\bibitem [{\citenamefont {Liu}\ \emph {et~al.}(2018)\citenamefont {Liu},
  \citenamefont {Asthana}, \citenamefont {Cheng},\ and\ \citenamefont
  {Mukherjee}}]{liu2018unitary}%
  \BibitemOpen
  \bibfield  {author} {\bibinfo {author} {\bibfnamefont {J.}~\bibnamefont
  {Liu}}, \bibinfo {author} {\bibfnamefont {A.}~\bibnamefont {Asthana}},
  \bibinfo {author} {\bibfnamefont {L.}~\bibnamefont {Cheng}},\ and\ \bibinfo
  {author} {\bibfnamefont {D.}~\bibnamefont {Mukherjee}},\ }\bibfield  {title}
  {\bibinfo {title} {Unitary coupled-cluster based self-consistent polarization
  propagator theory: A third-order formulation and pilot applications},\
  }\href@noop {} {\bibfield  {journal} {\bibinfo  {journal} {J. Chem. Phys.}\
  }\textbf {\bibinfo {volume} {148}},\ \bibinfo {pages} {244110} (\bibinfo
  {year} {2018})}\BibitemShut {NoStop}%
\bibitem [{\citenamefont {Davidson}(1975)}]{davidsorq1975theiterative}%
  \BibitemOpen
  \bibfield  {author} {\bibinfo {author} {\bibfnamefont {E.}~\bibnamefont
  {Davidson}},\ }\bibfield  {title} {\bibinfo {title} {The iterative
  calculation of a few of the lowest eigenvalues and corresponding eigenvectors
  of large real-symmetric matrices},\ }\href@noop {} {\bibfield  {journal}
  {\bibinfo  {journal} {J. Comput. Phys}\ }\textbf {\bibinfo {volume} {17}},\
  \bibinfo {pages} {87} (\bibinfo {year} {1975})}\BibitemShut {NoStop}%
\bibitem [{\citenamefont {Liu}(1978)}]{liu1978simultaneous}%
  \BibitemOpen
  \bibfield  {author} {\bibinfo {author} {\bibfnamefont {B.}~\bibnamefont
  {Liu}},\ }\bibfield  {title} {\bibinfo {title} {The simultaneous expansion
  method for the iterative solution of several of the lowest eigenvalues and
  corresponding eigenvectors of large real-symmetric matrices},\ }\href@noop {}
  {\bibfield  {journal} {\bibinfo  {journal} {Num. Alg. Chem.}\ ,\ \bibinfo
  {pages} {49}} (\bibinfo {year} {1978})}\BibitemShut {NoStop}%
\bibitem [{\citenamefont {Tretiak}\ \emph {et~al.}(2009)\citenamefont
  {Tretiak}, \citenamefont {Isborn}, \citenamefont {Niklasson},\ and\
  \citenamefont {Challacombe}}]{tretiak2009representation}%
  \BibitemOpen
  \bibfield  {author} {\bibinfo {author} {\bibfnamefont {S.}~\bibnamefont
  {Tretiak}}, \bibinfo {author} {\bibfnamefont {C.~M.}\ \bibnamefont {Isborn}},
  \bibinfo {author} {\bibfnamefont {A.~M.}\ \bibnamefont {Niklasson}},\ and\
  \bibinfo {author} {\bibfnamefont {M.}~\bibnamefont {Challacombe}},\
  }\bibfield  {title} {\bibinfo {title} {Representation independent algorithms
  for molecular response calculations in time-dependent self-consistent field
  theories},\ }\href@noop {} {\bibfield  {journal} {\bibinfo  {journal} {J.
  Chem. Phys.}\ }\textbf {\bibinfo {volume} {130}},\ \bibinfo {pages} {054111}
  (\bibinfo {year} {2009})}\BibitemShut {NoStop}%
\bibitem [{\citenamefont {Matthews}\ and\ \citenamefont
  {Stanton}(2016)}]{matthews2016new}%
  \BibitemOpen
  \bibfield  {author} {\bibinfo {author} {\bibfnamefont {D.~A.}\ \bibnamefont
  {Matthews}}\ and\ \bibinfo {author} {\bibfnamefont {J.~F.}\ \bibnamefont
  {Stanton}},\ }\bibfield  {title} {\bibinfo {title} {A new approach to
  approximate equation-of-motion coupled cluster with triple excitations},\
  }\href@noop {} {\bibfield  {journal} {\bibinfo  {journal} {J. Chem. Phys.}\
  }\textbf {\bibinfo {volume} {145}},\ \bibinfo {pages} {124102} (\bibinfo
  {year} {2016})}\BibitemShut {NoStop}%
\bibitem [{\citenamefont {McClean}\ \emph
  {et~al.}(2020{\natexlab{b}})\citenamefont {McClean}, \citenamefont {Rubin},
  \citenamefont {Sung}, \citenamefont {Kivlichan}, \citenamefont
  {Bonet-Monroig}, \citenamefont {Cao}, \citenamefont {Dai}, \citenamefont
  {Fried}, \citenamefont {Gidney}, \citenamefont {Gimby}, \citenamefont
  {Gokhale}, \citenamefont {Häner}, \citenamefont {Hardikar}, \citenamefont
  {Havl{\'{\i}}{\v{c}}ek}, \citenamefont {Higgott}, \citenamefont {Huang},
  \citenamefont {Izaac}, \citenamefont {Jiang}, \citenamefont {Liu},
  \citenamefont {McArdle}, \citenamefont {Neeley}, \citenamefont {O'Brien},
  \citenamefont {O'Gorman}, \citenamefont {Ozfidan}, \citenamefont {Radin},
  \citenamefont {Romero}, \citenamefont {Sawaya}, \citenamefont {Senjean},
  \citenamefont {Setia}, \citenamefont {Sim}, \citenamefont {Steiger},
  \citenamefont {Steudtner}, \citenamefont {Sun}, \citenamefont {Sun},
  \citenamefont {Wang}, \citenamefont {Zhang},\ and\ \citenamefont
  {Babbush}}]{openfermion}%
  \BibitemOpen
  \bibfield  {author} {\bibinfo {author} {\bibfnamefont {J.~R.}\ \bibnamefont
  {McClean}}, \bibinfo {author} {\bibfnamefont {N.~C.}\ \bibnamefont {Rubin}},
  \bibinfo {author} {\bibfnamefont {K.~J.}\ \bibnamefont {Sung}}, \bibinfo
  {author} {\bibfnamefont {I.~D.}\ \bibnamefont {Kivlichan}}, \bibinfo {author}
  {\bibfnamefont {X.}~\bibnamefont {Bonet-Monroig}}, \bibinfo {author}
  {\bibfnamefont {Y.}~\bibnamefont {Cao}}, \bibinfo {author} {\bibfnamefont
  {C.}~\bibnamefont {Dai}}, \bibinfo {author} {\bibfnamefont {E.~S.}\
  \bibnamefont {Fried}}, \bibinfo {author} {\bibfnamefont {C.}~\bibnamefont
  {Gidney}}, \bibinfo {author} {\bibfnamefont {B.}~\bibnamefont {Gimby}},
  \bibinfo {author} {\bibfnamefont {P.}~\bibnamefont {Gokhale}}, \bibinfo
  {author} {\bibfnamefont {T.}~\bibnamefont {Häner}}, \bibinfo {author}
  {\bibfnamefont {T.}~\bibnamefont {Hardikar}}, \bibinfo {author}
  {\bibfnamefont {V.}~\bibnamefont {Havl{\'{\i}}{\v{c}}ek}}, \bibinfo {author}
  {\bibfnamefont {O.}~\bibnamefont {Higgott}}, \bibinfo {author} {\bibfnamefont
  {C.}~\bibnamefont {Huang}}, \bibinfo {author} {\bibfnamefont
  {J.}~\bibnamefont {Izaac}}, \bibinfo {author} {\bibfnamefont
  {Z.}~\bibnamefont {Jiang}}, \bibinfo {author} {\bibfnamefont
  {X.}~\bibnamefont {Liu}}, \bibinfo {author} {\bibfnamefont {S.}~\bibnamefont
  {McArdle}}, \bibinfo {author} {\bibfnamefont {M.}~\bibnamefont {Neeley}},
  \bibinfo {author} {\bibfnamefont {T.}~\bibnamefont {O'Brien}}, \bibinfo
  {author} {\bibfnamefont {B.}~\bibnamefont {O'Gorman}}, \bibinfo {author}
  {\bibfnamefont {I.}~\bibnamefont {Ozfidan}}, \bibinfo {author} {\bibfnamefont
  {M.~D.}\ \bibnamefont {Radin}}, \bibinfo {author} {\bibfnamefont
  {J.}~\bibnamefont {Romero}}, \bibinfo {author} {\bibfnamefont {N.~P.~D.}\
  \bibnamefont {Sawaya}}, \bibinfo {author} {\bibfnamefont {B.}~\bibnamefont
  {Senjean}}, \bibinfo {author} {\bibfnamefont {K.}~\bibnamefont {Setia}},
  \bibinfo {author} {\bibfnamefont {S.}~\bibnamefont {Sim}}, \bibinfo {author}
  {\bibfnamefont {D.~S.}\ \bibnamefont {Steiger}}, \bibinfo {author}
  {\bibfnamefont {M.}~\bibnamefont {Steudtner}}, \bibinfo {author}
  {\bibfnamefont {Q.}~\bibnamefont {Sun}}, \bibinfo {author} {\bibfnamefont
  {W.}~\bibnamefont {Sun}}, \bibinfo {author} {\bibfnamefont {D.}~\bibnamefont
  {Wang}}, \bibinfo {author} {\bibfnamefont {F.}~\bibnamefont {Zhang}},\ and\
  \bibinfo {author} {\bibfnamefont {R.}~\bibnamefont {Babbush}},\ }\bibfield
  {title} {\bibinfo {title} {{OpenFermion}: the electronic structure package
  for quantum computers},\ }\href {https://doi.org/10.1088/2058-9565/ab8ebc}
  {\bibfield  {journal} {\bibinfo  {journal} {Quantum Science and Technology}\
  }\textbf {\bibinfo {volume} {5}},\ \bibinfo {pages} {034014} (\bibinfo {year}
  {2020}{\natexlab{b}})}\BibitemShut {NoStop}%
\bibitem [{\citenamefont {Aleksandrowicz}\ \emph {et~al.}(2019)\citenamefont
  {Aleksandrowicz}, \citenamefont {Alexander}, \citenamefont {Barkoutsos},
  \citenamefont {Bello}, \citenamefont {Ben-Haim}, \citenamefont {Bucher},
  \citenamefont {Cabrera-Hern{\'a}ndez}, \citenamefont {Carballo-Franquis},
  \citenamefont {Chen}, \citenamefont {Chen} \emph
  {et~al.}}]{aleksandrowicz2019qiskit}%
  \BibitemOpen
  \bibfield  {author} {\bibinfo {author} {\bibfnamefont {G.}~\bibnamefont
  {Aleksandrowicz}}, \bibinfo {author} {\bibfnamefont {T.}~\bibnamefont
  {Alexander}}, \bibinfo {author} {\bibfnamefont {P.}~\bibnamefont
  {Barkoutsos}}, \bibinfo {author} {\bibfnamefont {L.}~\bibnamefont {Bello}},
  \bibinfo {author} {\bibfnamefont {Y.}~\bibnamefont {Ben-Haim}}, \bibinfo
  {author} {\bibfnamefont {D.}~\bibnamefont {Bucher}}, \bibinfo {author}
  {\bibfnamefont {F.~J.}\ \bibnamefont {Cabrera-Hern{\'a}ndez}}, \bibinfo
  {author} {\bibfnamefont {J.}~\bibnamefont {Carballo-Franquis}}, \bibinfo
  {author} {\bibfnamefont {A.}~\bibnamefont {Chen}}, \bibinfo {author}
  {\bibfnamefont {C.-F.}\ \bibnamefont {Chen}}, \emph {et~al.},\ }\bibfield
  {title} {\bibinfo {title} {Qiskit: An open-source framework for quantum
  computing},\ }\href@noop {} {\bibfield  {journal} {\bibinfo  {journal}
  {Accessed on: Mar}\ }\textbf {\bibinfo {volume} {16}} (\bibinfo {year}
  {2019})}\BibitemShut {NoStop}%
\bibitem [{\citenamefont {Asthana}\ \emph {et~al.}(2022)\citenamefont
  {Asthana}, \citenamefont {Grimsley}, \citenamefont {Kumar}, \citenamefont
  {Abraham},\ and\ \citenamefont {Mayhall}}]{giteom}%
  \BibitemOpen
  \bibfield  {author} {\bibinfo {author} {\bibfnamefont {A.}~\bibnamefont
  {Asthana}}, \bibinfo {author} {\bibfnamefont {H.~R.}\ \bibnamefont
  {Grimsley}}, \bibinfo {author} {\bibfnamefont {A.}~\bibnamefont {Kumar}},
  \bibinfo {author} {\bibfnamefont {V.}~\bibnamefont {Abraham}},\ and\ \bibinfo
  {author} {\bibfnamefont {N.~J.}\ \bibnamefont {Mayhall}},\ }\bibfield
  {title} {\bibinfo {title} {Adapt-vqe:
  https://github.com/asthanaa/adapt-vqe},\ }\href@noop {} {\bibfield  {journal}
  {\bibinfo  {journal} {Github repository}\ } (\bibinfo {year}
  {2022})}\BibitemShut {NoStop}%
\bibitem [{\citenamefont {Hirata}\ \emph {et~al.}(2000)\citenamefont {Hirata},
  \citenamefont {Nooijen},\ and\ \citenamefont {Bartlett}}]{hirata2000high}%
  \BibitemOpen
  \bibfield  {author} {\bibinfo {author} {\bibfnamefont {S.}~\bibnamefont
  {Hirata}}, \bibinfo {author} {\bibfnamefont {M.}~\bibnamefont {Nooijen}},\
  and\ \bibinfo {author} {\bibfnamefont {R.~J.}\ \bibnamefont {Bartlett}},\
  }\bibfield  {title} {\bibinfo {title} {High-order determinantal
  equation-of-motion coupled-cluster calculations for ionized and
  electron-attached states},\ }\href@noop {} {\bibfield  {journal} {\bibinfo
  {journal} {Chem. Phys. Lett.}\ }\textbf {\bibinfo {volume} {328}},\ \bibinfo
  {pages} {459} (\bibinfo {year} {2000})}\BibitemShut {NoStop}%
\bibitem [{\citenamefont {Dutta}\ \emph {et~al.}(2018)\citenamefont {Dutta},
  \citenamefont {Vaval},\ and\ \citenamefont {Pal}}]{dutta2018lower}%
  \BibitemOpen
  \bibfield  {author} {\bibinfo {author} {\bibfnamefont {A.~K.}\ \bibnamefont
  {Dutta}}, \bibinfo {author} {\bibfnamefont {N.}~\bibnamefont {Vaval}},\ and\
  \bibinfo {author} {\bibfnamefont {S.}~\bibnamefont {Pal}},\ }\bibfield
  {title} {\bibinfo {title} {Lower scaling approximation to eom-ccsd: A
  critical assessment of the ionization problem},\ }\href@noop {} {\bibfield
  {journal} {\bibinfo  {journal} {Int. J. Quantum Chem.}\ }\textbf {\bibinfo
  {volume} {118}},\ \bibinfo {pages} {e25594} (\bibinfo {year}
  {2018})}\BibitemShut {NoStop}%
\bibitem [{\citenamefont {Fan}\ \emph {et~al.}(2021)\citenamefont {Fan},
  \citenamefont {Liu}, \citenamefont {Li},\ and\ \citenamefont
  {Yang}}]{fan2021equation}%
  \BibitemOpen
  \bibfield  {author} {\bibinfo {author} {\bibfnamefont {Y.}~\bibnamefont
  {Fan}}, \bibinfo {author} {\bibfnamefont {J.}~\bibnamefont {Liu}}, \bibinfo
  {author} {\bibfnamefont {Z.}~\bibnamefont {Li}},\ and\ \bibinfo {author}
  {\bibfnamefont {J.}~\bibnamefont {Yang}},\ }\bibfield  {title} {\bibinfo
  {title} {Equation-of-motion theory to calculate accurate band structures with
  a quantum computer},\ }\href@noop {} {\bibfield  {journal} {\bibinfo
  {journal} {J. Phys. Chem. Lett.}\ }\textbf {\bibinfo {volume} {12}},\
  \bibinfo {pages} {8833} (\bibinfo {year} {2021})}\BibitemShut {NoStop}%
\bibitem [{\citenamefont {Liu}\ \emph {et~al.}(2022)\citenamefont {Liu},
  \citenamefont {Matthews},\ and\ \citenamefont {Cheng}}]{liu2022quadratic}%
  \BibitemOpen
  \bibfield  {author} {\bibinfo {author} {\bibfnamefont {J.}~\bibnamefont
  {Liu}}, \bibinfo {author} {\bibfnamefont {D.~A.}\ \bibnamefont {Matthews}},\
  and\ \bibinfo {author} {\bibfnamefont {L.}~\bibnamefont {Cheng}},\ }\bibfield
   {title} {\bibinfo {title} {Quadratic unitary coupled-cluster singles and
  doubles scheme: Efficient implementation, benchmark study, and formulation of
  an extended version},\ }\href@noop {} {\bibfield  {journal} {\bibinfo
  {journal} {J. Chem. Theory Comput.}\ }\textbf {\bibinfo {volume} {18}},\
  \bibinfo {pages} {2281} (\bibinfo {year} {2022})}\BibitemShut {NoStop}%
\bibitem [{\citenamefont {Hodecker}\ \emph {et~al.}(2022)\citenamefont
  {Hodecker}, \citenamefont {Dempwolff}, \citenamefont {Schirmer},\ and\
  \citenamefont {Dreuw}}]{hodecker2022theoretical}%
  \BibitemOpen
  \bibfield  {author} {\bibinfo {author} {\bibfnamefont {M.}~\bibnamefont
  {Hodecker}}, \bibinfo {author} {\bibfnamefont {A.~L.}\ \bibnamefont
  {Dempwolff}}, \bibinfo {author} {\bibfnamefont {J.}~\bibnamefont
  {Schirmer}},\ and\ \bibinfo {author} {\bibfnamefont {A.}~\bibnamefont
  {Dreuw}},\ }\bibfield  {title} {\bibinfo {title} {Theoretical analysis and
  comparison of unitary coupled-cluster and algebraic-diagrammatic construction
  methods for ionization},\ }\href@noop {} {\bibfield  {journal} {\bibinfo
  {journal} {J. Chem. Phys.}\ }\textbf {\bibinfo {volume} {156}},\ \bibinfo
  {pages} {074104} (\bibinfo {year} {2022})}\BibitemShut {NoStop}%
\bibitem [{\citenamefont {Datta}\ \emph
  {et~al.}(1993{\natexlab{b}})\citenamefont {Datta}, \citenamefont
  {Mukhopadhyay},\ and\ \citenamefont {Mukherjee}}]{datta1993consistent}%
  \BibitemOpen
  \bibfield  {author} {\bibinfo {author} {\bibfnamefont {B.}~\bibnamefont
  {Datta}}, \bibinfo {author} {\bibfnamefont {D.}~\bibnamefont
  {Mukhopadhyay}},\ and\ \bibinfo {author} {\bibfnamefont {D.}~\bibnamefont
  {Mukherjee}},\ }\bibfield  {title} {\bibinfo {title} {Consistent propagator
  theory based on the extended coupled-cluster parametrization of the ground
  state},\ }\href@noop {} {\bibfield  {journal} {\bibinfo  {journal} {Phys.
  Rev. A}\ }\textbf {\bibinfo {volume} {47}},\ \bibinfo {pages} {3632}
  (\bibinfo {year} {1993}{\natexlab{b}})}\BibitemShut {NoStop}%
\bibitem [{\citenamefont {Huggins}\ \emph {et~al.}(2021)\citenamefont
  {Huggins}, \citenamefont {McClean}, \citenamefont {Rubin}, \citenamefont
  {Jiang}, \citenamefont {Wiebe}, \citenamefont {Whaley},\ and\ \citenamefont
  {Babbush}}]{huggins2021efficient}%
  \BibitemOpen
  \bibfield  {author} {\bibinfo {author} {\bibfnamefont {W.~J.}\ \bibnamefont
  {Huggins}}, \bibinfo {author} {\bibfnamefont {J.~R.}\ \bibnamefont
  {McClean}}, \bibinfo {author} {\bibfnamefont {N.~C.}\ \bibnamefont {Rubin}},
  \bibinfo {author} {\bibfnamefont {Z.}~\bibnamefont {Jiang}}, \bibinfo
  {author} {\bibfnamefont {N.}~\bibnamefont {Wiebe}}, \bibinfo {author}
  {\bibfnamefont {K.~B.}\ \bibnamefont {Whaley}},\ and\ \bibinfo {author}
  {\bibfnamefont {R.}~\bibnamefont {Babbush}},\ }\bibfield  {title} {\bibinfo
  {title} {Efficient and noise resilient measurements for quantum chemistry on
  near-term quantum computers},\ }\href@noop {} {\bibfield  {journal} {\bibinfo
   {journal} {npj Quant. Info.}\ }\textbf {\bibinfo {volume} {7}},\ \bibinfo
  {pages} {1} (\bibinfo {year} {2021})}\BibitemShut {NoStop}%
\bibitem [{\citenamefont {Kumar}\ \emph {et~al.}(2022)\citenamefont {Kumar},
  \citenamefont {Asthana}, \citenamefont {Masteran}, \citenamefont {Valeev},
  \citenamefont {Zhang}, \citenamefont {Cincio}, \citenamefont {Tretiak},\ and\
  \citenamefont {Dub}}]{kumar2022accurate}%
  \BibitemOpen
  \bibfield  {author} {\bibinfo {author} {\bibfnamefont {A.}~\bibnamefont
  {Kumar}}, \bibinfo {author} {\bibfnamefont {A.}~\bibnamefont {Asthana}},
  \bibinfo {author} {\bibfnamefont {C.}~\bibnamefont {Masteran}}, \bibinfo
  {author} {\bibfnamefont {E.~F.}\ \bibnamefont {Valeev}}, \bibinfo {author}
  {\bibfnamefont {Y.}~\bibnamefont {Zhang}}, \bibinfo {author} {\bibfnamefont
  {L.}~\bibnamefont {Cincio}}, \bibinfo {author} {\bibfnamefont
  {S.}~\bibnamefont {Tretiak}},\ and\ \bibinfo {author} {\bibfnamefont {P.~A.}\
  \bibnamefont {Dub}},\ }\bibfield  {title} {\bibinfo {title} {Accurate quantum
  simulation of molecular ground and excited states with a transcorrelated
  hamiltonian},\ }\href@noop {} {\bibfield  {journal} {\bibinfo  {journal}
  {arXiv preprint arXiv:2201.09852}\ } (\bibinfo {year} {2022})}\BibitemShut
  {NoStop}%
\bibitem [{\citenamefont {Motta}\ \emph
  {et~al.}(2020{\natexlab{c}})\citenamefont {Motta}, \citenamefont {Gujarati},
  \citenamefont {Rice}, \citenamefont {Kumar}, \citenamefont {Masteran},
  \citenamefont {Latone}, \citenamefont {Lee}, \citenamefont {Valeev},\ and\
  \citenamefont {Takeshita}}]{Motta_CTF12_quantum}%
  \BibitemOpen
  \bibfield  {author} {\bibinfo {author} {\bibfnamefont {M.}~\bibnamefont
  {Motta}}, \bibinfo {author} {\bibfnamefont {T.~P.}\ \bibnamefont {Gujarati}},
  \bibinfo {author} {\bibfnamefont {J.~E.}\ \bibnamefont {Rice}}, \bibinfo
  {author} {\bibfnamefont {A.}~\bibnamefont {Kumar}}, \bibinfo {author}
  {\bibfnamefont {C.}~\bibnamefont {Masteran}}, \bibinfo {author}
  {\bibfnamefont {J.~A.}\ \bibnamefont {Latone}}, \bibinfo {author}
  {\bibfnamefont {E.}~\bibnamefont {Lee}}, \bibinfo {author} {\bibfnamefont
  {E.~F.}\ \bibnamefont {Valeev}},\ and\ \bibinfo {author} {\bibfnamefont
  {T.~Y.}\ \bibnamefont {Takeshita}},\ }\bibfield  {title} {\bibinfo {title}
  {Quantum simulation of electronic structure with a transcorrelated
  hamiltonian: improved accuracy with a smaller footprint on the quantum
  computer},\ }\href@noop {} {\bibfield  {journal} {\bibinfo  {journal} {Phys.
  Chem. Chem. Phys.}\ }\textbf {\bibinfo {volume} {22}},\ \bibinfo {pages}
  {24270} (\bibinfo {year} {2020}{\natexlab{c}})}\BibitemShut {NoStop}%
\end{thebibliography}%
\end{document}